\documentclass[a4paper,titlepage,12pt]{article}
\pdfoutput=1
\usepackage{graphicx} 
\usepackage[utf8]{inputenc}
\usepackage{geometry}
\usepackage[english]{babel}
\usepackage{amsmath, amsfonts, graphicx}
\usepackage{mathtools}
\usepackage{graphicx}
\usepackage{float}
\usepackage{mciteplus}
\usepackage{bbold}
\usepackage{pstricks}
\usepackage{xcolor}
\usepackage[labelfont=bf,font={small}]{caption}
\usepackage{pgfplots}
\usepackage{empheq}
\usepackage{tikz}
\usetikzlibrary{positioning, arrows}
\usetikzlibrary{external}
\usepackage{footmisc}
\usepackage{multirow}
\usepackage{url}
\usepackage[breakable, theorems, skins]{tcolorbox}
\usepackage{enumerate}
\usepackage{physics}
\usepackage{bm}
\usepackage{braket}
\usepackage{epstopdf}
\usepackage{hyperref}
\usepackage[OT2,T1]{fontenc}
\DeclareSymbolFont{cyrletters}{OT2}{wncyr}{m}{n}
\DeclareMathSymbol{\Sha}{\mathalpha}{cyrletters}{"58}
\date{June 2024}

\definecolor{darkblue}{RGB}{0,0,139}

\hypersetup{
    colorlinks,
    citecolor=blue,
    filecolor=blue,
    linkcolor=blue,
    urlcolor=blue
}

	\numberwithin{equation}{section}
\begin{document}

\begin{titlepage}

\setcounter{page}{1} \baselineskip=15.5pt \thispagestyle{empty}

\vfil

${}$
\vspace{1cm}

\begin{center}

\def\thefootnote{\fnsymbol{footnote}}
\begin{changemargin}{0.05cm}{0.05cm} 
\begin{center}
{\Large \bf Quantum group origins of edge states in double-scaled SYK}
\end{center} 
\end{changemargin}

~\\[1cm]
{Andreas Belaey,\footnote{{\protect\path{andreas.belaey@ugent.be}}} Thomas G. Mertens,\footnote{{\protect\path{thomas.mertens@ugent.be}}} Thomas Tappeiner,\footnote{{\protect\path{thomas.tappeiner@ugent.be}}}}
\\[0.3cm]
{\normalsize { \sl Department of Physics and Astronomy
\\[1.0mm]
Ghent University, Krijgslaan, 281-S9, 9000 Gent, Belgium}}\\[3mm]

\end{center}

 \vspace{0.2cm}
\begin{changemargin}{01cm}{1cm} 
{\small  \noindent 
\begin{center} 
\textbf{Abstract}
\end{center} 
Double-scaled SYK (DSSYK) is known to have an underlying quantum group theoretical description. We precisely pinpoint the quantum group structure, improving upon earlier work in the literature. This allows us to utilize this framework for bulk gravitational applications. We explain bulk discretization in DSSYK from the underlying irreducibility of the representations. We derive trumpet and brane amplitudes using character insertions of the quantum group, simplifying earlier calculations. Most importantly, we factorize the bulk Hilbert space dual to DSSYK in the quantum group description using a complete set of edge degrees of freedom living at a bulk entangling surface. An analogous treatment for $\mathcal{N}=1$ DSSYK is provided in the same quantum group theoretical framework.
}
\end{changemargin}
 \vspace{0.3cm}
\vfil
\begin{flushleft}
\today
\end{flushleft}

\end{titlepage}

\tableofcontents

\setcounter{footnote}{0}

\section{Introduction}

Factorization of the bulk Hilbert space is of considerable interest in field theory and gravity alike. The obstruction in particular to factorization in gauge theory and gravity has led to the introduction of an extended Hilbert space and edge sectors, see e.g. \cite{Buividovich:2008gq,Casini:2013rba,Donnelly:2014gva,Lin:2018bud,Ghosh:2015iwa} for early work in gauge theories. 
Within such effective quantum theories, these edge states produce auxiliary quantum numbers and superselection sectors in the Hilbert space, and effectively incorporate charges to which the gauge theory couples without explicitly introducing microscopic matter degrees of freedom. Ultimately, their deeper meaning would have to come from a deeper UV complete microscopic theory, but it is to some extent surprising that the IR theory knows about them at all \cite{Harlow:2015lma}. Nonetheless, when applied to pure gravity itself and suitably interpreted, they allow for a counting of states that matches the Bekenstein-Hawking entropy (see e.g. \cite{McGough:2013gka,Lin:2018xkj,Blommaert:2018iqz,Mertens:2022ujr,Wong:2022eiu} for work along these lines).

In recent years, the double-scaling limit of the SYK model has attracted considerable attention \cite{Berkooz:2018qkz,Berkooz:2018jqr,Lin:2022rbf,Jafferis:2022wez,Susskind:2022bia,Bhattacharjee:2022ave,Blommaert:2023opb,Blommaert:2023wad,Susskind:2023hnj,Mukhametzhanov:2023tcg,Berkooz:2023cqc,Okuyama:2023bch,Lin:2022nss,Berkooz:2022mfk,Goel:2023svz,Narovlansky:2023lfz,Verlinde:2024zrh,Berkooz:2024evs,Lin:2023trc,Verlinde:2024znh,Almheiri:2024ayc,Almheiri:2024xtw,Bossi:2024ffa,Xu:2024hoc,Xu:2024gfm,Heller:2024ldz,Tietto:2025oxn} (see \cite{Berkooz:2024lgq} for a recent review) as a model that is some sense ``halfway'' between the usual effective gravitational quantum theories (like JT gravity, 3d gravity, and Liouville gravities including the Virasoro minimal string \cite{Collier:2023cyw}), and truly microscopic models with a finite number of degrees of freedom. Crucially, it implements a bounded (albeit continuous) energy spectrum. This boundedness in turn is related to a discretization of the geometry that emerges in the dual gravitational bulk, usually phrased in terms of the integer chord number \cite{Berkooz:2018qkz,Berkooz:2018jqr,Lin:2022rbf}.\footnote{The recent complex Liouville string \cite{Collier:2024kmo} also has a bounded energy spectrum on the disk, but no apparent discretization in the bulk.} Finding independent perspectives on the origin of this discreteness is valuable, and we will present an additional one in this work.

An important goal is then to understand in more detail how this model factorizes in the gravitational bulk and understand what changes in the edge mode sector. This is the main goal of our current work. In order to realize this goal, we will use the fact that the bulk of DSSYK has a gravitational description as a gauged version of sine dilaton gravity \cite{Blommaert:2024ydx,Blommaert:2024whf,Blommaert:2025avl}. This model can be rewritten in its first order formulation as a Poisson sigma model
\cite{Blommaert:2023wad,Blommaert:2024ydx},\footnote{Poisson-sigma models and their relation to dilaton gravities have been studied extensively in the past, see e.g. \cite{Schaller:1994es,Ikeda:1993aj,Ikeda:1993fh,Cattaneo:2001bp,Grumiller:2003ad} for some early references.} which upon quantization is fully governed by a suitable $q$-deformation of the $\mathfrak{sl}(2,\mathbb{R})$ Lie algebra.\footnote{There is a closely related story for Liouville gravity, its relation to sinh dilaton gravity, and its interpretation as a Poisson sigma model, see e.g. \cite{Mertens:2020hbs,Mertens:2020pfe,Fan:2021bwt,Kyono:2017pxs}.} This perspective has allowed us in the past \cite{Blommaert:2023opb} to recognize the two-boundary wavefunction
\begin{align}
\label{eq:grawa}
\frac{q^n H_n(\cos\theta \vert q^2)}{(q^2;q^2)_n}
\end{align}
as a representation matrix element. Within any model whose wavefunctions are described in this group-theoretical language, there is an obvious natural factorization map:
\begin{align}
\label{equaitonin}
    R_{ab}(g_1\cdot g_2)=\sum_s R_{as}(g_1)R_{sb}(g_2) ,\qquad g_1,g_2\in G, \qquad a,b,s=1,\dots, \text{dim}(R),
\end{align} 
which more abstractly is the co-product on the dual Hopf algebra. We have in the past explored this factorization map in various gravity models \cite{Blommaert:2018iqz,Mertens:2022ujr,Wong:2022eiu,Mertens:2025ydx}. \\

In order to implement this strategy concretely in the context of DSSYK, we delve more deeply into the details of the underlying q-deformed group-theoretical structure, complementing earlier work \cite{Berkooz:2022mfk,Blommaert:2023opb}. Our main results can be summarized as follows:
\begin{itemize}
\item 
Compared to the conventional SU$_q(1,1)$ description with real $0<q<1$, we reproduce the gravitational wavefunctions of DSSYK as principal series matrix elements of the SL$^+
_q(2,\mathbb{R})$ quantum group in the same range of real $q$. This resolves the following puzzle: if the DSSYK Hilbert space was truly just described by the SU$_q(1,1)$ principal series representations, then the $q \to 1^-$ limit would yield SU$(1,1) \simeq \text{SL}(2,\mathbb{R})$ as the analogous classical group for JT gravity. This cannot be true, since JT gravity is related to a slight refinement of this group, studied independently in \cite{Blommaert:2018oro,Iliesiu:2019xuh}. Instead, our more specific real form enforces positivity of the underlying structure on the gravitational matrix elements, analogous to the positive SL$^+(2,\mathbb{R})$ semigroup description of JT gravity \cite{Blommaert:2018iqz,Blommaert:2018oro}.
\item On a mathematical level, we induce the spin-$j$ principal series (co)representations of this quantum group in terms of a q-deformed M\"obius transformations 
\begin{align}
\label{eq:mobiusintro}
    f(x) \mapsto \frac{\left(xa+q^{-j+1/2}c\right)_q^j\left(d+xq^{-1/2}b\right)_{q^{-1}}^j}{x^{j}}f\left((xq^{-j+1/2}b+d)^{-1}(xa+q^{j-1/2}c)\right),
\end{align}
where $a,b,c,d$ are the $q$-coordinates on the SL$_q(2,\mathbb{R})$ group manifold precisely defined in subsection \ref{sec:Hopfduality}, and $(x + y)^n_q\equiv(x+y)(x+qy)\dots (x+ q^{n-1}y)$, suitably analytically continued to non-integer $n$.
\item
A useful identity (that we derive) for the gravitational wavefunction \eqref{eq:grawa} is:
\begin{align}
\label{eq:whittin}
\frac{H_n(\cos\theta | q^2)}{(q^2;q^2)_n} =\frac{q^{-2n\frac{i\theta}{\log q^2}}}{(q^2;q^2)_{\infty}^2(1-q^2)}  \int_0^{\infty/1} d_{q^2} x \, x^{\frac{2i\theta}{\log q^2}-1} E_{q^2}\left(-\frac{q^2}{1-q^2}x\right) E_{q^2}\left(-\frac{q^2}{1-q^2}\frac{q^{2n}}{x}\right).
\end{align}
This is the DSSYK deformation of the modified Bessel function identity
\begin{align}
\label{eq:besselide}
    K_{2ik}(2e^{-\ell/2}) = e^{ i\ell k}\int_0^\infty dx\,  x^{2ik-1} e^{-x}e^{-e^{-\ell}/x},
\end{align}
of relevance in JT gravity. This identity has a representation theoretic interpretation in terms of decomposing a mixed parabolic matrix element into its bra and ket Whittaker vector components. The $x$-index here is an auxiliary parameter. 
The novelty here is that the $x$-index for the DSSYK case \eqref{eq:whittin} has become discretized, much like the $\ell$ length parameter becoming the discrete chord label $n$.
\item 
Discretization of the bulk gravitational spacetime arises in this language from demanding irreducibility of the principal series representations. Unlike the case of the modular double quantum algebra \cite{Faddeev:1999fe}, the relevant quantum algebra representation is only irreducible on a discrete lattice, which in turn leads to a discretized bulk geometry and chord number $n$.
\item 
Characters of suitable representations of the relevant quantum group are computed, and matched to single trumpet and end-of-the-world brane amplitudes in DSSYK, providing a simple and structurally promising approach to computing these.
\item
Using the $q$-Hermite identity
\begin{align}\label{eq:hermiteidentity}
\frac{H_n(\cos\theta \vert q^2)}{(q^2;q^2)_n} = \int_{-\pi}^{\pi} \frac{ds}{2\pi} \frac{e^{-isn}}{(e^{is\pm i\theta};q^2)_\infty},
\end{align}
we can suggestively interpret this formula as a factorization property \eqref{equaitonin} across a bulk entangling surface in the gravitational dual to DSSYK, by writing:
\begin{align}
   \hspace{-0.4cm}\frac{q^n H_n(\cos\theta \vert q^2)}{(q^2;q^2)_n} &= \int_{-\pi}^{\pi}ds \, \frac{1}{\sqrt{2\pi}}\frac{e^{-isn_1}}{(qe^{is-i\theta};q^2)_\infty} \times
  \frac{1}{\sqrt{2\pi}}\frac{e^{-isn_2}}{(qe^{is-i\theta};q^2)_\infty}, \quad n=n_1+n_2,
\end{align}
where the two factors in the integral are to be interpreted as single-sided (or black hole) gravitational wavefunctions $R_{\mathfrak{i}s}(g_1)$ and $R_{s\mathfrak{i}}(g_2)$ respectively in the above notation (where $\mathfrak{i}$ denotes a fixed boundary index following notation in previous work). The $s$-label has the meaning of an edge label living on the factorization surface itself. Pictorially, this equation corresponds to:
\begin{align}
\raisebox{-0.45\height}{\includegraphics[height=2cm]{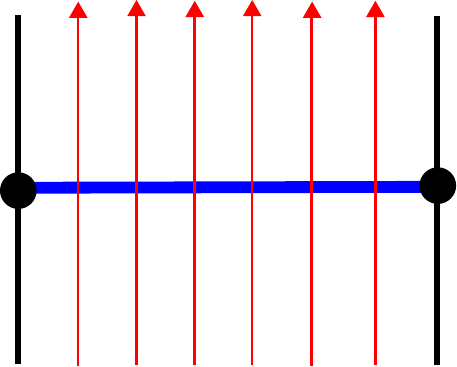}}\;&=\int_{-\pi}^{\pi} ds\ \raisebox{-0.47\height}{\includegraphics[height=2.5cm]{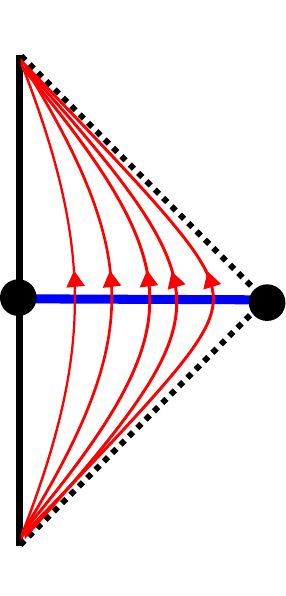}}
\;s\; \raisebox{-0.47\height}{\includegraphics[height=2.5cm]{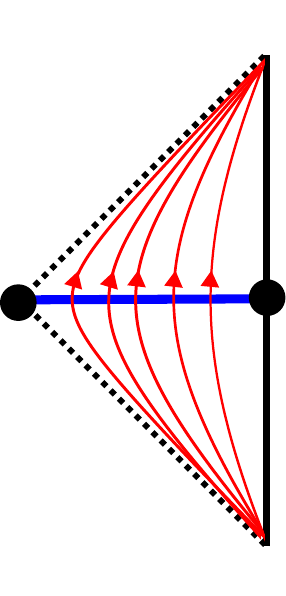}},
\end{align}
where we denoted the spatial slice in blue, and have also drawn the flow of the relevant modular Hamiltonians. It is amusing to interpret these flowlines also directly as chord lines in the chord diagrammatic evaluation of DSSYK \cite{Berkooz:2018qkz,Berkooz:2018jqr,Lin:2022rbf}. The total number of chords $n$ piercing through the blue spatial slice on the left is split into $n_1$ and $n_2$, with $n=n_1+n_2$ in an arbitrary way.  The earlier $\log(x)$-index (introduced in equation \eqref{eq:whittin}) is the conjugate to this $s$-index. Hence the periodicity of the $s$-index is related to the discreteness of the $x$-index in exactly the same way that the momentum quantum number is periodic (in the $1^{st}$ Brillouin zone) on a spatial lattice in condensed matter physics (e.g. in the tight-binding approximation of a 1d atomic array).
\end{itemize}

The remainder of this work is organized as follows. In subsection \ref{s:obs}, next to defining the various special functions we need, we derive an identity for the $q$-Hermite polynomials that will be crucial in the remainder of this work. \textbf{Section} \ref{s:techcor} is the technical core of this paper where we refine the quantum group theoretical structure within DSSYK, as first proposed in \cite{Berkooz:2018jqr} and made concrete later in \cite{Berkooz:2022mfk,Blommaert:2023opb}. We provide a short summary in subsection \ref{s:sum}. These technical developments are needed in order to develop several gravitational consequences, which we do in \textbf{section} \ref{s:gravapp} that follows. In order to illustrate the generality of our techniques, we generalize various results in \textbf{section} \ref{s:Nis1} to the case of $\mathcal{N}=1$ DSSYK. We end this work with some speculation and future directions in the concluding \textbf{section} \ref{s:concl}. Various technical calculations are provided in the appendices. In particular appendix \ref{sec:AppendixA} develops the principal series corepresentations from equation \eqref{eq:mobiusintro}. This can be read independently from the physical applications in the main text.

\subsection{Observation: $q$-Hermite integral identity}
\label{s:obs}

We start by rewriting the known two-sided wavefunction of DSSYK in terms of a new integral identity, which will suggestively yield the building blocks of the representation-theoretic decomposition.

We start from the generating function of the $q$-Hermite polynomials $H_n(\cos\theta\vert q^2)$:
\begin{align}
    \sum_{n=0}^\infty H_n(\cos\theta\vert q^2)\frac{t^n}{(q^2;q^2)_n}=\frac{1}{(te^{\pm i\theta};q^2)_\infty}, \qquad n\in \mathbb{Z},\;\,\, \theta\in [0,\pi], \;\,\,0<q<1,
\end{align}
in terms of the $q$-Pochhammer symbols, defined as\footnote{Note that, unlike the classical Pochhammer symbol, we can extended this to an infinite convergent product in the range $0<q<1$. }
\begin{align}
    (a;q)_n \equiv \prod_{j=0}^{n-1}(1-aq^j).
\end{align}
Using a unit circle contour integral in the complex $t$-plane (taking the poles on the circle outside of the contour), we can project onto a single term in the expansion
\begin{align}\label{eq:hermiteidentity}
\frac{H_n(\cos\theta \vert q^2)}{(q^2;q^2)_n} = \frac{1}{2\pi}\int_{-\pi}^{\pi} dp \frac{e^{-ipn}}{(e^{ip\pm i\theta};q^2)_\infty}.
\end{align}
The standard $q$-gamma function is defined for $0<q<1$ as:
\begin{align}\label{eq:qpochammergamma}
\Gamma_{q^2}(x) \equiv (1-q^2)^{1-x}\frac{(q^2;q^2)_{\infty}}{(q^{2x};q^2)_{\infty}},
\end{align}with $\Gamma_{q^2}(1)=1$ and $\lim_{q\rightarrow1^-}\Gamma_{q^2}(x)=\Gamma(x)$. Using the definition \eqref{eq:qpochammergamma}, this function is found to satisfy a recursive property in terms of the $q$-number $[x]_{q^2}$:\begin{align}\label{eq:qgamma}
    \Gamma_{q^2}(x+1)=[x]_{q^2}\Gamma_{q^2}(x),\qquad \qquad [x]_{q^2} \equiv \frac{q^{2x}-1}{q^2-1}.
\end{align}
By the above definition, the poles of $\Gamma_{q^2}(x)$ are at non-positive integers. Also note that the combination $\Gamma_{q^2}(x)(1-q^2)^x$ is periodic in imaginary steps of $x$ with period $2\pi i/\log q^2$. Introducing the $q$-deformed exponential
\begin{equation}\label{eq:qexponential}
E_{q^2}(x) \equiv \left(-(1-q^2)x;q^2\right)_\infty= \sum_{j=0}^{+\infty} q^{j(j-1)}\frac{x^j}{[j]_{q^2}!}
\end{equation}
and the definition of the improper Jackson integral\footnote{The notation $\infty/A$ is needed to signal the precise discretization used on the right hand side of this expression. This notation was introduced in \cite{de2003integral} and corrected the earlier textbook \cite{KC02}.}
\begin{align}
\label{eq:improperJackson}
    \int_0^{\infty/A} d_{q^2}x\;f(x) \equiv (1-q^2)\sum_{n\in \mathbb{Z}}\frac{q^{2n}}{A}f\left(\frac{q^{2n}}{A}\right),
\end{align}
we find the known integral identity
\begin{align}
\Gamma_{q^2}(x) = \int_0^{\infty/(1-q^2)} d_{q^2} t \, t^{x-1} E_{q^2}\left(-q^2t\right) .
\end{align} 
This identity satisfies the same recursion relation as \eqref{eq:qpochammergamma} and $\Gamma_{q^2}(1)=1$.
Making the substitution $t\rightarrow (1-q^2)t$, we rewrite the $q$-gamma function as:\footnote{The combination $\Gamma_{q^2}(x)(1-q^2)^x$ will frequently appear throughout this paper and is related to the $q$-Pochhammer symbol through (\ref{eq:qpochammergamma}). Using the definition of the $q$-beta function for $s,t>0$ 
\begin{align}
    B_{q^2}(t,s)=\int_0^1d_{q^2}x \;x^{t-1}(q^2x;q^2)_{s-1},
\end{align}
we can also relate this combination with the integral expression of the $q$-beta function as 
\begin{align}
    B_{q^2}(x,\infty)=\int_0^{1}d_{q^2}t\; t^{x-1} E_{q^2}\left(-\frac{q^2}{1-q^2}t\right)=(1-q^2)^x\Gamma_{q^2}(x),
\end{align} 
where the $q$-exponential \eqref{eq:qexponential} is identically zero for $t>1$ in the improper Jackson integral, leading to the final identification with \eqref{eq:qbeta}.} 
\begin{align}
\label{eq:qbeta}
  \Gamma_{q^2}(x)(1-q^2)^x=\int_{0}^{\infty/1} d_{q^2}t\;t^{x-1}E_{q^2}\left(-\frac{q^2}{1-q^2}t\right).
\end{align}
Inserted in \eqref{eq:hermiteidentity}, we can hence rewrite
\begin{align}
\frac{H_n(\cos\theta | q^2)}{(q^2;q^2)_n} =& \int_0^{2\pi} \frac{dp}{2\pi} \frac{e^{-ipn}}{(q^2;q^2)_{\infty}^2(1-q^2)^2} \\&\times\int_0^{\infty/1} d_{q^2} x d_{q^2} y \,x^{\frac{ip+i\theta}{\log q^2}-1} y^{\frac{ip-i\theta}{\log q^2}-1} E_{q^2}\left(-\frac{q^2}{1-q^2}x\right) E_{q^2}\left(-\frac{q^2}{1-q^2}y\right). \nonumber
\end{align} 
The Jackson integrals discretize $x$ and $y$ as $x=q^{2j}$ and $y=q^{2k}$ for $j,k \in \mathbb{Z}$. The $p$-integral can be done as:\footnote{The Jackson $q$-integrals in this expression do not absolutely converge, so a priori a swapping of the $p$-integral with the $x$- and $y$-integrals is not allowed. One can fix this by adding a $x^\epsilon y^\epsilon$ regulator with $\epsilon \to 0^+$ in the summand. After the $p$-integral this leads to a $q^{2n\epsilon}$ leftover factor, which is $x$-independent and can hence be cleanly taken to 1. All convergence statements have direct analogues in the $q\to 1^-$ limit, where the Gamma function integral $\Gamma(it) = \int_0^{+\infty}dx \, x^{it-1}e^{-x}$ needs a regulator $x^\epsilon$ for absolute convergence (i.e. to makes sense of the Fourier transform of $\Gamma(it)$ as $\lim_{\epsilon \to 0^+} \Gamma(it+\epsilon)$), but the final integral \eqref{eq:besselide} does not require a regulator $x^\epsilon$ anymore.}
\begin{align}
\int_0^{2\pi} \frac{dp}{2\pi} e^{-ipn} e^{ipj}e^{ipk} = \delta_{k,n-j},
\end{align}
leading to $y=q^{2(n-j)} = q^{2n} /x$.
We hence finally obtain a new integral identity for the $q$-Hermite polynomials\footnote{This Jackson integral is absolutely convergent. We directly checked this final identity numerically as well with Mathematica.} 
\begin{align}
\label{eq:whitt}
\boxed{
\frac{H_n(\cos\theta | q^2)}{(q^2;q^2)_n} =\frac{q^{-2n\frac{i\theta}{\log q^2}}}{(q^2;q^2)_{\infty}^2(1-q^2)}  \int_0^{\infty/1} d_{q^2} x \, x^{\frac{2i\theta}{\log q^2}-1} E_{q^2}\left(-\frac{q^2}{1-q^2}x\right) E_{q^2}\left(-\frac{q^2}{1-q^2}\frac{q^{2n}}{x}\right).}
\end{align}
In the following sections, we will pin down the proper quantum group theoretic structure that reproduces this integral as a representation matrix element evaluated in the principal series representation. Identifying this with the known gravitational wavefunction of DSSYK \cite{Berkooz:2018jqr,Berkooz:2018qkz, Lin:2022rbf}, then recovers the quantum group structure governing DSSYK. 

Mathematically, we will see that in the classical $q\rightarrow 1^-$ limit, this integral reduces to the gravitational wavefunction of JT gravity in terms of the Bessel-$K$ functions. The group theory structure of DSSYK then exactly limits to the SL$^+(2,\mathbb{R})$ semigroup structure of JT gravity   \cite{Blommaert:2018iqz, Blommaert:2018oro}.

\section{Refined quantum group structure of double-scaled SYK}
\label{s:techcor}

We define the $q$-deformed Hopf algebra of $U(\mathfrak{sl}_2)$ and its real forms. Unlike the conventional description in terms of $U_q(\mathfrak{su}(1,1))$ with real $0<q<1$, we determine the relevant structure of the gravitational implementation of DSSYK in terms of the $U_q(\mathfrak{sl}(2,\mathbb{R}))$ real form in the same range of $q$. We find this to be the more convenient real form to directly implement the gravitational restriction to smooth geometries. The price we pay is that the interpretation of unitarity on this Hopf algebra requires the introduction of a twisted star structure. The choice of the precise Hopf (sub)algebra fixes the bulk discretization of the gravitational description of DSSYK. We similarly introduce the appropriate dual as a real form of the SL$_q(2)$ coordinate algebra. We then induce the principal series generators from the principal series corepresentations of the coordinate algebra. These irreducible representations physically propagate in the gravitational bulk of DSSYK. The mixed parabolic matrix elements in these representations finally recover the two-sided gravitational wavefunctions of DSSYK via \eqref{eq:whitt}. 

\subsection{The quantum deformation of $U(\mathfrak{sl}_2)$}

The standard definition the associative algebra $U_q(\mathfrak{sl}_2)$ \cite{Klimyk:1997eb} at generic $q$ is given as the quotient of the free algebra $\mathbb{C}\langle E^\pm, K, K^{-1}\rangle$ by the ideal generated via the relations\footnote{In the literature this is called a Drinfeld-Jimbo algebra and this type of deformation can be easily extended to arbitrary semi-simple lie algebras.} 
\begin{align}
 \label{eq:alegbragenerators2}
     KE^\pm=q^{\pm 2}E^\pm K, \qquad KK^{-1}=K^{-1}K=1, \qquad [E^+,E^-]=\frac{K-K^{-1}}{q-q^{-1}},
 \end{align}
where the product and unit are inherited from the free algebra. This is a Hopf algebra with the coproduct $\Delta$:\footnote{For completeness, the antipode $S$ and counit $\epsilon$ are given by 
\vspace{-0.2cm}
\begin{equation*}
     S(K)=K^{-1}, \quad S(E^-)=-K^{-1}E^-, \quad S(E^+)=-E^+K, \quad \epsilon(K)=1, \quad \epsilon(E^-)=\epsilon(E^+)=0.
\end{equation*}\vspace{-0.7cm}}$^{,}$\footnote{The coproduct of the generators is fixed by compatibility with the conventions of the Gauss-Euler decomposition discussed in section \ref{sec:Hopfduality}. }
 \begin{equation}
     \label{eq:coproduct2}
    \Delta(E^+)=E^+\otimes K^{-1}+1\otimes E^+, \quad \Delta(E^-)=E^-\otimes 1+K\otimes E^-, \quad \Delta(K)=K\otimes K.
\end{equation}
This algebra can be viewed as a quantum deformation of the universal enveloping algebra $U(\mathfrak{sl}_2)$ and is similarly spanned by a basis of monomials of the form $\{(E^-)^nK^\ell(E^+)^m\vert \ell\in \mathbb{Z}, \;n,m\in \mathbb{N}_0\}$.
This can easily be seen by taking the classical limit $q\rightarrow1^-$. To take this limit, it is useful to define $q\equiv e^{-h}$ ($h>0$) and $K\equiv q^{2H}$ which leads to the algebra relations 
\begin{align}\label{eq:Halgeba}
    [H,E^\pm]=\pm E^\pm, \qquad [E^+,E^-]=\frac{q^{2H}-q^{-2H}}{q-q^{-1}}=\frac{\sinh(2hH)}{\sinh(h)}.
\end{align}
In the $h\rightarrow 0$ limit, these algebra relations constitute the familiar algebra relations of $\frak{sl}_2$:
\begin{align}
     [H,E^\pm]=\pm E^\pm, \qquad [E^+,E^-]=2H.
\end{align}
We can define the principal series generators of $U_q(\mathfrak{sl}_2)$ following \cite{Ponsot:1999uf,Ponsot:2000mt,Kharchev:2001rs}. We will derive these generator expression from the associated principal series corepresentation of the coordinate Hopf algebra in section \ref{sec:principalseries} below. 
These generators act on the space of square integrable functions on the real line $ f(x)\in L^2(\mathbb{R}) $ as
\begin{align}
\label{eq:principalseriesgenerators2}
    K&=R_{q^2},\qquad E^-= q^{-1/2}\frac{q^jR_{q^2}-q^{-j}}{x(q-q^{-1})},\qquad E^+=-xq^{-1/2} \frac{q^{-j}-q^jR_{q^{-2}}}{q-q^{-1}},
\end{align}
in terms of the scaling operator $R_{q^2}$ defined as $R_{q^2}f(x) \equiv f(q^2x)$. In terms of the $q$-derivative operator $D_{q^2}\equiv \frac{R_{q^2}-1}{x(q^2-1)}$, these can be recast as 
\begin{align}
    K&=R_{q^2},\quad E^-=q^{j+1/2}D_{q^2}+q^{1/2-j}x^{-1}[j]_{q^2},\quad   E^+=-q^{-3/2 + j}x^2D_{q^{-2}}-xq^{1/2-j}[j]_{q^2}.
\end{align} 
One can check that these generators satisfy the algebra relations \eqref{eq:alegbragenerators2}. We will show later that these representations are unitary under our choice of real form exactly when $j$ takes the form $j = -\frac{1}{2} + \frac{i\theta}{\log q^2}$ where $\theta\in[0,\pi]$. While these expressions look a bit unfamiliar in the DSSYK context, we illustrate in Appendix \ref{sec:AppendixD} that with a different carrier space, these take the form of those presented in \cite{Burban:1992ys,groenevelt} and used in \cite{Berkooz:2018jqr,Jafferis:2022wez}. Note that the range of the coordinate $x$ can be consistently restricted to half of the real line under the consecutive action of the generators. We hence restrict to the function space $L^2(\mathbb{R}^+)$ with $x>0$.

The Casimir operator, defined to be central in the Hopf algebra i.e $[\mathcal{C}_2,X]=0$ for all $X \in U_q(\frak{sl}_2)$,\footnote{It can be shown that the center of $U_q(\mathfrak{sl}_2)$ is fully generated by this operator if $q^2\neq 0,1$.} plays a crucial role in the realization of DSSYK in terms of quantum group theory. It takes the explicit form   
\begin{align}
\label{eq:generalcasimir}
    \mathcal{C}&=E^-E^+ +\frac{qK+q^{-1}K^{-1}}{(q-q^{-1})^2}=E^+E^-+\frac{qK^{-1}+q^{-1}K}{(q-q^{-1})^2}.
\end{align} 
Since the principal series representations should be irreducible, the Casimir evaluated in these representations should be constant by Schur's lemma, which we check by simply plugging in the explicit form of the principal series generators in \eqref{eq:generalcasimir} leading to: 
\begin{align}
\mathcal{C}= \frac{2\cos\theta}{(q-q^{-1})^2},
\label{eq:casimireigenvalue}
\end{align}
which takes unique values only if $\theta\in[0,\pi]$ and where one directly recognizes that the Casimir evaluated on the principal series representations reproduces the eigenspectrum of the Hamiltonian in DSSYK \cite{Berkooz:2018jqr}.

The Hopf algebra $U_q(\mathfrak{sl}_2)$ has, up to equivalence classes, three real forms with well defined classical counterparts,\footnote{In addition there are two real forms (that will play no role in this paper) with $q \in i \mathbb{R} $ that do not yield classical real forms \cite{Klimyk:1997eb}. } denoted  $U_q(\mathfrak{sl}(2,\mathbb{R}))$, $U_q(\mathfrak{su}(1,1))$ and $U_q(\mathfrak{su}(2))$, characterized respectively by different ranges of $q$ and the star relations
\begin{alignat}{4}
\label{eq:realforms}
    U_q(\mathfrak{sl}(2,\mathbb{R}))&:\, (E^+)^\dag=-E^+, \;\; &&(E^-)^\dag=-E^-, \;\; &&K^\dag=K, \, &&|q|=1.\\\label{eq:su(1,1)star} 
    U_q(\mathfrak{su}(1,1))&:\, (E^+)^\dag=-q^{-1}K^{-1}E^-, \quad &&(E^-)^\dag=-q^{-1}KE^+, \quad &&K^\dag=K, \, &&q\in \mathbb{R} .\\
    U_q(\mathfrak{su}(2))&:\, (E^+)^\dag=q^{-1}K^{-1}E^-, \quad &&(E^-)^\dag=q^{-1}KE^+, \quad &&K^\dag=K, \qquad &&q\in \mathbb{R}. 
\end{alignat} 
In the limit $q\rightarrow1^-$, these become the familiar $*$-relations of the classical $\mathfrak{sl}_2$ real forms.
Since the principal series representations, which propagate in the gravitational bulk, are not unitary in the case of $U_q(\mathfrak{su}(2))$ we neglect this object in most gravitational applications.
Having $q$ take values on the unit circle $|q|=1$ describes amplitudes of Liouville quantum gravity with the associated real form $U_q(\mathfrak{sl}(2,\mathbb{R}))$.\footnote{To be more precise, the amplitudes of Liouville quantum gravity are reproduced from the representation theory of the modular double quantum algebra $U_q(\mathfrak{sl}(2,\mathbb{R})) \otimes U_{\tilde{q}}(\mathfrak{sl}(2,\mathbb{R}))$ \cite{Fan:2021bwt}.} On the other hand, since DSSYK is governed by real values of the deformation parameter in the range $0<q<1$, DSSYK is commonly ascribed to the $U_q(\mathfrak{su}(1,1))$ structure.

\subsubsection{Restriction to smooth geometries - the positive semigroup}

However, we know that DSSYK amplitudes cannot be described by pure $U_q(\mathfrak{su}(1,1))$, since the classical limit $q \to 1^-$  would then recover the amplitudes of SU$(1,1)$ BF gauge theory, which is \emph{not} gravity. In particular, one still has to constrain the gauge theory description to Teichm\"uller space to constrain to regular Euclidean geometries, which can be achieved algebraically by restricting to the semigroup  $\text{SL}^+(2,\mathbb{R})$. 
Similarly, we have shown that the restriction to the positive semigroup persists in the group-theoretic description of $\mathcal{N}=1,2$ JT supergravity models \cite{Fan:2021wsb, Belaey:2024dde} by imposing the same positivity restriction on the gravitational SL$(2,\mathbb{R})$ subblock of the respective orthosymplectic supergroups. In more detail, in the classical limit $q\rightarrow1^-$, we can expand the generators \eqref{eq:principalseriesgenerators2} around $q=1-h+\mathcal{O}(h^2)$ with $h\ll1$, to obtain the symmetric Borel-Weil realization of $\mathfrak{sl}(2,\mathbb{R})$. In particular
\begin{align}
    H\rightarrow x\partial_x, \qquad E^-&\rightarrow \frac{j}{x}+\partial_x, \qquad E^+\rightarrow-x^2\partial_x+jx.\label{eq:parabolicsl2R}
\end{align}
These are precisely the differential generators of $\mathfrak{sl}(2,\mathbb{R})$ deduced from the exponentiated (symmetrical) principal series representation 
\begin{align}
    (g\cdot f)(x)=\frac{(ax+c)^j(bx+d)^j}{x^j}f\left(\frac{ax+c}{bx+d}\right),
\label{eq:classprincseries}
\end{align} 
under a finite SL$^+(2,\mathbb{R})$ element 
\begin{align}
    g=\begin{pmatrix}
        a&b\\c&d
    \end{pmatrix},\qquad ad-bc=1, \qquad a,b,c,d>0.
\end{align}
At the level of the principal series representations acting on square integrable functions $L^2(\mathbb{R})$, this positivity is compatible with a positive range of the carrier space coordinate $x\in \mathbb{R}^+$.

A well-known isomorphism relates the real forms SU$(1,1)$ and SL$(2,\mathbb{R})$ via \cite{Vilenkin}:
\begin{equation}
    g\in \text{SL}(2,\mathbb{R}) \,\, \longleftrightarrow \,\, h \in \text{SU}(1,1),\;\;h=t^{-1}gt,\;\;
    \label{eq:isomorphism}
\end{equation}
where $t=\frac{1}{\sqrt{2}}\left(\begin{array}{cc}
1 & i \\
i & 1 \\ 
 \end{array}\right)$ 
 is an element of the complexified SL$(2,\mathbb{C})$. The requirement of positivity of the SL$^+(2,\mathbb{R})$ group elements then can be tracked through this isomorphism to the equivalent positivity on the semigroup SU$^+(1,1)$: 
 \begin{align}
     \text{SU}^+(1,1) \equiv \left\{ \begin{pmatrix}
         a&b\\\overline{b}& \overline{a}
     \end{pmatrix}\;\Big|\; |a|^2-|b|^2=1, \; |\Im(a)|<\Re(b), \; |\Im(b)|<\Re(a)\right\}.
     \label{eq:susemigroup}
 \end{align} 
 The real carrier space coordinate $x$ for SL$^+(2,\mathbb{R})$ can be transformed to a complex coordinate $z$ on the unit circle by a Cayley transform of the form 
 \begin{align}
     x\mapsto z\equiv -i\frac{x+i}{x-i}.
 \end{align} 
 The restriction of $x$ to the positive half line $[0,\infty[$, restricts the complex $z$ coordinate to the right half-circle with phase angle in the range $\left]-\pi/2,\pi/2\right]$:
 \begin{align}
\raisebox{-0.42\height}{\includegraphics[height=3cm]{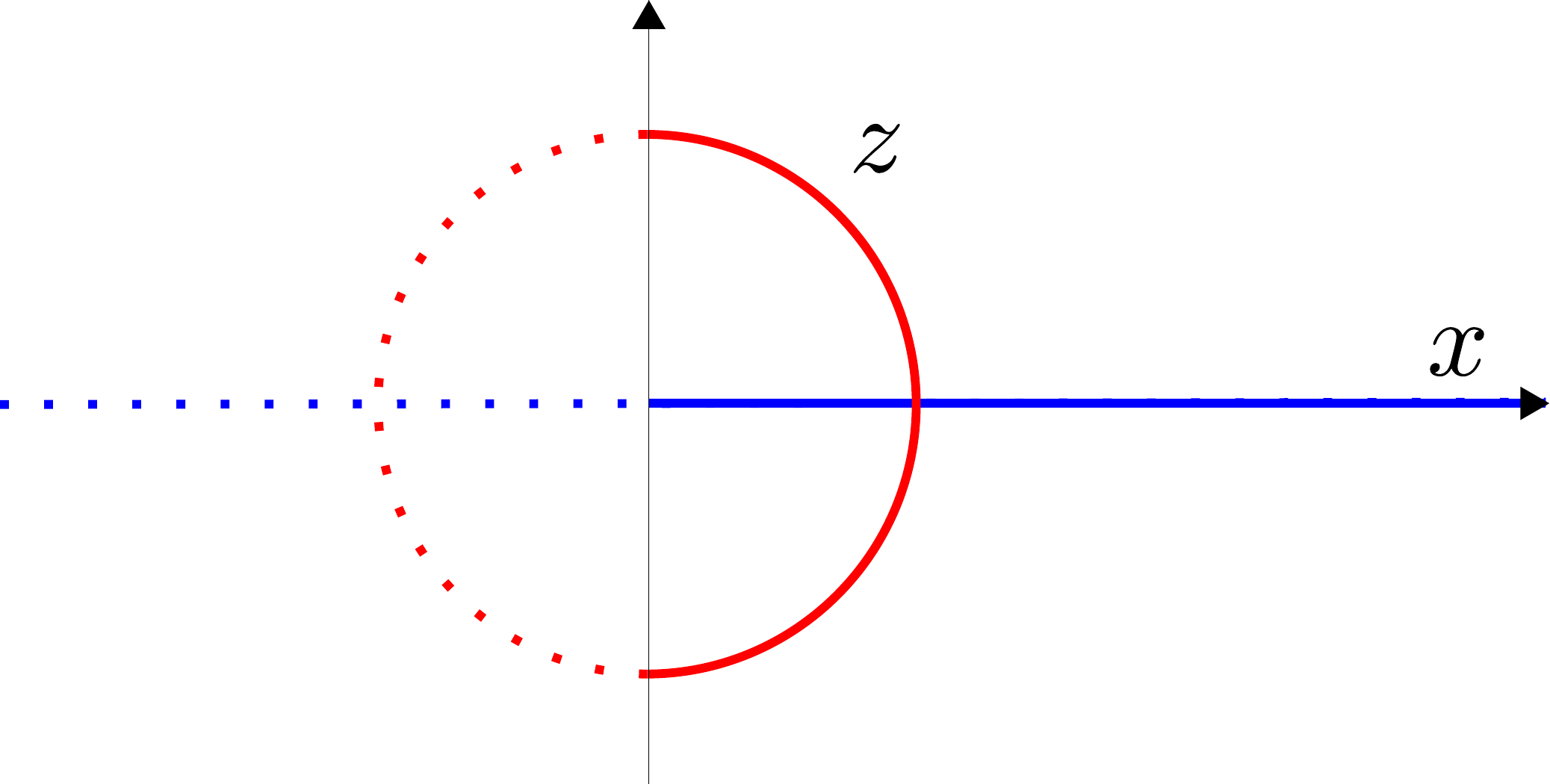}}
\end{align}
 
 The principal series representation \eqref{eq:classprincseries} then leads to an equivalent one in SU$(1,1)$ notation, mapping the complex $z$ coordinate as $z\mapsto \frac{az+\overline{b}}{bz+\overline{a}}$. The conditions of the semigroup element \eqref{eq:susemigroup} can readily be shown to map the right semicircle onto itself under this M\"obius transformation.

To achieve positivity in the $q$-deformed setting on the representations of $U_q(\mathfrak{su}(1,1))$, one first needs to transfer this positivity directly through an isomorphism analogous to \eqref{eq:isomorphism}. 
However, we will find it more convenient to impose the positivity of the gravitational structure directly in the SL$_q(2,\mathbb{R})$ gravitational real form. Imposing positivity directly at the SU$_q(1,1)$ level, would require a deeper understanding of the Cayley transform in the quantum-deformed setting, which is beyond the scope of this work. The price we will pay for this, is that we have to reinterpret the resulting structure as a \emph{twisted} Hopf $*$-algebra as we will make more clear further on.

\subsubsection{Gravitational real form}

 The Hopf algebra generated by the elements \eqref{eq:principalseriesgenerators2}: 
\begin{align}
    K&=R_{q^2},\qquad E^-= q^{-1/2}\frac{q^jR_{q^2}-q^{-j}}{x(q-q^{-1})},\qquad E^+=-xq^{-1/2} \frac{q^{-j}-q^jR_{q^{-2}}}{q-q^{-1}},
\end{align}
relates a function $f(x)$ evaluated at a given $x_0$ to the function evaluated at $q^{2n} x_0$ only, where $n\in \mathbb{Z}$; schematically:
\begin{align}
\raisebox{-0.42\height}{\includegraphics[height=3.3cm]{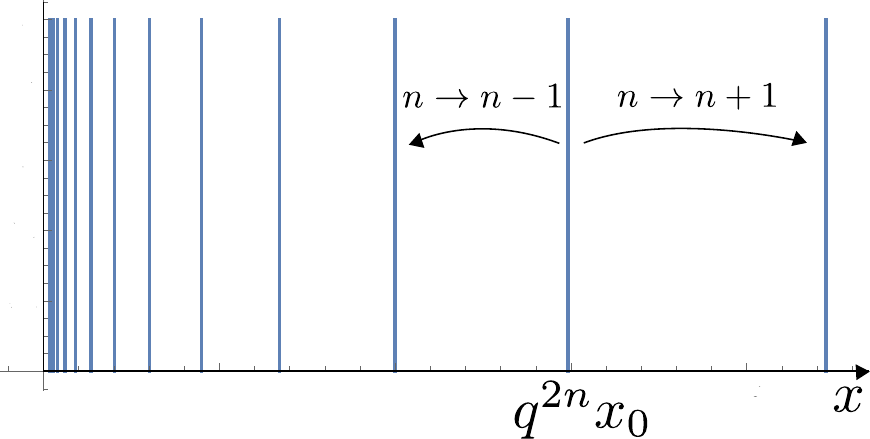}}
\end{align}
The principal series realization of the monomial basis spanning the universal enveloping algebra via 
\begin{equation}
\mathcal{A} = \text{span}\{(E^-)^\ell K^{n}(E^+)^m\vert l,m\in \mathbb{N}_0,n\in \mathbb{Z} \}
\end{equation}
then only contains integer powers of $R_{q^2}$.
The principal series generators therefore act irreducibly on the invariant $q^2$-discretized grid, denoted $\mathbb{R}_{q^2}$ 
\begin{equation}
\mathbb{R}_{q^2}\equiv \{x=q^{2n}\vert n\in \mathbb{Z}\}.
\end{equation}
In terms of an exponentiated coordinate $x=e^t$, this only allows discrete shifts $t \to t +n \log q^2$. The full representation space $L^2(\mathbb{R}^+)$ hence is reducible and decomposes into those with fixed seed coordinate $t$ as:
\begin{equation}\label{eq:L+}
L^2(\mathbb{R}^+) = \int_{\bigoplus  {t\in [0,2\log q[}} dt \, L^2(t+2\log q \cdot\mathbb{Z}),
\end{equation}
where the spaces $L^2(t+2\log q \cdot \mathbb{Z})$ contain no further invariant subspaces under the above action. Note that this is very different from the modular double quantum algebra $U_q(\mathfrak{sl}(2,\mathbb{R})) \otimes U_{\tilde{q}}(\mathfrak{sl}(2,\mathbb{R}))$  \cite{Faddeev:1999fe,Ponsot:1999uf,Ponsot:2000mt,Kharchev:2001rs,Bytsko:2002br,Bytsko:2006ut} where one adds a second copy of the same algebra with a different parameter $\tilde{q}$, with the choice of incommensurate values of $\sqrt{\log q}$ and $\sqrt{\log \tilde{q}}$, making the initial representation generated by \eqref{eq:principalseriesgenerators2} irreducible immediately.

Note that there exists another $q$-deformed Hopf algebra, usually denoted $\breve{U}_q(\mathfrak{sl}_2)$, defined by the algebra relations\footnote{Note that at the level of the $(E^-,E^+,H)$ generators, both algebras imply the infinitesimal algebra (\ref{eq:Halgeba}), where $\Breve{U}_q(\mathfrak{sl}_2)$ follows from the identification $K=q^{H}$ and $U_q(\mathfrak{sl}_2)$ follows by setting $K=q^{2H}$. The coproducts obviously still differ. } 
\begin{align}
    &KE^\pm=q^{\pm 1}E^\pm K, \qquad KK^{-1}=K^{-1}K=1, \qquad [E^+,E^-]=\frac{K^2-K^{-2}}{q-q^{-1}}.\label{eq:algebrarelations}\\
    &\Delta(E^-)=E^-\otimes K^{-1}+K\otimes E^-, \; \Delta(E^+)=E^+\otimes K^{-1}+K\otimes E^+,\; \Delta(K)=K\otimes K. 
\end{align} 
The homomorphism of Hopf algebras $\varphi: U_q(\mathfrak{sl}_2)\rightarrow \breve{U}_q(\mathfrak{sl}_2)$ defined by
\begin{align}
\label{eq:homomorphism}
    \varphi(E^+)=E^+K^{-1}, \qquad \varphi(E^-)=KE^-, \qquad \varphi(K)=K^2.
\end{align} 
shows that the image of $U_q(\mathfrak{sl}_2)$ under $\varphi$ forms a (Hopf)-subalgebra of $\breve{U}_q(\mathfrak{sl}_2)$. The corresponding principal series generators can then be found from \eqref{eq:principalseriesgenerators2} as 
\begin{align}
    K&=R_q,\qquad E^-= q^{1/2}\frac{q^jR_q-q^{-j}R_{q^{-1}}}{x(q-q^{-1})},\qquad E^+=-x q^{-1/2}\frac{q^{-j}R_q-q^jR_{q^{-1}}}{q-q^{-1}}.
\end{align} 
This implies that the \emph{irreducibility} of the principal series representation of $\breve{U}_q(\mathfrak{sl}(2,\mathbb{R}))$  requires restriction to the discretized target space of $q$-separated points $\mathbb{R}_q\equiv \{x=q^n\vert n\in \mathbb{Z}\}$, whereas those of $U_q(\mathfrak{sl}(2,\mathbb{R}))$ act on the larger grid $\mathbb{R}_{q^2}$. Looking at the integral expression of the known gravitational wavefunctions of DSSYK \eqref{eq:whitt}, we observe that the measure relates points on $\mathbb{R}_{q^2}$. Therefore, the choice of the smaller Hopf (sub)algebra $U_q(\mathfrak{sl}_2)$ \eqref{eq:alegbragenerators2} reproduces the gravitational structures of DSSYK. Note that in order to make this conclusion, it is important to use the homomorphism \eqref{eq:homomorphism} as a mapping that preserves the entire Hopf algebra structure. Mapping only the algebra relations by substituting $K\rightarrow K^2$ would not yield the required $q^2$-discretization in the parabolic $E^\pm$ generators.

Going back to the (sub)algebra $U_q(\mathfrak{sl}_2)$, we define the real form of interest here by committing to an inner product on $L^2(\mathbb{R}^+)$. On the irreducible subspaces, the inner product is defined in terms of a $q^2$-discretized Jackson integral: 
\begin{align}
\label{eq:innerproduct}
    \braket{f\vert g}=\int_0^{\infty/1} \frac{d_{q^2}x}{x}\;\overline{f(x)}g(x) = (1-q^2)\sum_{n\in \mathbb{Z}}q^{2n}\overline{f\left(q^{2n}\right)}g(q^{2n}),
\end{align}
where the upper bound indicates the embedding of the $q^2$-discretized grid $\mathbb{R}_{q^2}$ along the real line (including the representative point $q^0=1$), according to the definition \eqref{eq:improperJackson}. In the classical $q\rightarrow1^-$ limit, this inner product becomes that of the SL$^+(2,\mathbb{R})$ gravitational real form describing JT gravity.

Setting $j\equiv -\frac{1}{2}+\frac{i\theta}{\log q^2}$ with $\theta\in \mathbb{R}$, we can easily calculate the adjoint action of the principal series generators \eqref{eq:principalseriesgenerators2}.
We denote the real form in the range of real $0<q<1$, together with these star relations as $U_{0<q<1}(\mathfrak{sl}(2,\mathbb{R}))$:\footnote{We will often drop the explicit range in $q$ in the rest of the text for notational convenience, and let it be implied from  context.}
\begin{align}\label{eq:ourstar}
     U_q(\mathfrak{sl}(2,\mathbb{R})):\;(E^+)^\dag=-q^{-1}KE^+, \;\; (E^-)^\dag=-q^{-1}K^{-1}E^-, \;\; K^\dag=K^{-1}, \;\; 0<q<1.
\end{align}  
One can easily check that these $*$-relations are compatible with the algebra relations \eqref{eq:alegbragenerators2} in the range of real $0<q<1$: 
\begin{align}    [E^+,E^-]^\dag&=\left(\frac{K-K^{-1}}{q-q^{-1}}\right)^\dag\quad\leftrightarrow \quad [E^-,E^+]=\frac{K^{-1}-K}{q-q^{-1}},
    \end{align}
    and
    \begin{align}
        (KE^\pm)^\dag&=q^{\pm2} \left(E^\pm K\right)^\dag \quad \leftrightarrow \quad -E^\pm K^{-1}=-q^{\pm2} K^{-1} E^\pm.
    \end{align}
The reader may be slightly perplexed by this choice of star since the option of $K^\dag=K^{-1}$ does not appear in the list \eqref{eq:realforms}.\footnote{Note that in terms of $K=q^{2H}$, both ranges of real $0<q<1$ and $|q|=1$ imply that the Cartan generator is automatically anti-hermitian $H^\dag=-H$.} The reason is that this is not a Hopf star in the formal definition but rather a \textit{twisted Hopf star} \cite{Coquereaux:1999va,Buffenoir:1999,Majid:1992bz}. The difference between these two is found in the manner in which the star is extended over tensor products as follows. For this paragraph only, let $*$ denote a true Hopf star and denote a twisted Hopf star by $*_\tau$.
Normally we assume that a star is an anti-homomorphism with respect to the algebra product i.e. $(ab)^* = b^* a^*$ while the star is extended to tensor products by acting on both factors independently such that $(a\otimes b)^* = a^* \otimes b^*$. Although this choice is natural, since generally $a,b$ can be chosen from different algebras $\mathcal{A},\mathcal{B}$, in the case where $\mathcal{A} = \mathcal{B}$ a different choice is possible. We define a star to be twisted if it satisfies
\begin{equation}
    \left( a \otimes b \right)^{*_\tau}  \equiv b^{*_\tau} \otimes a^{*_\tau}, 
 \end{equation}
i.e. the star \textit{twists} the tensor product.\footnote{In the case of Hopf algebras an equivalent definition is given by, instead of twisting the action of the star on tensor products, twisting the condition stars have to satisfy with respect to the coproduct. Explicitly, while $\Delta(a^*) = \Delta^*(a)$ for Hopf stars we can define twisted stars to  satisfy $\Delta(a^*) =\Delta^\text{cop}(a)^* \equiv \tau(\Delta(a)^*) $ where $\tau$ is defined to swap the factors of the tensor product.} 
The star operation defined in (\ref{eq:ourstar}) is exactly a star of the coproduct \eqref{eq:coproduct2} of $U_q(\mathfrak{sl}_2)$ in this sense. It was noted in \cite{Coquereaux:1999va}, that twisting of stars, in certain cases can be seen as an analytic continuation in $q$ of the standard Hopf stars. In the following the formal differences between twisted and non-twisted star structures will not play a major role.

\subsubsection{Hopf duality and the quantum group $\text{
SL}^+_{0<q<1}(2,\mathbb{R})$}\label{sec:Hopfduality}
Until now we have mostly focused on the local algebraic structure encoded by the quantum deformed universal enveloping algebra however generally a choice of global quantum group is needed fully describe our system. For example the question which representations show up in the basis decomposition of the Hilbert space requires knowledge of the Plancherel measure on the ``quantum group manifold'' which equivalently specifies the spectral density of the model. 
Classically this choice of explicit group can be encoded through the exponential map and an explicit choice of quotient map.
For Hopf algebras this notion is generalized under the name of Hopf duality. The natural Hopf dual to $U_q(\mathfrak{sl}(2,\mathbb{R}))$ is the quantum deformed coordinate algebra  $\mathcal{O}_q(\text
{SL}(2,\mathbb{R})) \equiv \text{SL}_q(2,\mathbb{R})$ \cite{Klimyk:1997eb}, which one may want to understand as the algebra of functions on the ``quantum group manifold''.\footnote{In the remaining text we will sometimes, in a slight imprecise manner use the name quantum group manifold to refer to this object.}  
This object is defined to be generated by arbitrary words in the generators $a,b,c,d$ that are normally arranged in the form of a $2\times 2$ matrix, 
\begin{equation}
\label{eq:generalmatrix}
    g \equiv \begin{pmatrix}
        a & b \\ c & d
    \end{pmatrix},
\end{equation}
for convenience and satisfy the  relations: 
\begin{align}
\label{eq:coordinatealgebra}
   ab = qba, \quad cd = qdc, \quad ac = qca, \quad bd &= qdb, \quad ad - da = (q - q^{-1})bc \\
    \det{}_q &\equiv ad - qbc = 1.
\end{align}
The last of these relations is the $q$-deformed version of restricting to unit determinant and is allowed since $\det\{\}_q$ is central. 
The coproduct is defined to implement the standard matrix product
\begin{equation}
\label{eq:coproductcoordinates}
    \Delta \left(g\right) \equiv \begin{pmatrix}
        a & b \\ c &d 
    \end{pmatrix} \otimes \begin{pmatrix}
        a & b \\ c &d 
    \end{pmatrix} = \begin{pmatrix}
    a \otimes a + b \otimes c & 
    a  \otimes b + b \otimes  d \\
    c \otimes  a + d \otimes  c & c \otimes b + d \otimes  d
    \end{pmatrix},
\end{equation}
which can equivalently be written as $\Delta(g_{ij}) = \sum_k g_{ik} \otimes g_{kj}$.
As is the case for JT-gravity, the correct Plancherel measure, reproducing the density of states of DSSYK is found by the quantum deformed version of restricting to the positive sub-semigroup. This is implemented by a formal restriction to positive self-adjoint operators $a,b,c,d$, following the detailed analysis in \cite{Ponsot:1999uf,Ip:2013}. This real form is denoted as $\text{
SL}^+_{0<q<1}(2,\mathbb{R})$.\footnote{ We note that the definition of the co-representation does not require a choice of star on the dual quantum group. This choice of star structure on the dual turns out to be rather subtle. We explore this in Appendix \ref{app:A2}.}

The relation of this quantum group and $U_q(\frak{sl}_2)$ is now realized in terms of a Hopf duality. Formally two algebras $\mathcal A$ and $\mathcal{U}$ are defined to be dual if there exists a bilinear $\braket{\cdot , \cdot}:\mathcal{U}\times \mathcal{A} \to \mathbb{C}$ such that 
\begin{align}
    \braket{X,g_1 g_2} &= \braket{\Delta(X),g_1 \otimes g_2}, \qquad\braket{X,1_\mathcal{A}} =\epsilon_\mathcal{U}(X), \\
    \braket{XY,g} &= \braket{X \otimes Y, \Delta(g)},
    \qquad \braket{1_\mathcal{U},g_1} = \epsilon_\mathcal{A}(g_1),
    \label{eq:hopfduality}
\end{align} 
for all $g_1,g_2 \in \mathcal{A}$, and $X,Y \in \mathcal{A}$.
In our case this bilinear can be defined by specifying how it acts on the generators of both algebras and takes the form
\begin{align}
\braket{K,a} = q, \quad \braket{K,d} = q^{-1},\quad 
\braket{E^+,b} =1,  \quad \braket{E^-,c} = 1,
\end{align} 
where these relations can be read as the $q$-deformed equivalent of saying that $E$ induces infinitesimal changes in $b$ and so on. 
This abstract definition can be brought into a more familiar form due to an argument by Fronsdal and Galindo \cite{Fronsdal:1991gf} by rewriting the duality as a generalization of the Gauss-Euler decomposition, making the interpretation as a generalized exponential map more explicit. This is achieved by introducing two (parabolic) coordinates $\gamma, \beta$ and a (hyperbolic) coordinate $\phi$ and parameterizing $g$ via
\begin{align}
    g=\begin{pmatrix}
        a&b\\c&d
    \end{pmatrix}= \begin{pmatrix}
        e^\phi & e^\phi \gamma \\ \beta e^\phi& e^{-\phi}+\beta e^\phi\gamma
    \end{pmatrix},
    \label{eq:gecoords}
\end{align}
where the commutation relations \eqref{eq:coordinatealgebra} are encoded in the relations 
\begin{align}
    e^\phi\gamma=q\gamma e^\phi, \qquad e^\phi\beta=q\beta e^\phi, \qquad [\beta, \gamma]=0,
\end{align}
of the new coordinates and  $\det_q(g)\equiv 1$ is automatically satisfied. 
The pairing in these coordinates then takes the form
\begin{align}
    \braket{H,\phi}=1/2, \qquad \braket{E^-,\beta}=\braket{E^+,\gamma}=1.
\end{align}
We can now view the Hopf duality as formally defining an action of the deformed universal enveloping algebra on the $q$-deformed space of functions on the group manifold. Classically one function contained in this space is the matrix valued function defining the fundamental representation of the group in terms of a choice of exponential coordinates on the group manifold (e.g. the coordinates given in \eqref{eq:gecoords}). This is the function that evaluated on the classical group element just produces the corresponding matrix in the representation in terms of these coordinates. In the classical setting, where the elements of the universal enveloping algebra are realized as differential operators, the duality is then encoded as $\braket{X_1^nX_2^m ..., t_{ij}} \equiv \partial_{t_1}^n \partial_{t_2}^m ... t_{ij}(ge^{t_1 X_1 }e^{t_2 X_2} ...)\vert_{g = e, t_i =0}$, where $t_{ij}$ are the matrix functions in e.g. the fundamental representation \cite{Klimyk:1997eb}. It is the quantum deformation of this map that leads to the Gauss-Euler decomposition given below in equation \eqref{eq:generalGE} with the change that the evaluation of this map at a single point is not well defined since the components of this matrix are no longer commutative \cite{Fronsdal:1991gf}.    
One can derive the explicit form of this map by using the definition of Hopf duality \eqref{eq:hopfduality} to induce the coproduct \eqref{eq:coproduct2} on one algebra from the product relations in the dual and vice versa. 

Following this argument one finds that this function can be expressed in terms of both generators and coordinates and takes the explicit form 
\begin{align}\label{eq:generalGE}
    g=e_{q^{-2}}^{\beta E^-}e^{2\phi H} e_{q^2}^{\gamma E^+} = e_{q^{-2}}^{\beta E^-} K^{-n} e_{q^2}^{\gamma E^+}.
\end{align}
and further forms an explicit corepresentation of the matrix quantum group \cite{Fronsdal:1991gf}. Here the quantum analogs of the parabolic subgroups are expanded in terms of specific $q$-deformed exponentials, defined as 
\begin{align}
    e_{q^2}^x\equiv \sum_{n=0}^\infty \frac{x^n}{[n]_{q^2}!}.
\end{align}

\subsubsection{Aside: Principal series (co)-representations from quantum M\"obius transformations}
\label{sec:principalseries}
From the perspective of the $q \to 1$ limit it is natural to interpret representations of $U_q(\mathfrak{sl}(2,\mathbb{R}))$ as infinitesimal, or since $n$ is discrete, minimal step-size versions of finite transformations along some carrier space where the finite transformations are to be interpreted as the action of the quantum deformed group $\text{SL}^+_q(2,\mathbb{R})$ that we just defined above. As such one may ask what the corresponding finite transformations correspond to for the principal series representations on the discrete halfline $\mathbb{R}_{q^2}$. We claim that these can be found by a generalization of the well-known realization of principal series representations of $\text{SL}(2,\mathbb{R})$ in terms of M\"obius transforms defined in (\ref{eq:classprincseries}). Specifically let us define the principal series corepresentation with label $j$ to act on the function space generated by monomials on the (positive) discretized line $\mathbb{R}_{q^2}$ via 
\begin{equation}
\boxed{
     f(x) \mapsto\phi(f(x)) \equiv N^{-j/2} (x z_{11} + z_{21})^j(xz_{12} + z_{22})^jx^{-j} f\left(\frac{xz_{11} + z_{21}}{xz_{12} + z_{22}} \right) N^{-j/2}}\, ,
     \label{eq:corep}
\end{equation}
where $z_{ij}$ are suitable coordinate functions on the quantum deformed coordinate ring \cite{Ip:2013}, such that the expression is independent of ordering and $xz_{ij} + z_{kl} \equiv x \otimes z_{ij} + 1\otimes z_{kl}$.
It is easily checked that this indeed furnishes a corepresentation of $\text{SL}_q(2)$ as we show in Appendix \ref{sec:AppendixA} where we also introduce the appropriate coordinates and definition of a corepresentation in detail. 
In terms of the generators $a,b,c,d$ the corepresentation can be rewritten as 
\begin{align}
\label{eq:pseriesgr}
    \phi(f(x)) = \left(xa+q^{-j+1/2}c\right)_q^j\left(d+xq^{-1/2}b\right)_{q^{-1}}^jx^{-j}f\left((xq^{-j+1/2}b+d)^{-1}(xa+q^{j-1/2}c)\right)
\end{align}
by using the commutation relations for $N$ given in (\ref{eq:commN}). Here we use the $q$-deformation of the binomial formula
\begin{align}
    (x + y)^n_q\equiv(x+y)(x+qy)\dots (x+ q^{n-1}y),
\end{align} 
and its the 
standard analytic continuation to arbitrary values of the exponent $\alpha$ defined via
\begin{align}
\left(x+y\right)^\alpha_q\equiv \frac{\left(x+y\right)^\infty_q}{\left(x+q^\alpha y\right)_q^\infty}.
\end{align}
One can see from \eqref{eq:pseriesgr} that having $a,b,c,d$ positive operators is consistent with $x>0$, either on a grid, or on the continuous real line $\mathbb{R}^+$. It remains to verify that the defined corepresentations are indeed dual to the principal series representations on $U_q(\mathfrak{sl}(2))$ defined above. 
Here dual is defined as follows.
Given any corepresentation $\phi : V \to V \otimes \mathcal{A}$ and a pairing of Hopf algebras $\braket{\cdot, \cdot} : \mathcal{U} \times \mathcal{A} \to \mathbb{C}$ we can induce a representation $\hat{\phi}$ of $\mathcal{U}$ on $V$ defined by
\begin{equation}
    \hat{\phi}(X) v \equiv (\text{id} \otimes \braket{X, \cdot}) \circ (\phi(g) v),
\end{equation}
where $v \in V, g \in \mathcal{A}$ and $X \in \mathcal{U}$.
We call a representation $\psi$ dual to a corepresentation $\phi$ if $\psi \simeq \hat{\phi}$  up to an equivalence of representations. 

Note that this notion of duality makes explicit how the choice of dual quantum group affects our model. Generally given any representation $\psi$ of the quantum deformed universal enveloping algebra there is no guarantee that this can be induced from a corepresentation of a specific dual Hopf algebra.\footnote{A trivial example of this is given by writing down the Hopf duality between $U_q(\mathfrak{g})$ and the trivial Hopf algebra consisting only of the identity element. In this case, due to the degeneracy of the pairing, only the trivial representation can be induced. However even for non degenerate pairings only certain representations can be induced \cite{Klimyk:1997eb}.} In classical Lie groups the equivalent statement is that representations of the universal enveloping algebra form representations of a specific group $G$ under the exponential map only if they are compatible with the quotient map used to define $G$. 

Deriving the explicit action of the generators 
$E^+,E^-,K$ on $f(x) \in L^2(\mathbb{R}_{q^2})$ is a somewhat tedious algebraic exercise and is presented in detail in Appendix \ref{sec:AppendixB}. There, using the repeated application of the defining properties of the Hopf algebra and the Hopf duality we derive that indeed 
\begin{align}
        K(f(x))  &\equiv \text{id} \otimes \braket{K,\cdot}(\phi(f(x)) =  R_{q^{2}} f(x) ,\\
        K_0 f(x) &\equiv \text{id} \otimes \braket{K_0,\cdot}(\phi(f(x))= f(x),\\
        E^+(f(x)) & \equiv \text{id} \otimes \braket{E^+,\cdot}(\phi(f(x)) = - x q^{-1/2} \frac{q^{-j} - q^{ j}R_{q^{-2}}}{q - q^{-1}} f(x),\\
        E^-(f(x)) &\equiv \text{id} \otimes \braket{E^-,\cdot}(\phi(f(x)) = \frac{q^{-1/2}}{x}\frac{q^jR_{q^2} - q^{- j}}{q - q^{-1}} f(x),
    \end{align}
reproducing the principal series generators defined in (\ref{eq:alegbragenerators2}).

\subsection{Gravitational matrix element}

Our proposal, including the twisted star, leads to a real form that is defined on real $q$, has unitary principal series representations and further allows for a natural implementation of the gravitational restriction to regular geometries. Furthermore, we can easily implement asymptotic boundary conditions. In the classical limit to the JT gravity regime, governed by SL$^+(2,\mathbb{R})$, these boundary conditions are implemented by evaluating the matrix element between two appropriate eigenstates. 
To be more precise, we consider a right parabolic eigenstate $\ket{\phi^+}$ associated to the right boundary which diagonalizes the action of $E^+$, and a left parabolic (conjugate) eigenstate $\bra{\phi^-}$ associated to the left boundary, which diagonalizes the right adjoint action of $E^-$. Evaluating the (quantum) group element $g$ in the continuous principal series representation between two such eigenstates then reproduces the two-sided gravitational wavefunction in terms of a \emph{mixed} parabolic matrix element, schematically depicted as 
\begin{align}
\raisebox{-0.45\height}{\includegraphics[height=1.7cm]{Figures/gravprop.pdf}}=\braket{\phi^-\vert g\vert \phi^+}
\end{align} 
In the case of the $U_q(\mathfrak{sl}(2,\mathbb{R}))$ $*$-relations \eqref{eq:ourstar}, the parabolic generators are proportional to themselves after the action of the involution $(E^\pm)^\dag=-q^{-1}K^{\pm 1}E^\pm$. The left Whittaker ket eigenvector $\ket{\phi^-}$ is then also a proper eigenfunction of the left parabolic generator $E^-$.\footnote{In the case of $U_q(\mathfrak{su}(1,1))$, the left eigenvector would have to be identical to the right eigenvector in order to diagonalize the adjoint action of $E^-$ under the star \eqref{eq:su(1,1)star}. In that case, we would talk about a ``right parabolic'' matrix element.}

This ``bottom-up'' analysis complements the earlier one presented in \cite{Blommaert:2023opb,Berkooz:2022mfk}, where the Whittaker function was found in a ``top-down'' fashion as an eigenfunction of the reduced Casimir operator, which reduces to the transfer matrix of DSSYK. Since the principal series generators explicitly diagonalize the Casimir \eqref{eq:casimireigenvalue}, the mixed parabolic matrix element in these representations is by construction a consistent solution. We present a Hamiltonian reduction argument of the full regular Casimir operator towards $q$-Liouville quantum mechanics in Appendix \ref{sec:Hamiltonianreduction}. 

\subsubsection{Whittaker vectors; or gravitational boundary eigenstates}
The Whittaker vectors $\phi^\pm_j(x)$ which implement these boundary conditions, are eigenfunctions on the real line which diagonalize the respective left and right parabolic generators $E^\pm$ evaluated in the principal series representations labeled by $j$. We propose the following definitions of the Whittaker vectors for $U_q(\mathfrak{sl}(2,\mathbb{R}))$:
\begin{align}
    \braket{x\vert \phi_j^+}\equiv\phi_j^+(x)&= \frac{1}{(q^2,q^2)_\infty\sqrt{1-q^2}}\,x^jE_{q^2}\left(-\frac{q^2}{1-q^2}\frac{1}{x}\right)\label{eq:rightwhit},\\ \braket{x\vert \phi_j^-}\equiv\phi^-_j(x)&= \frac{1}{(q^2,q^2)_\infty\sqrt{1-q^2}}\, x^{-j}E_{q^2}\left(-\frac{q^2}{1-q^2}x\right)\label{eq:leftwhit}.
\end{align}
The constant ($j$-independent) prefactor fixes the normalization of the Whittaker function later on. Using the definition of the $q$-deformed exponentials \eqref{eq:qexponential}, these Whittaker vectors diagonalize the generators $E^\pm$ \eqref{eq:principalseriesgenerators2} up to the action of the Cartan group element $K$:\footnote{The generalization of the $q$-deformed case has the option to consider a one-parameter family of extensions,
given by diagonalizing the parabolic generators up to the action of the Cartan generator $E^\pm \phi_j^\pm(x)\sim  K^{\pm \alpha} \phi_j^\pm(x)$. The concrete choice of $\alpha=-1$ has to be made to match with DSSYK, and can be motivated directly from various perspectives. Most importantly, within the first-order formulation of the bulk dilaton gravity model, these classical currents are directly related to the boundary values of the metric and dilaton field. The above combination is directly shown to correspond to fixed asymptotic fall-off boundary conditions at the boundary \cite{Blommaert:2023wad}.} 
\begin{align}
    E^+ \phi^+_j(x)&= -\frac{q^{-j-1/2}x^j}{(q^2,q^2)_\infty\sqrt{1-q^2}}\sum_{n=1}^\infty \frac{q^{n(n-1)}}{[n]_{q^2}!}\frac{1-q^{2n}}{q-q^{-1}}\left(-\frac{q^2}{1-q^2}\right)^nx^{-n+1}.
\end{align} 
Using \eqref{eq:qgamma} and shifting $n\rightarrow n+1$ yields 
\begin{align}
    E^+\phi_j^+(x)&= -\frac{q^{5/2}}{1-q^2}q^j K^{-1}\phi_j^+(x)\label{eq:rightaction}.
\end{align}
Similarly, the left action of $E^-$ is diagonalized up to the action of the Cartan generator: 
\begin{align}
    E^-\phi_j^-(x)&=-\frac{q^{5/2}}{1-q^2}q^jK\;\phi_j^-(x).
\end{align}  
The adjoint right action of $E^-$ on the other hand is exactly diagonalized using the star relations of $U_q(\mathfrak{sl}(2,\mathbb{R}))$ \eqref{eq:ourstar}, leading to  
\begin{align}
    \bra{\phi_j^-} E^-\equiv \left((E^-)^\dag\ket{\phi_j^-}\right)^\dag=\left(-q^{-1}K^{-1}E^-\ket{\phi_j^-}\right)^\dag =\frac{q^{1/2}}{1-q^2} q^{-j} \bra{\phi_j^-} \label{eq:adjointleftaction}
\end{align}
inside the inner product.

\subsubsection{Whittaker function; or two-boundary gravitational wavefunction}

Since the irreducible modules of $U_q(\mathfrak{sl}(2,\mathbb{R}))$ relate $q^2$-separated points, the Whittaker vectors  take unique solutions on the $\mathbb{R}_{q^2}$-line. 

In the enveloping algebra, hyperbolic elements are spanned by integer powers $K^{-n}$ of $K$. We can equivalently introduce the coordinate $\phi$ using the dictionary 
\begin{align}\label{eq:dictionary}
    \phi\equiv n|\log q|=-n\log q,
\end{align} 
to write the hyperbolic group element as an exponential $K^{-n}=e^{2\phi H}$. Following intuition of JT gravity, the hyperbolic label $\phi$ can be identified (up to a factor of 2) with the geodesic distance $\ell \equiv 2\phi$ between the two boundaries of the thermofield double state.
The Cartan element $H$ then generates the geodesic separation from one boundary to the other. The Whittaker function can then be constructed by propagating horizontally between the two boundaries from the right Whittaker vector to the left Whittaker vector along a geodesic distance $2\phi$, using the action of the Cartan generator: 
\begin{align}
   \raisebox{-0.45\height}{\includegraphics[height=1.7cm]{Figures/gravprop.pdf}}&=\braket{\phi_j^-\vert K^{-n}\vert \phi_j^+}=\int_0^{\infty/1}\frac{d_{q^2}x}{x} \braket{\phi_j^-\vert x}K^{-n}\braket{x\vert\phi_j^+}.
\end{align} 
Using the explicit form of the Whittaker vectors \eqref{eq:rightwhit}, \eqref{eq:leftwhit}, and the form of $j\equiv -1/2+i\theta/\log q^2$ required for unitarity, yields explicitly
\begin{align}
   \braket{\phi_j^-\vert K^{-n}\vert \phi_j^+} = 
   \frac{q^n q^{-\frac{2in\theta}{\log q^2}}}{(q^2,q^2)^2_\infty(1-q^2)}\int_0^{\infty/1} d_{q^2}x\;
 x^{\frac{2i\theta}{\log q^2}-1}E_{q^2}\left(-\frac{q^2}{1-q^2}x\right)E_{q^2}\left(-\frac{q^2}{1-q^2}\frac{q^{2n}}{x}\right)\label{eq:whittaker}.
\end{align}
Using the identity \eqref{eq:whitt}, we can finally evaluate the Whittaker function, 
\begin{align}
\label{eq:finalchordwhitt}
    \boxed{\braket{\phi_j^-\vert K^{-n}\vert \phi_j^+}=\frac{q^n}{(q^2;q^2)_n}H_n\left(\cos\theta\vert q^2\right),}
\end{align}
in terms of the known two-sided gravitational wavefunction of DSSYK.

\subsection{Summary}
\label{s:sum}

In conclusion of this section, the integral identity \eqref{eq:whitt} of the known gravitational wavefunctions of DSSYK allows us to pin down a convenient $q$-deformed real form which reproduces this answer as a constrained principal series matrix element (Whittaker function), in terms of the $U_q(\mathfrak{sl}(2,\mathbb{R}))$ algebra, but continued to real $0<q<1$. The latter differs in the star relations from the more conventional $U_q(\mathfrak{su}(1,1))$ quantum algebra. The resulting structure has the interpretation of a twisted star Hopf algebra.  Using a description in terms of $U_q(\mathfrak{sl}(2,\mathbb{R}))$ allows an easy implementation of the restriction to smooth geometries by committing to the positive semigroup, which has a correct classical limit to JT gravity. 
A novelty feature for the $q$-deformed representation theory is the discretization of the irreducible subspaces, which in turn guides us to choose the smaller Hopf (sub)algebra. The implementation of the gravitational boundary conditions is achieved by choosing eigenstates which diagonalize the parabolic generators $E^\pm$ respectively.
These more detailed features of the precise quantum group are invisible at the level of solving the Casimir eigenvalue equation.

\section{Gravitational applications}
\label{s:gravapp}
In this section, we apply our exact results obtained in the previous section to better understand the gravitational bulk dual of DSSYK. We start by shedding new light on some important properties of DSSYK (length discretization and length positivity), after which we describe single trumpet and brane amplitudes using our quantum group techniques. Our main application is presentated in subsection \ref{s:edges} where we explicitly construct an edge state factorization of the DSSYK gravitational Hilbert space.

\subsection{Length  discreteness in $U_q(\mathfrak{sl}(2,\mathbb{R}))$}
Two of the defining features of double-scaled SYK are the discreteness and positivity of the gravitational length quantum number. These conditions are obvious from the chord construction, and of vital importance in understanding both the thermodynamics and bulk interpretation of the model. As such it is natural to study under what guise these features show up in the group theoretic description of DSSYK. 

We have already discussed the emergent discreteness of the target space coordinate $x$ by reducing it to the irreducible subspaces of the principal series representation. The fine-graining of the grid depends on the choice of the specific $q$-deformed algebra (eg. $U_q(\mathfrak{sl}_2)$ vs $\breve{U}_q(\mathfrak{sl}_2)$). 

Discreteness of lengths, or equivalently the hyperbolic parameter, is equivalently implemented by the choice of quantum group. Indeed the natural identification of $n = -\frac{\phi}{\log q }$, where $\phi$ parameterizes translations along $H$ on the group manifold via $e^{2 \phi H} \equiv K^{-n}$, immediately implies that the ``quantum group manifold'' is discretized along this direction. This is captured in the $q$-deformed universal enveloping algebra $U_q(\mathfrak{sl}(2,\mathbb{R}))$ 
by defining the algebra to be generated by monomials in the parabolic generators $E^-,E^+$ and the group-like Cartan elements  $K = q^{2H}, K^{-1}$ as 
\begin{equation}
\label{eq:hopfal}
\mathcal{A} = \text{span}\{(E^-)^\ell K^{n}(E^+)^m\vert l,m\in \mathbb{N}_0,n\in \mathbb{Z} \}.
\end{equation}
Here the explicit inclusion of the ``exponentiated'' grouplike element $K$ is what encodes the discreteness of the chord number and can be understood as the non-existence of an infinitesimal translation along $n$ or equivalently $\phi$.  
The form of the Gauss-Euler decomposition \eqref{eq:generalGE} projects down to the algebra spanned by monomials of the exponentiated generator $K=q^{2H}$ upon discretizing the bulk length in terms of $n$ according to \eqref{eq:dictionary}.

Equivalently, if we explicitly evaluate the Cartan generator in the principal series representation as $K=R_{q^2}$ and commit to an irreducible subspace isomorphic to $\mathbb{R}_{q^2}$, it is obvious that the group exponentiation $e^{2\phi H}=R_{q^{-2n}}$ can only involve integer powers of $K$ to preserve the discrete grid. 

\subsection{Length positivity for boundary amplitudes}
Length positivity of the bare geodesic length $\ell_{\text{bare}}=2\phi_{\text{bare}}$ is a feature that has to be imposed by hand on any gravitational system, including JT gravity and DSSYK in particular. In the latter case positivity of the bare geodesic length is replaced by positivity of the chord number. This can be seen by studying the classical limit of DSSYK to JT gravity. Let us briefly review this limit.

This identification is found through e.g. taking the classical $q\rightarrow 1^-$ limit of the two sided wavefunction. Explicitly  we introduce a new coordinate $t=x/(1-q^2)$  in  \eqref{eq:whittaker} which in the classical limit leads to 
\begin{align}
   \braket{\phi_j^-\vert e^{2\phi H}\vert \phi_j^+}\, \rightarrow \,\, \frac{1-q^2}{(q^2;q^2)_\infty} e^{2\phi_rj} \int_0^\infty dt\;t^{2j}e^{-t}e^{-e^{-2\phi_r}/t},
\end{align}
where the discretized Jackson integral limits to a continuous Riemann integral. Note that the integration bounds are automatically constrained to the positive half-line in the classical limit. We have defined the renormalized gravitational length as
\begin{align}
    \frac{q^{2n}}{(1-q^2)^2}\equiv e^{-2\phi_r},
\end{align}
leading to the identification
\begin{align}
    \phi_r=-n\log q+\log(1-q^{2})\label{eq:ide tification}
\end{align} 
in the $q\to 1^-$ limit. We can directly compare this to the JT gravity two-sided wavefunction: 
\begin{align}
    \braket{\phi_{j,\nu}^-\vert e^{2\phi H}\vert \phi_{j,\lambda}^+}=\frac{e^{2\phi_r j}}{\nu}\int_0^\infty dt t^{2j} e^{-t}e^{-e^{-2\phi_r}/t}= e^{-\phi_r}K_{2ik}(2e^{-\phi_r}),
\end{align}
in terms of the renormalized variable $\sqrt{\nu\lambda}e^{-\phi}\rightarrow e^{-\phi_r}$, and the continuous momentum label $k$ defined from unitarity of the spin $j=-1/2+ik$. The Whittaker vectors diagonalizing the classical Borel-Weil $\mathfrak{sl}(2,\mathbb{R})$ generators \eqref{eq:parabolicsl2R} are given for general boundary eigenvalues as 
\begin{align}
    \braket{x\vert \phi_{j,\lambda}^+}= \phi_{j,\lambda}^+(x)=x^{j}e^{-\lambda/x}, \qquad \braket{x\vert \phi_{j,\nu}^-}= \phi_{j,\nu}^-(x)=\;x^{-j}e^{-\nu x},
\end{align}
satisfying $E^+\phi_{j,\lambda}^+(x)=-\lambda\; \phi_{j,\lambda}^+(x)$ and $E^-\phi_{j,\nu}^-(x)=-\nu\; \phi_{j,\nu}^-(x)$.
In the case of JT gravity, the Brown-Henneaux asymptotic AdS$_2$ boundary conditions are implemented in terms of the UV regulator $\epsilon$ 
\begin{align}
    \nu=\lambda=\frac{1}{\epsilon},  \qquad \phi_r=\phi+\log\epsilon\label{eq:renormparameter}.
\end{align}
Geometrically, this leads to the well-known renormalization of the bare geodesic length $\ell\equiv 2\phi$ which reaches the AdS boundary $\ell_{\text{bare}}=\ell_{\text{ren}}-\log \epsilon^2$.
Compared to the eigenvalues of DSSYK Brown-Henneaux boundary conditions in the classical limit, we can identify 
\begin{align}
    \nu=\lambda=\frac{1}{1-q^2},
\end{align} 
which leads to an identification of the UV regulator in the classical limit $q\rightarrow1^-$ with the underlying discreteness of the microscopic theory \begin{align}
    \epsilon=1-q^2\approx 2|\log q|.
\end{align}Using the dictionary of the bare hyperbolic parameter \eqref{eq:dictionary}, the renormalized parameter \eqref{eq:renormparameter} matches with the earlier identification \eqref{eq:ide tification}. Intuitively, the classical UV regulator matches with the multiplicative separation between two neighboring points of the underlying discrete grid.

Going back to the case of DSSYK this, in practice, implies that the exponentiation in the Gauss-Euler decomposition $e^{2\phi H}$ should be a priori constrained to $\phi\geq 0$, where this restriction should again be viewed as a condition on the eigenvalues of the operator. For lengths that reach the asymptotic boundary on either side, the bare lengths that show up in the gravitational wavefunction are renormalized, using \eqref{eq:renormparameter}, by the implementation of the gravitational boundary conditions. In the limit $\epsilon\rightarrow 0$ where we reach the AdS boundary, the renormalized parameter can take any value on the real line $\phi_{\text{ren}}\in \mathbb{R}$. Effectively, taking the gravitational coset on either side, the UV boundary conditions can be a priori absorbed in the definition of hyperbolic parameter in the Gauss-Euler decomposition, and extend its range from the positive line to the entire real axis. 

Within a more microscopic theory of quantum gravity, such as DSSYK, the bare lengths $\phi$ are discretized in terms of integer steps of the fundamental length scale $|\log q|$. Reaching the asymptotic boundary does not lead to a vanishing UV parameter $\epsilon=1-q^2$ in the finite $q$ regime. As such, the physical quantities are the bare lengths ``$n$'' which are \emph{not} renormalized and should be constrained a priori to be positive $n\geq0$. We can see that this is consistent in the construction of the Whittaker function \eqref{eq:whittaker} which automatically vanishes for negative lengths $n<0$.\footnote{Using the definition of the $q$-deformed exponential $E_{q^2}(.)$ \eqref{eq:qexponential}, we have for negative $m\leq 0$: 
\begin{align}
    E_{q^2}\left(-\frac{q^{2m}}{1-q^2}\right)\equiv (q^{2m},q^2)_\infty =  0,\qquad \text{if }m\leq0,
\end{align}
since for negative $m$ one has $\left(q^{2m};q^2\right)_\infty=\prod_{i=0}^\infty (1-q^{2m+2i})=0$ as the factor with $i= \abs{m}$ vanishes. Therefore, in the argument of the Jackson integral \eqref{eq:whittaker} which discretizes $x$ as $x=q^{2j}$, $j\in \mathbb{Z}$, we have that 
\begin{align}
    E_{q^2}\left(-\frac{q^{2(j+1)}}{1-q^2}\right) E_{q^2}\left(-\frac{q^{2(n - j +1)}}{1-q^2}\right),
\end{align}
can only be non-zero if $j\geq 0$ since otherwise the first exponential vanishes. If further $n<0$ we have that $n  - j + 1 \leq 0  $ and therefore the second exponential vanishes. If on the other hand $n\geq 0$, there exists a range $j=0,\dots, n$ where both exponentials survive and the Whittaker function is non-zero.} Thus, the only possible non-vanishing solutions of the gravitational wavefunctions can have positive bulk discretizations $n\geq 0$.

\subsection{Trumpets and branes in DSSYK from quantum group characters}
\label{s:trumbran}
While the method of chords is an extremely powerful tool for deriving exact DSSYK amplitudes on the disk, the formal extension of these tools to amplitudes including defects has thus far been somewhat elusive. Group theory seems to naturally include a description of such amplitudes in terms of character insertions in the path integral, according to our investigation in $\mathcal{N}=0,1,2,4$ JT (super)gravity in \cite{Belaey:2023jtr}. We apply these methods to the computation of the single trumpet and an end of the world brane amplitude which was first proposed by \cite{Okuyama:2023byh} and recently derived from the sine dilaton formulation of DSSYK in \cite{Blommaert:2025avl}. \\

\textbf{\emph{Trumpets}} \\
The calculation follows the equivalent amplitude in JT-gravity where the single trumpet can be produced via the insertion of a principal series character \cite{Mertens:2019tcm}. This character produces a defect that can be classified in terms of the monodromy around  the boundary circle. A macroscopic puncture producing the trumpet can be produced by the evaluation of the principal series character $\chi_j(g)$ on a hyperbolic group element i.e. $g \sim \text{diag}(e^{b/2},e^{-b/2})$.
The hyperbolic group parameter $b$ is then identified as the geodesic length along the neck of the trumpet. This calculation is now readily generalized to DSSYK by replacing the group theoretic character with its respective quantum group theoretic counterpart, schematically:
 \begin{align}
    Z_{\text{Trumpet}}(\beta, n) = \raisebox{-0.42\height}{\includegraphics[height=1.8cm]{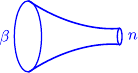}}=\int_0^{\pi}d\theta \;e^{-\beta E(\theta)}\textcolor{blue}{\chi_\theta(n)} .
    \end{align}
We define the character of the quantum algebra as:
\begin{equation}
\label{eq:charhy}
\chi_j(n) \equiv \text{Tr}_\chi(T_j(K^{-n})),
\end{equation}
for the representation $T_j$. In this definition, we focus on the subalgebra of \eqref{eq:hopfal} of elements generated solely by the hyperbolic generator $K$. Note that this only involves integer powers of $K$. Classically, every element in the hyperbolic conjugacy class is conjugate to one generated by the generator $K$. Such a definition has been presented in the past for compact quantum groups in \cite{Alvarez-Gaume:1988izd,Alvarez-Gaume:1989blj}.\footnote{The definition of character of a quantum group is not unique.  We define a character as the (regularized) sum of diagonal elements of the representation matrix which leads to a match with the canonical answer calculated in \cite{Blommaert:2025avl}. A more systematic treatment of what role differently defined characters (see e.g. \cite{Mudrov2007,Isaev:1991pg}) play in DSSYK will not be pursued here.} 

It remains to compute the explicit forms of both the highest weight and principal series character in the quantum group. For this calculation a few properties of this amplitude will turn out to be crucial.  We will further use that the defined representations can be constructed (on the deformed universal enveloping algebra) in a similar manner to classical representation theory of $\mathfrak{sl}(2,\mathbb{C})$ via raising and lowering operators \cite{Burban:1992ys} which we briefly outline in Appendix \ref{sec:AppendixD}. The character can then, whenever convergent, be computed as a series of (exponentiated) eigenvalues of the Cartan element.

The computation of the principal series character is somewhat more involved since the naive trace does not converge. This, however, should not be a surprise since we are dealing with infinite dimensional representations where characters are to be understood in the distributional sense and thus regularization may be required. We implement this indirectly via the following workaround.

As discussed in Appendix \ref{sec:AppendixD}, at integer values of $j$ the (analytically continued) principal series representations $T_j$ decompose into a direct sum of highest weight $T^-_j$, lowest weight $T^+_j$ and finite dimensional $T^{\text{FD}}_j$ representations:
\begin{equation}
\label{eq:relrep}
    T_j \simeq T^+_j \oplus T^\text{FD}_j \oplus T^-_j.
\end{equation}
As such the character satisfies
\begin{equation}
\label{eq:chardeco}
    \chi_j(n) = \chi^+_j(n) + \chi^{\text{FD}}_j(n) + \chi^-_j(n).
\end{equation}
Analytically continuing from $j \in \mathbb{N}$ to $j = -\frac{1}{2} + ik = -\frac{1}{2} + i\frac{\theta}{\log q^2}$ then yields the principal series character. Each of the three quantities on the right hand side of \eqref{eq:chardeco} can now be computed separately as follows.
\begin{itemize}
\item
Since $T^{\text{FD}}_j$ is finite dimensional, the character trivially converges trivially leading to 
\begin{align}
 \chi_j^{\text{FD}}(n) = \sum_{m = -j}^j q^{-2nm} = \frac{q^{n(2j + 1)} - q^{-n(2j + 1)}}{q^n - q^{-n}}.
\end{align}
\item
Since the highest weight discrete series representation is spanned by the states $\ket{m}$ with $m < -j$ we find
\begin{align}
    \chi^-_j(n) \equiv \text{Tr}(K^{-n}) = \hspace{-0.4cm}\sum_{m = -(j + 1)}^{-\infty}\hspace{-0.2cm} \braket{m \vert K^{-n} \vert m } = \hspace{-0.4cm}\sum_{m = -(j + 1)}^{-\infty}\hspace{-0.2cm} q^{-2mn} = \frac{q^{2n(j + 1)}}{1 - q^{2n}} = \frac{q^{n(2j+1)}}{q^{-n}-q^n},
\end{align}
where we have used that the states $\ket{m}$ form an eigenbasis of $K$ with $K\ket{m} = q^{2m}\ket{m}$ and the trace thus simply reduces to a geometric series.  
\item
The lowest weight character $\chi^+_j(n)$ needs to be regularized as the series does not converge. We will implement this implicitly by instead show that on hyperbolic elements $\chi^+_j(n) = \chi_j^-(n)$. This follows from a $q$-generalization of Bargmann's automorphism \cite{Vilenkin}. 
The map $\phi: U_q(\frak{sl}_2) \to U_q(\frak{sl}_2)$ defined by $\phi(K) = K $ and $ \phi(E^{\pm}) = - E^{\pm}$ is a Hopf-automorphism of $U_q(\frak{sl}_2)$.\footnote{On the dual quantum group this corresponds to sending $\begin{bmatrix}
    a & b \\ c &d  
\end{bmatrix} \mapsto \begin{bmatrix}
    a & -b \\ -c &d  
\end{bmatrix}. $}
Applying this automorphism to the relations defining $T_j$ specified in \eqref{eq:modulessu111} - \eqref{eq:modulesu1F} while simultaneously relabeling $n \to -n$ maps $T_j^+ \mapsto T_j^{-}$ and vice versa such that $T^+_j(g) = T^-_j(\phi(g))$. Further since $\phi$ acts trivially on $K$ we find that for all of elements of the form $K^{-n}$ we get $T^+_j(K^{-n}) = T^-_j(K^{-n}) \implies \chi_j^+(n) = \chi^-_j(n)$ showing the equality of the characters. 
\end{itemize}

Combining these results we find the principal series character to be given by
\begin{align}
\label{eq:pcch}
    \chi_j(n) = 2 \chi_j^-(n) + \chi_j^\text{FD}(n) &=  \frac{q^{n(2j + 1)} + q^{-n(2j + 1)}}{q^{-n} - q^{n}} 
        = \frac{\cos(n\theta)}{\sinh (n \log q)},
\end{align}
where in the last equality we have used $j  = -\frac{1}{2} + i\frac{\theta}{\log q^2}$ and the fact that $q < 1$. The denominator in the above formula plays the role of the Weyl denominator, and should be stripped off in actual calculations of single trumpet amplitudes. We denote this character as $\chi_\theta(n) = \cos n \theta$ and obtain
 \begin{align}
Z_{\text{Trumpet}}(\beta, n) = \frac{1}{\pi}\int_0^{\pi}d\theta \;e^{\beta \cos\theta} \cos(n\theta) = I_n(\beta),
\end{align}
matching known results \cite{Okuyama:2023byh,Blommaert:2025avl}, in terms of the modified Bessel function of the first kind $I_n(.)$.

Note that the discretization of $n$ follows completely from our choice of quantum group, restricting us to only integer powers of $K$ in \eqref{eq:charhy}. So within our framework, this discretization is one and the same as all of the previous discretizations and these do not require separate proofs. 

The (normalized) principal series characters $\chi_\theta(n)$ satisfy completeness and orthogonality relations as:
\begin{align}
\label{eq:compl}
\frac{1}{\pi}\sum_{n=-\infty}^{+\infty} \cos (n\theta) \cos (n\theta') &= \delta(\theta-\theta'), \qquad \theta \in [0,\pi]\, , \\
\int_0^\pi d\theta \cos (n\theta) \cos (m\theta) &= \frac{\pi}{2} (\delta_{n,m} + \delta_{n,m}\delta_{n,0}) \ .
\end{align}
The denominator of \eqref{eq:pcch} is suggestively interpretable as the Weyl denominator. Indeed, 
for SL$(2,\mathbb{R})$ 
$
\vert \Delta(t)\vert^2 = \sinh^2\phi$. If we now realize that in the $q$-deformed set-up we have the relation
\begin{equation}
K^n = e^{2n\log q H} =e^{2\phi H}
\end{equation}
we obtain $
\vert \Delta(t)\vert = \sinh (n\log q)$, allowing us to reinterpret the completeness relation \eqref{eq:compl} as
\begin{align}
\frac{1}{\pi}\sum_{n=-\infty}^{+\infty} \vert \Delta(n)\vert^2 \chi_j(n) \chi_{j'}(n) &= \delta(\theta-\theta'), \qquad \theta \in [0,\pi]\, ,
\end{align}
in terms of the original characters \eqref{eq:pcch}, but now including the ``Jacobian'' in a discretized Weyl integration formula. This morally matches with the discussion on the (Jackson) discretized Haar measure for this quantum group in section 3.3 in \cite{Blommaert:2023opb}. \\

\textbf{\emph{Branes}} \\
As demonstrated explicitly in \cite{Belaey:2023jtr, Blommaert:2021etf} for applications of JT gravity, the contribution of the degrees of freedom of an EOW brane particle is produced via a character insertion $\chi^-_j(b) = \Tr_j (e^{bH})$ in the highest weight discrete series representation. Here the highest weight $j$ specifies the mass $\mu$ of the probe particle described by the EOW brane in a way to be specified below.
This leads to the amplitude for the EOW brane amplitude:
 \begin{align}
        Z_{\text{EOW}}(\beta, j) = \raisebox{-0.42\height}{\includegraphics[height=1.8cm]{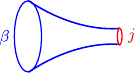}}=\sum_{n=0}^\infty \int_0^{\pi}d\theta \;e^{-\beta E(\theta)}\textcolor{blue}{\chi_\theta(n)} \textcolor{red}{\chi^-_j(n)} \nonumber,
    \end{align}
where the geodesic length along the neck of the trumpet is now discretized, as expected from the discrete nature of the chord number $n$. The explicit final formula matches that of \cite{Okuyama:2023byh}:
 \begin{align}
        Z_{\text{EOW}}(\beta, j) =\sum_{n=0}^\infty \int_0^{\pi}d\theta \;e^{\beta \cos\theta} \cos(n\theta) \frac{q^{2n(j + 1)}}{1 - q^{2n}},
\end{align}
upon identifying the brane tension $\mu = j+1/2$.

\subsection{Edge states in DSSYK }
\label{s:edges}
In lower-dimensional gauge theory models, Hilbert space factorization can be done explicitly by using the defining property of a representation: 
\begin{align}\label{equaiton}
    R_{ab}(g_1\cdot g_2)=\sum_s R_{as}(g_1)R_{sb}(g_2) ,\qquad g_1,g_2\in G, \qquad a,b,s=1,\dots, \text{dim}(R).
\end{align} 
By rewriting the two-sided gravitational amplitudes of DSSYK in terms of suitable representation matrices of a $q$-deformed gauge theory, we can interpret the above expression as the factorization of the two-sided Hilbert space into the product of one-sided gravitational wavefunctions with the edge labels $s$ living at the entangling surface of a black hole horizon according to one-sided observers. As in applications in JT gravity \cite{Blommaert:2018iqz} and Liouville gravity \cite{Mertens:2022aou}, this label $s$ should be a ``hyperbolic'' index, to obtain a complete set of states in the gravity model.
Discretizing the carrier space coordinate in terms of $x=q^{2n}$, we can write down properly normalized hyperbolic eigenstates with real label $s$
\begin{align}
    \braket{x\vert s}=\frac{x^{\frac{is}{\log q^2}}}{\sqrt{2\pi(1-q^2)}}=\frac{e^{ins}}{\sqrt{2\pi(1-q^2)}}\equiv  \braket{n\vert s}.
\end{align} 
These states readily diagonalize the Cartan generator $K$ as 
\begin{align}
    K \braket{n\vert s}=e^{is} \braket{n\vert s}.
\end{align} 
We can check the normalization of these states with respect to the conventional inner product \eqref{eq:innerproduct} as: 
\begin{align}
    \braket{s_1\vert s_2}&=\int_0^{\infty/1} \frac{d_{q^2}x}{x}\braket{s_1\vert x}\braket{x\vert s_2}=\frac{1}{2\pi}\sum_{n=-\infty}^\infty e^{in(s_2-s_1)}
    =\sum_{n=-\infty}^\infty\delta(s_1-s_2+2\pi n).
\end{align}
The $2\pi$ periodicity is a remnant of the discreteness of the dual space $\braket{n\vert s+2\pi}=\braket{n\vert s}$.
We can determine an unambiguous normalization by restricting the hyperbolic label to the first Brillouin zone: $-\pi<s<\pi$, such that $\braket{s_1\vert s_2}=\delta(s_1-s_2)$.\footnote{At $q\rightarrow1^-$, we obtain the classical SL$(2,\mathbb{R})$ hyperbolic eigenstates $\braket{x\vert s} = \frac{1}{\sqrt{2\pi}} x^{is}$, where now $s$ extends along the entire real line.}  The hyperbolic matrix element evaluated in a hyperbolic basis describes the interior wavefunction with no holographic boundaries, and is given by
\begin{align}
\raisebox{-0.46\height}{\includegraphics[height=2.5cm]{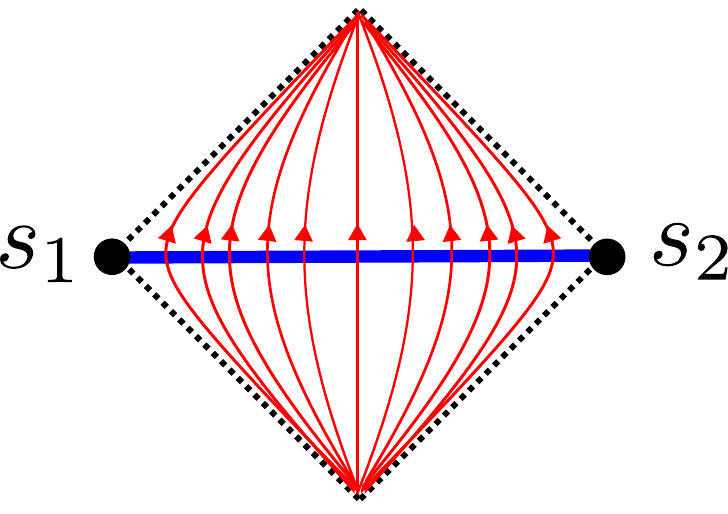}}\hspace{0cm} \,\, = \braket{s_1\vert K^{-n}\vert s_2}&=e^{-is_2n}\;\delta(s_1-s_2).\label{eq:hyperbolicCartan}
\end{align}
Besides being orthogonal, the hyperbolic eigenstates are also complete in the first periodic zone.  We thus have the generic Fourier decomposition with respect to the Jackson integral: 
\begin{align}
    f(x)&=\frac{1}{\sqrt{2\pi(1-q^2)}}\int_{-\pi}^\pi ds\; x^{\frac{is}{\log q^2}}F(s), \quad
    F(s)=\frac{1}{\sqrt{2\pi(1-q^2)}}\int_0^{\infty/1} \frac{d_{q^2}y}{y}y^{-\frac{is}{\log q^2}}f(y).
\end{align}
Applied to the right Whittaker vector $\braket{x\vert \phi^+}$ \eqref{eq:rightwhit}, we can determine the basis transformation to the hyperbolic basis states as the Fourier transform 
\begin{align}
    \braket{s\vert \phi^+_j}&=\int_0^{\infty/1} \frac{d_{q^2}x}{x}\braket{s\vert x}\braket{x\vert \phi^+_j}\\ &=\frac{1}{\sqrt{2\pi}(1-q^2)(q^2;q^2)_\infty}\int_0^{\infty/1} \frac{d_{q^2}x}{x}x^{j-is/\log q^2}E_{q^2}\left(-\frac{q^2}{1-q^2}\frac{1}{x}\right),
\end{align}
substituting $x\rightarrow 1/x$ and using the integral representation of the $q$-gamma function \eqref{eq:qbeta} yields 
\begin{align}
\label{eq:rightbosonicedge}
    \braket{s\vert \phi^+_j}=\frac{1}{\sqrt{2\pi}(1-q^2)(q^2;q^2)_\infty}\Gamma_{q^2}\left(\frac{is-i\theta}{\log q^2} +\frac{1}{2}\right)(1-q^2)^{i(s-\theta)/\log q^2+1/2}.
\end{align} 
Using the definition of the $q$-gamma function in terms of the $q$-Pochhammer symbol \eqref{eq:qpochammergamma}, this is written more compactly as 
\begin{align}
s \hspace{0.1cm}
\raisebox{-0.46\height}{\includegraphics[height=2.8cm]{Figures/rightedge.pdf}}\hspace{0cm}=\braket{s\vert \phi^+_j} & =\frac{1}{\sqrt{2\pi}}\frac{1}{(qe^{is-i\theta};q^2)_\infty}.
\end{align}
The inverse Fourier transform rewrites the right Whittaker vector in terms of an edge state decomposition 
\begin{align}
    \braket{x\vert \phi^+_j}&=\int_{-\pi}^\pi ds \braket{x\vert s}\braket{s\vert \phi_j^+}=\frac{(1-q^2)^{-3/2}\log q^2}{2\pi(q^2,q^2)_\infty}x^j \int_{\mathcal{C}}d\zeta \;x^{-i\zeta}(1-q^2)^{-i\zeta}\Gamma_{q^2}(-i\zeta),
\end{align}
where again $j = -1/2 + i \theta/\log q^2 $ and we have deformed the contour $\mathcal{C}$ to go from $[-\frac{\pi}{\log q^2}, \frac{\pi}{\log q^2}[$ while avoiding the pole of the Gamma function at $\zeta=0$ from above. Using the recursive property $\Gamma_{q^2}(-i\zeta+1)=[-i\zeta]_{q^2}\Gamma_{q^2}(-i\zeta)$, one can easily check that this expression correctly  diagonalizes the right parabolic generator $E^+$ up to the action of $K$ as \eqref{eq:rightaction}. 

Similarly, applied to the (adjoint) left Whittaker vector $\braket{\phi_j^-\vert x}$ \eqref{eq:leftwhit}, we can determine the Fourier transformation with respect to the hyperbolic basis 
\begin{align}\label{eq:leftbosonicedge}
    \braket{\phi^-_j\vert s}&=\int_0^{\infty/1} \frac{d_{q^2}x}{x}\braket{\phi_j^-\vert x}\braket{x\vert s}\\&=\frac{1}{\sqrt{2\pi}(1-q^2)(q^2;q^2)_\infty}\Gamma_{q^2}\left(\frac{is+i\theta}{\log q^2}+\frac{1}{2}\right)(1-q^2)^{i(s+\theta)/\log q^2+1/2},
\end{align}
leading to 
\begin{align}
    \raisebox{-0.46\height}{\includegraphics[height=2.8cm]{Figures/leftedge.pdf}}\hspace{0.1cm} s=\braket{\phi_j^-\vert s}=\frac{1}{\sqrt{2\pi}}\frac{1}{(qe^{is+i\theta};q^2)_\infty},
\end{align} 
and the edge state decomposition of the left Whittaker vector 
\begin{align}
    \braket{x\vert \phi_j^-}=\frac{(1-q^2)^{-3/2}\log q^2}{2\pi(q^2,q^2)_\infty}x^{-j}\int_{\mathcal{C}}d\zeta\; x^{i\zeta} \Gamma_{q^2}(-i\zeta)(1-q^2)^{-i\zeta},
\end{align}
where the contour $\mathcal{C}$ is defined in the same way. Using these integral expressions for the Whittaker vectors, the Whittaker function can equivalently be written as 
\begin{align}
    &\braket{\phi_j^-\vert K^{-n}\vert \phi_j^+}=\int_0^\infty \frac{d_{q^2}x}{x}\braket{\phi_j^-\mid x}K^{-n}\braket{x\mid \phi_j^+}\\&=\frac{q^n}{2\pi (1-q^2)^2(q^2;q^2)^2_\infty}\int_{-\pi}^\pi dse^{-isn}(1-q^2)^{\frac{2is}{\log q^2}}\label{eq:gammawhittakerexp}
    \Gamma_{q^2}\left(\frac{is + i\theta }{\log q^2}\right)\Gamma_{q^2}\left(\frac{is - i\theta}{\log q^2}\right),
\end{align}
where the contour now avoids the poles at $s=\pm \theta$ from below. Using the recursive properties of the $q$-gamma function, one can easily check that this function indeed satisfies the $q$-Liouville recurrence equation \eqref{eq:bosonicrecursion} or equivalently that this is indeed a solution of the DSSYK transfer matrix as required by consistency.

The hyperbolic basis transformations determine the one-sided gravitational wavefunction with one holographic boundary \begin{align}
\label{eq:onewf1}
    \braket{s\vert K^{-n}\vert \phi^+_j} &=\int_{-\pi}^\pi ds_1\braket{s\vert K^{-n}\vert s_1}\braket{s_1\vert \phi^+_j}= \frac{1}{\sqrt{2\pi}}\frac{e^{-ins}}{(qe^{is-i\theta};q^2)_\infty}, \\
    \label{eq:onewf2}
    \braket{\phi^-_j\vert K^{-n}\vert s} &=\int_{-\pi}^\pi ds_1\braket{\phi^-_j\vert s_1}\braket{s_1\vert K^{-n}\vert s}= \frac{1}{\sqrt{2\pi}}\frac{e^{-ins}}{(qe^{is+i\theta};q^2)_\infty}.
\end{align}
Note that we require $n$ in these expressions to be a non-negative integer, in order for these one-sided wavefunctions to be proper representation matrices of our quantum group $U_q(\mathfrak{sl}(2,\mathbb{R}))$.

The two-sided gravitational wavefunction of DSSYK can then be factorized along both horizons using these edge state amplitudes 
\begin{align}
\raisebox{-0.45\height}{\includegraphics[height=1.8cm]{Figures/gravprop.pdf}}\;&=\int_{-\pi}^{\pi} ds_1\int_{-\pi}^{\pi} ds_2\ \raisebox{-0.47\height}{\includegraphics[height=2.6cm]{Figures/leftedge.pdf}}\hspace{0.1cm}\raisebox{-0.47\height}{\includegraphics[height=2.5cm]{Figures/hyperbolic.pdf}}\hspace{0.1cm}\raisebox{-0.47\height} {\includegraphics[height=2.6cm]{Figures/rightedge.pdf}},\\
  \qquad  \braket{\phi^-_j\vert K^{-n}\vert \phi^+_j}&= \frac{1}{2\pi}\int_{-\pi}^\pi ds_1\int_{-\pi}^\pi ds_2\frac{e^{-in_1 s_1}}{(qe^{is+i\theta};q^2)_\infty}e^{-in_2s_2}\delta(s_1-s_2)\frac{e^{-in_3s_2}}{(qe^{is-i\theta};q^2)_\infty},
\end{align}
which is a rewriting of \eqref{eq:gammawhittakerexp}, where we are free to partition $n=n_1+n_2+n_3$ in three integers in any way we want.
We can rewrite this expression in terms of the defining $q$-Hermite identity \eqref{eq:hermiteidentity} by shifting $s\rightarrow s-i\log q$:\footnote{We also perform a contour shift along an imaginary direction. One can drop the vertical pieces since the integrand itself is $2\pi$-periodic.}
\begin{align}
    \braket{\phi^-_j\vert K^{-n}\vert \phi^+_j}
    =\frac{q^n}{2\pi}\int_{-\pi}^{\pi}dp\frac{e^{-inp}}{(e^{ip- i\theta};q^2)_\infty(e^{ip+ i\theta};q^2)_\infty}=\frac{q^n}{(q^2;q^2)_n}H_n(\cos\theta \vert q^2).
\end{align}  
Therefore, we can physically interpret the integral identity \eqref{eq:hermiteidentity} as an edge state factorization of the two-sided gravitational wavefunction along an entangling surface.

In the case of BF gauge theory with a compact gauge group, the Euclidean disk amplitude (with coset boundary conditions) can be written using a one-sided angular slicing as $\text{Tr}e^{-\beta H}$ over a one-sided Hilbert space, as:
\begin{align}
\label{eq:BFangular}
\raisebox{-0.48\height}{\includegraphics[height=2.5cm]{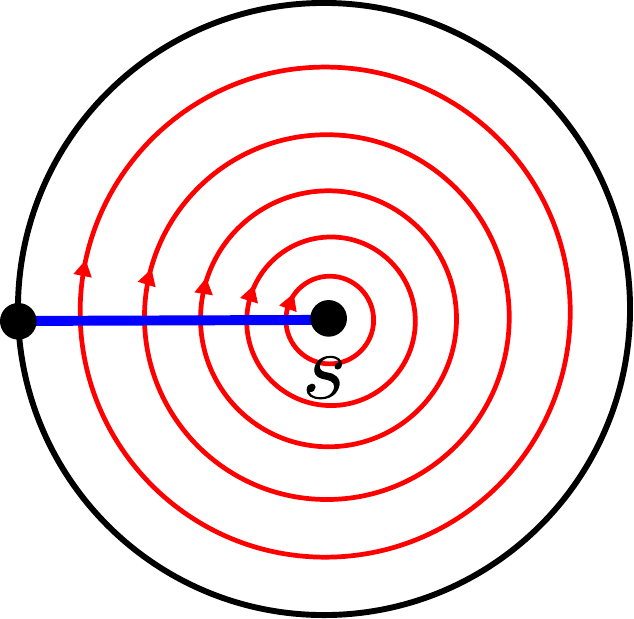}} = \sum_{R} \text{dim}R  e^{-\mathcal{C}_R \beta} = \frac{1}{V_G} \int_G dg \sum_R \sum_{s=1}^{\text{dim}R}\text{dim}R \, e^{-\mathcal{C}_R \beta} \, R_{\mathfrak{i}s}(g) R_{s \mathfrak{i}}(g^{-1}).
\end{align}
This last way of writing is implementing $\text{Tr}e^{-\beta H}$ with the one-sided coordinate space orthonormal wavefunctions
\begin{equation}
\langle g \vert R,\mathfrak{i},s \rangle \equiv \sqrt{\text{dim}R} \, R_{\mathfrak{i}s}(g).
\end{equation}
Such an angular slicing is physically important, since it allows us to visualize the degrees of freedom at the black hole horizon, edge states labeled by $s$ here, at the tip of the cigar. Demanding the result of this calculation is the same as that of the two-sided evaluation of the disk amplitude corresponds to implementing a shrinkability constraint on the point-like defect in the middle of the Euclidean disk, i.e. the point is not really there geometrically \cite{Donnelly:2018ppr,Hung:2019bnq,Blommaert:2018iqz,Mertens:2022ujr}. 

We claim that it is possible to rewrite the DSSYK disk partition function in the same way. In order to do that, we note that in \eqref{eq:BFangular} the $G$-integral is trivial, and one can reduce $G$ to solely a maximal torus $T$ by equivalently rewriting
\begin{align}
\label{eq:onesug}
\frac{1}{V_T}\int_T dh \sum_R  \sum_{s=1}^{\text{dim}R} \text{dim}R\, e^{-\mathcal{C}_R \beta} \,  \, R_{\mathfrak{i}s}(h) R_{s \mathfrak{i}}(h^{-1}),
\end{align}
where we decomposed $G \simeq G/T \times T$.

Utilizing the above one-sided gravitational wavefunctions \eqref{eq:onewf1}, \eqref{eq:onewf2}, one can then write down the DSSYK analogue of \eqref{eq:onesug}:
\begin{align}
\frac{1}{V_T}\sum_{n=0}^{+\infty}\int_{0}^{\pi}d\theta\, \int_{-\pi}^{\pi}ds\, (e^{\pm 2i\theta};q^2)_\infty e^{-\beta E(\theta)} \frac{1}{\sqrt{2\pi}}\frac{e^{-isn}}{(qe^{is-i\theta};q^2)_\infty} \times \frac{1}{\sqrt{2\pi}}\frac{e^{+isn}}{(qe^{is+i\theta};q^2)_\infty},
\end{align}
where $V_T = \sum_{n=0}^{+\infty}1$ is the analogue of the volume of the maximal torus on the discretized quantum group manifold. Due to the non-compactness of both the group and the representation of interest, there are as usual divergences appearing in these equations, indicating that the underlying model is not fully microscopic. 
This expression can be readily found to agree with the DSSYK disk partition function as:
\begin{align}
&\frac{\left(\sum_{n=0}^{+\infty}1\right)}{V_T}\int_{0}^{\pi}d\theta (e^{\pm 2i\theta};q^2)_\infty e^{-\beta E(\theta)} \frac{1}{2\pi}\int_{-\pi}^{\pi}ds \frac{1}{(qe^{is-i\theta};q^2)_\infty}\frac{1}{(qe^{is+i\theta};q^2)_\infty}  \nonumber \\
&=\int_{0}^{\pi}d\theta\, (e^{\pm 2i\theta};q^2)_\infty e^{-\beta E(\theta)} = Z_{\text{DSSYK}}(\beta).
\end{align}
In the first line, the sum over $n$ factorizes and cancels with $V_T$. In the second line, we evaluated the integral over $s$ using the $q$-Hermite integral identity \eqref{eq:hermiteidentity} for the case $n=0$.

The angular slicing itself corresponds to inserting a shrinkable boundary in the middle of the Euclidean disk. In a closed channel slicing, this can be thought of as a brane boundary, called an entanglement (or $E$-) brane, studied and defined in various contexts in \cite{Donnelly:2016jet,Donnelly:2018ppr,Hung:2019bnq,Donnelly:2020teo}. This brane state is essentially the unit group element (or trivial holonomy) state, which can be expanded in the irrep basis $\{\ket{\theta}; \theta \in [0,\pi]\}$ as:\footnote{This equality is formal, since it depends on volume regulators for non-compact groups (see e.g. appendix B.2 of \cite{Mertens:2022ujr}).}
\begin{equation}
\ket{e} \sim \int_{0}^{\pi} d\theta \, (e^{\pm 2i \theta};q^2)_{\infty} \, \ket{\theta}.
\end{equation}

\section{Extension to $\mathcal{N}=1$ DSSYK}
\label{s:Nis1}
We can use the lessons of the bosonic representation theory to deduce the gravitational wavefunctions of $\mathcal{N}=1$ DSSYK by the representation theory of the $q$-deformed orthosymplectic OSp$_q(1|2,\mathbb{R})$ quantum group \cite{Kulish:1989sv}. See appendix B of \cite{Blommaert:2023opb} for a recent reference of the $q$-deformed coordinate algebra in this context.

\subsection{The gravitational real form of $\mathcal{N}=1$ DSSYK}
The $\breve{U}_q(\mathfrak{osp}(1|2,\mathbb{R}))$ quantum algebra consists of the Cartan generator $K$ and two fermionic generators $F^-,F^+$, satisfying the (opposite) relations 
\begin{align}
    KF^\pm=q^{\pm 1/2}F^\pm K, \qquad \{F^+,F^-\}=-\frac{K^2-K^{-2}}{8(q^{1/2}-q^{-1/2})}.
\end{align}
Again identifying $K=q^H$ and $q=e^{-h}$ ($h>0$), leads to the infinitesimal $\breve{U}_h(\mathfrak{osp}(1|2,\mathbb{R}))$ version of the algebra 
\begin{align}
\label{eq:infinitsuper}
    [H,F^\pm]=\pm\frac{1}{2}F^\pm, \qquad \{F^+,F^-\}=-\frac{\sinh 2hH}{8\sinh h/2}.
\end{align}
In the classical $h\rightarrow0$ limit, this becomes the familiar opposite $U(\mathfrak{osp}(1|2,\mathbb{R}))$ algebra 
\begin{align}
    [H,F^\pm]=\pm\frac{1}{2}F^\pm, \qquad \{F^+,F^-\}=-\frac{1}{2}H.
\end{align}
A compatible $\mathbb{Z}_2$-graded coproduct is\footnote{For graded Hopf algebras $\mathcal{A}_i, \;i\in \mathbb{N}_0$, with $\mathcal{A}_i\mathcal{A}_j\subseteq \mathcal{A}_{i+j}$, the product $\cdot$ of two objects $a\otimes a_i\in \mathcal{A}\otimes \mathcal{A}_i$, $a_j'\otimes a'\in \mathcal{A}_j\otimes \mathcal{A}$ is defined by $a\otimes a_i\cdot a_j'\otimes a'=(-1)^{ij}aa_j'\otimes a_ia'$. }
\begin{align}
    \Delta(F^-)=F^-\otimes K^{-1}+K\otimes F^-, \,\, \Delta(F^+)=F^+\otimes K^{-1}+K\otimes F^+, \,\, \Delta(K)=K\otimes K.
\end{align}
There exists an ``sCasimir'' operator which commutes with all bosonic generators while anticommuting with all fermionic generators, given by \begin{align}
    Q=\frac{q^{1/2}K^2-q^{-1/2}K^{-2}}{4(q-q^{-1})}+2F^-F^+=\frac{q^{1/2}K^{-2}-q^{-1/2}K^2}{4(q-q^{-1})}-2F^+F^-.
\end{align}  The quadratic Casimir is obtained by squaring the sCasimir, yielding an operator which commutes with all combinations of generators  
\begin{align}
\label{eq:supercasimir}
    \mathcal{C}=Q^2=\frac{qK^4+q^{-1}K^{-4}-2}{16(q-q^{-1})^2}-\frac{(qK^2+q^{-1}K^{-2})F^-F^+}{2(q^{1/2}+q^{-1/2})}-4(F^-)^2(F^+)^2.
\end{align}
We can define a consistent set of principal series generators of $\breve{U}_q(\mathfrak{osp}(1|2,\mathbb{R}))$ on the superline $(x\vert \vartheta)\in \mathbb{R}^{1|1}$ with a real bosonic coordinate $x$ and a fermionic Grassmann number $\vartheta$ as 
\begin{align}
    K&=R_q,\\
    F^+&=\frac{x^{1/2}}{2\sqrt{2}}\left(\frac{q^{-j}R_q-q^{j}R_{q^{-1}}}{q^{1/2}-q^{-1/2}}\partial_{\vartheta}+\frac{q^{-j}R_q+q^{j}R_{q^{-1}}}{q^{1/2}+q^{-1/2}}\vartheta\right),\\
    F^-&=-\frac{x^{-1/2}}{2\sqrt{2}}\left(\frac{q^{j}R_q-q^{-j}R_{q^{-1}}}{q^{1/2}-q^{-1/2}}\partial_\vartheta+\frac{q^{j}R_q+q^{-j}R_{q^{-1}}}{q^{1/2}+q^{-1/2}}\vartheta\right).
\end{align}
The bosonic generators $E^\pm$ are defined in the enveloping algebra through $E^\pm\equiv \mp4(F^\pm)^2$, leading to
\begin{align}
    E^+&=-\frac{x}{2}\left(\frac{q^{-2j+1/2}R_{q^2}-q^{2j-1/2}R_{q^{-2}}+q^{1/2}-q^{-1/2}}{q-q^{-1}}-\frac{2}{q^{1/2}+q^{-1/2}}\vartheta\partial_\vartheta\right),\\
    E^-&=\frac{1}{2x}\left(\frac{q^{2j-1/2}R_{q^2}-q^{-2j+1/2}R_{q^{-2}}+q^{-1/2}-q^{1/2}}{q-q^{-1}}+\frac{2}{q^{1/2}+q^{-1/2}}\vartheta\partial_\vartheta\right).
\end{align}
The irreducible representations of $\breve{U}_q(\mathfrak{osp}(1|2,\mathbb{R}))$ act again on $q$-separated points of the bosonic coordinate $x$, while the discretization does not affect $\vartheta$. These generator expressions are the same as those constructed first for the modular double $U_q(\mathfrak{osp}(1|2,\mathbb{R})) \otimes U_{\tilde{q}}(\mathfrak{osp}(1|2,\mathbb{R}))$ in \cite{Hadasz:2013bwa,Pawelkiewicz:2013wga}.

With respect to the inner product for even functions on $L^2(\mathbb{R}^{1|1}_{q^2})$ 
\begin{align}
\label{eq:supermeasure1}
    \braket{f\vert g}\equiv \int \frac{dx\,d\vartheta}{x}\; \overline{f(x,\vartheta)} g(x,\vartheta),
\end{align}
the generators satisfy the twisted star Hopf algebra relations in the range of real $0<q<1$
\begin{align}
    K^\dag=K^{-1}, \qquad (F^\pm)^\dag=F^\pm, \qquad (E^\pm)^\dag=-E^\pm,
\end{align}
provided we set $j\equiv -\frac{1}{4}+\frac{i\theta}{\log q^2}$.\footnote{We use the order preserving convention for complex conjugation of products of Grassmann variables.} 

To go the relevant gravitational form, the bosonic gravitational sector needs to be constrained with $q^2$-separations in $x$. Following the logic of the bosonic case, we again need to constrain to a subalgebra $U_q(\mathfrak{osp}(1|2,\mathbb{R}))$ using the homomorphism $\varphi:U_q(\mathfrak{osp}(1|2,\mathbb{R}))\rightarrow \breve{U}_q(\mathfrak{osp}(1|2,\mathbb{R}))$, defined by 
\begin{align}
    \varphi(F^+)=F^+K^{-1}, \qquad \varphi(F^-)=KF^-, \qquad \varphi(K)=K^2,
\end{align} 
leading to the following algebra relations and $\mathbb{Z}_2$-graded coproduct 
\begin{align}
    \{F^+,F^-\}&=-\frac{K-K^{-1}}{8(q^{1/2}-q^{-1/2})}, \qquad KF^\pm=q^{\pm}F^\pm K,\\ \Delta(F^-)&=F^-\otimes 1+K\otimes F^-, \quad \Delta(F^+)=F^+\otimes K^{-1}+1\otimes F^+, \quad \Delta(K)=K\otimes K,\nonumber
\end{align}
Upon identifying $K=q^{2H}$, we arrive at the same infinitesimal superalgebra \eqref{eq:infinitsuper}. 

Using the homomorphism $\varphi$, the principal series generators of $U_q(\mathfrak{osp}(1|2,\mathbb{R}))$ are given by the following differential operators on $\mathbb{R}^{1|1}$:
\begin{align}
    K&=R_{q^2},\\
    F^+&=\frac{1}{2\sqrt{2}}x^{1/2}\left(\frac{q^{-j}-q^{j}R_{q^{-2}}}{q^{1/2}-q^{-1/2}}\partial_{\vartheta}+\frac{q^{-j}+q^{j}R_{q^{-2}}}{q^{1/2}+q^{-1/2}}\vartheta\right),\\
    F^-&=-\frac{q^{-1/2}}{2\sqrt{2}}x^{-1/2}\left(\frac{q^{j}R_{q^2}-q^{-j}}{q^{1/2}-q^{-1/2}}\partial_\vartheta+\frac{q^{j}R_{q^2}+q^{-j}}{q^{1/2}+q^{-1/2}}\vartheta\right).
\end{align}
The bosonic parabolic generators $E^\pm$ are similarly defined through the homomorphism $\varphi(E^+)=q^{-1/2}E^+K^{-2}$ and $\varphi(E^-)=q^{1/2}K^2E^-$  in the universal enveloping algebra as: \begin{align}\label{eq:susybosonicparabolic}
    E^+&=-\frac{x}{2}\left(\frac{q^{-2j}+(1-q^{-1})R_{q^{-2}}-q^{2j-1}R_{q^{-4}}}{q-q^{-1}}-\frac{2(1-q^{-1})R_{q^{-2}}}{q-q^{-1}}\vartheta\partial_\vartheta\right),\\
    E^-&=\frac{1}{2qx}\left(\frac{q^{2j-1}R_{q^4}+(q^{-1}-1)R_{q^2}-q^{-2j}}{q-q^{-1}}-\frac{2(q^{-1}-1)R_{q^2}}{q-q^{-1}}\vartheta\partial_\vartheta\right).
\end{align} 
Since the principal series representations are irreducible, the sCasimir \eqref{eq:supercasimir} (after application of the homomorphism) is proportional to the $(-)^F\equiv 1-2\vartheta\partial_\vartheta $ differential operator, which flips the sign of the fermionic top part of a supernumber 
\begin{align}
\label{eq:scasimirprincipal}
    Q&=\frac{q^{1/2}K-q^{-1/2}K^{-1}}{4(q-q^{-1})}+2F^-F^+=\frac{\sin\theta}{2i(q-q^{-1})}(-)^F.
\end{align}On the other hand, squaring this generator, the Casimir is exactly diagonalized in these representations as \begin{align}\label{eq:principalseriescasimir}
    \mathcal{C}&=\frac{qK^2+q^{-1}K^{-2}-2}{16(q-q^{-1})^2}-\frac{(qK+q^{-1}K^{-1})F^-F^+}{2(q^{1/2}+q^{-1/2})}-4(F^-)^2(F^+)^2=-\frac{\sin^2\theta}{4(q-q^{-1})^2}.
\end{align}
The homomorphism extends the above star relations into 
\begin{gather}
    K^\dag=K^{-1}, \quad (F^+)^\dag=q^{-1/2}KF^+, \quad (F^-)^\dag=q^{-1/2}K^{-1}F^-,\\ 
    (E^-)^\dag=-q^{-2}K^{-2}E^-, \quad (E^+)^\dag =-q^{-2}K^2E^+.
\end{gather}
Since the scaling operators on the bosonic coordinate in the enveloping algebra are integer products of $q^{ 2}$ scalings, the irreducible subspaces act on the $q^2$-discretized superline  
\begin{equation}
\mathbb{R}_{q^2}^{1|1}=\{(x,\vartheta)\vert x=q^{2n}, n\in \mathbb{Z}\}.
\end{equation}
Since the bosonic coordinate is automatically constrained to one half of the real line, we denote the relevant gravitational real form as OSp$_q^+(1|2,\mathbb{R})$. The inner product on the irreducible subspace is then given by a Jackson integral\begin{align}\label{eq:supermeasure}
    \braket{f\vert g}\equiv \int \frac{d_{q^2}x\,d\vartheta}{x}\; \overline{f(x,\vartheta)} g(x,\vartheta).
\end{align} 

\subsection{Gravitational matrix element}
Next we compute the two-sided gravitational matrix element, in parallel with our earlier treatment in the bosonic case.

\subsubsection{Whittaker vectors}
To find the two-sided gravitational wavefunction in terms of a constrained principal series matrix element, we introduce the gravitational boundary eigenstates, or Whittaker vectors, that diagonalize the corresponding parabolic generators. In particular, the Whittaker vector $\braket{x,\vartheta\vert \phi_j^\pm}=\phi^\pm_j(x,\vartheta)$ should diagonalize the corresponding action of both $E^\pm$ and $F^\pm$ and satisfy the algebra relations between them. To achieve the latter, we introduce two Majorana fermions $\psi_1,\psi_2$, satisfying the Clifford algebra $\{\psi_i,\psi_j\}=2\delta_{ij}$, while anticommuting with all other independent Grassmann variables. The Whittaker vectors can then be expressed in terms of an edge state decomposition as 
\begin{align}
    \phi^+_j(x,\vartheta)&\equiv x^j\int_{\mathcal{C}}d\zeta x^{-i\zeta}(1-q^2)^{-2i\zeta}  \left(c_1 \Gamma^{\text{R}}_{q^2}(-i\zeta)-i\psi_2\vartheta c_2 \Gamma^{\text{NS}}_{q^2}(-i\zeta)\right),\\
   \phi^-_j(x,\vartheta)&\equiv x^{-j}\int_{\mathcal{C}}d\zeta x^{i\zeta}(1-q^2)^{-2i\zeta}\left(c_1 \Gamma^{\text{R}}_{q^2}(-i\zeta)-\psi_1 \vartheta c_2\Gamma^{\text{NS}}_{q^2}(-i\zeta)\right),
\end{align} 
where $c_1,c_2$ are two normalization coefficients whose ratio will be determined momentarily and the integration contour is defined as before $\mathcal{C}=[-\frac{\pi}{\log q^2}, \frac{\pi}{\log q^2}[$ while avoiding the poles at $\zeta=0$ and $\zeta=\pi/\log q^2$ for $\Gamma^{\text{NS}}_{q^{2}}$ and $\Gamma_{q^2}^{\text{R}}$ respectively. To shorten expressions in the following we have additionally introduced two $q$-deformed super gamma functions in terms of their bosonic counterpart: 
\begin{align}
\label{eq:Rgamma}
    \Gamma^{\text{R}}_{q^2}(x)&\equiv\Gamma_{q^2}\left(x+\frac{i\pi}{\log q^2}\right)\Gamma_{q^2}\left(x+\frac{1}{2}\right),\\
\label{eq:NSgamma}    
    \Gamma_{q^2}^{\text{NS}}(x)&\equiv\Gamma_{q^2}\left(x+\frac{i\pi}{\log q^2}+\frac{1}{2}\right)\Gamma_{q^2}(x).
\end{align}
Using the defining recursion relation of the regular $q$-deformed gamma function \eqref{eq:qgamma}, we determine the recursive relations of its supersymmetric analogues 
\begin{align}
    \Gamma^{\text{R}}_{q^2}\left(x+\frac{1}{2}\right)&=[x]^-_{q^2}\;\Gamma^{\text{NS}}_{q^2}\left(x\right), \qquad [x]^-_{q^2}=\frac{q^{2x}-1}{q^2-1},\\ 
    \Gamma^{\text{NS}}_{q^2}\left(x+\frac{1}{2}\right)&=-[x]^+_{q^2}\;\Gamma^{{\text{R}}}_{q^2}\left(x\right), \qquad [x]^+_{q^2}=\frac{q^{2x}+1}{q^2-1}.
\end{align}
Using the poles of the normal $q$-gamma function, the pole structure of $\Gamma^{\text{NS}}_{q^2}(x)$ is again at non-positive integers and negative half integers at $-i\pi/\log q^2$, whereas the poles of $\Gamma^{\text{R}}_{q^2}(x)$ are both at negative half integers and non-positive integers at $-i\pi/\log q^2$. Furthermore the integrand of the Whittaker vectors are $2\pi$ periodic since the supersymmetric $q$-gamma functions are products of the normal $q$ gamma function which are multiplied with appropriate powers of $(1-q^2)$ (up to constant prefactors), which relates them to $q$-Pochhammer symbols through \eqref{eq:qpochammergamma}. Therefore, we can deform the contour to $\mathcal{C}\rightarrow\mathcal{C}+i/2$ without encountering poles. 
\begin{align}
    F^+\phi_j^+(x,\vartheta)&=\frac{q^{-j}}{2\sqrt{2}}x^j \Bigg[\frac{i\psi_2c_2}{q^{1/2}-q^{-1/2}}\int_{\mathcal{C}}d\zeta x^{-i\zeta+1/2}(1-q^{2i\zeta})\Gamma_{q^2}^{\text{NS}}(-i\zeta)(1-q^2)^{-2i\zeta} \\&+\frac{c_1\vartheta}{q^{1/2}+q^{-1/2}}\int_{\mathcal{C}}d\zeta x^{-i\zeta+1/2}(1+q^{2i\zeta})\Gamma^{\text{R}}_{q^2}(-i\zeta)(1-q^2)^{-2i\zeta}\Bigg].
\end{align} 
The constant coefficients then need to satisfy 
\begin{align}
    \frac{c_2}{q^{1/2}-q^{-1/2}}&\equiv -\lambda c_1, \qquad \frac{c_1}{q^{1/2}+q^{-1/2}}\equiv \lambda c_2,
\end{align}
in order for $\phi_j^+(x,\theta)$ to be an eigenvector (upto action by $K^{\alpha}$) of $F^+$. In the range $0<q<1$ this is solved by
\begin{align}
\label{eq:defab}
    \lambda=\frac{1}{(q^{-1}-q)^{1/2}}, \qquad c_1=\left(\frac{q^{1/2}+q^{-1/2}}{q^{-1/2}-q^{1/2}}\right)^{1/4}, \qquad c_2=\left(\frac{q^{-1/2}-q^{1/2}}{q^{1/2}+q^{-1/2}}\right)^{1/4}.
\end{align}
Shifting $i\zeta\rightarrow i\zeta+1$ and using the above recursion properties, the Whittaker vectors diagonalize the parabolic generators with a fermionic eigenvalue: 
\begin{align}
    F^+\phi_j^+(x,\vartheta) &=\frac{q^{-j+1/2}}{2\sqrt{2}}\frac{i\psi_2}{\sqrt{q^{-2}-1}}x^j \Big(c_1 \int_{\mathcal{C}}d\zeta x^{-i\zeta}\Gamma^{\text{R}}_{q^2}(-i\zeta)(1-q^2)^{-2i\zeta}q^{2i\zeta} \nonumber\\
    &\qquad-i\psi_2\vartheta c_2 \int_{\mathcal{C}}d\zeta x^{-i\zeta}\Gamma^{\text{NS}}_{q^2}(-i\zeta)(1-q^2)^{-2i\zeta}q^{2i\zeta}\Big)\nonumber\\
    &=\frac{q^{j+1/2}}{2\sqrt{2}}\frac{i\psi_2}{\sqrt{q^{-2}-1}}K^{-1}\phi_j^+(x,\vartheta).
\end{align} 
Similarly, 
\begin{align}
    F^-\phi_j^-(x,\vartheta)&=\frac{q^j}{2\sqrt{2}}\frac{\psi_1}{\sqrt{q^{-2}-1}}\;K\phi_j^-(x,\vartheta).
\end{align}
Together with the Hermitian conjugate, the  latter implies the right action of 
\begin{align}
    \bra{\phi_j^-}F^-&=\left(q^{-1/2}K^{-1}F^-\ket{\phi_j^-}\right)^\dag =\frac{q^{-j-1}}{2\sqrt{2}\sqrt{q^{-2}-1}}\psi_1\bra{\phi_j^-}.
\end{align}
On the other hand, by squaring the fermionic generators (or using explicit action of the generators \eqref{eq:susybosonicparabolic}), the Whittaker vectors exactly diagonalize the respective bosonic generators, up to the action of the Cartan generator
\begin{align}
    E^+\phi_j^+(x,\vartheta)&=\frac{q^{2j+3}}{2(q-q^{-1})}K^{-2} \phi_j^+(x,\vartheta),\qquad E^-\phi_j^-(x,\vartheta)&=\frac{q^{2j+2}}{2(q-q^{-1})}K^2\phi_j^-(x,\vartheta).
\end{align}
Using the adjoint properties, the right adjoint action of $E^-$ is then given by
\begin{align}
    \bra{\phi_j^-}E^-&=-\frac{q^{-2j-1}}{2(q-q^{-1})}\bra{\phi_j^-}.
\end{align}

\subsubsection{Whittaker functions}
With respect to the measure on $L^2(\mathbb{R}^{1|1}_{q^2})$ \eqref{eq:supermeasure}, one can determine the Whittaker function to be proportional to 
\begin{align}
    \braket{\phi_j^-\vert K^{-n}\vert \phi_j^+}&=q^{n/2}\int_{-\pi}^\pi ds e^{ins} (1-q^2)^{-\frac{4is}{\log q^2}} \times \\
    &\hspace{-1cm}\times \left[i \psi_2\Gamma^{\text{R}}_{q^2}\left(\frac{i \theta - is}{\log q^2}\right)\Gamma^{\text{NS}}_{q^2}\left(-\frac{i \theta + is }{\log q^2}\right) +\psi_1\Gamma^{\text{NS}}_{q^2}\left(\frac{i\theta - is}{\log q^2}\right)
    \Gamma^{\text{R}}_{q^2}\left(-\frac{i \theta + is }{\log q^2}\right) \right]. \nonumber
\end{align}
This is the supersymmetric analogue of \eqref{eq:gammawhittakerexp}. Since the patch of the group manifold reached by the parametrization is connected to the identity, we refer to the above matrix element as the Ramond ($\mathbf{R}$) Whittaker function. The matrix element disconnected to the identity is refered to as the Neveu-Schwarz ($\mathbf{NS}$) Whittaker function, and is computed by the extra insertion of $(-)^F$, leading to a relative minus sign 
\begin{align}
    \braket{\phi_j^-\vert K^{-n}(-)^F\vert \phi_j^+}&=q^{n/2}\int_{-\pi}^\pi ds e^{ins} (1-q^2)^{-\frac{4is}{\log q^2}}\times \\
    &\hspace{-1.5cm}\times \left[-i \psi_2\Gamma^{\text{R}}_{q^2}\left(\frac{i \theta - is}{\log q^2}\right)\Gamma^{\text{NS}}_{q^2}\left(-\frac{i \theta + is }{\log q^2}\right)  +\psi_1\Gamma^{\text{NS}}_{q^2}\left(\frac{i\theta - is}{\log q^2}\right)
    \Gamma^{\text{R}}_{q^2}\left(-\frac{i \theta + is }{\log q^2}\right) \right] \nonumber.
\end{align}
Using the algebra relations and the left and right parabolic eigenvalues on the respective Whittaker vectors, one can calculate a recursion relation of the Whittaker function by inserting the action of the sCasimir: 
\begin{align}
    \frac{q^{1/2}T^n_{-1}-\left(q^{-1/2}+iq^{n+1/2}\psi_1\psi_2\right)T^n_1}{4(q-q^{-1})}\braket{\phi_j^-\vert K^{-n}\vert \phi_j^+}=\braket{\phi_j^-\vert QK^{-n}\vert \phi_j^+}.
\end{align}Inserting the value of the sCasimir in the principal series representations \eqref{eq:scasimirprincipal} relates the $\mathbf{R}$ Whittaker function to the $\mathbf{NS}$ Whittaker function as: 
\begin{align}
    \left(q^{1/2}T^n_{-1}-\left(q^{-1/2}+iq^{n+1/2}\psi_1\psi_2\right)T^n_1\right)\braket{\phi_j^-\vert K^{-n}\vert \phi_j^+}=-2i\sin \theta\braket{\phi_j^-\vert (-)^FK^{-n}\vert \phi_j^+}\nonumber.
\end{align}Analogously, the recursion relation of the $\mathbf{NS}$ Whittaker function in terms of the $\mathbf{R}$ Whittaker function is given by \begin{align}
   \left(q^{1/2}T^n_{-1}-\left(q^{-1/2}-iq^{n+1/2}\psi_1\psi_2\right)T^n_1\right) \braket{\phi_j^-\vert (-)^FK^{-n}\vert \phi_j^+}=-2i\sin \theta\braket{\phi_j^-\vert K^{-n}\vert \phi_j^+}\nonumber.
\end{align} These recursion relations form the supersymmetric analogues of the bosonic transfer matrix \eqref{eq:bosonicrecursion}. Using the shift properties of the supersymmetric $q$-gamma functions, one can verify that the above Whittaker functions explicitly satisfy these recursion relations. The successive action of the sCasimir first order recursion relations (or inserting  directly \eqref{eq:principalseriescasimir}) leads to the recursion relation of the quadratic Casimir on the Whittaker function in terms of the $\mathcal{N}=1$ $q$-Liouville difference equation: 
\begin{align}    &\left(qT^n_{-2}+(q^{-1}+q^n(q-1)i\psi_1\psi_2-q^{2n+2})T^n_{2}-2+q^n(q-1)i\psi_1\psi_2\right)\braket{\phi_j^-\vert K^{-n}\vert \phi_j^+}\nonumber\\&=-4\sin^2\theta\braket{\phi_j^-\vert K^{-n}\vert \phi_j^+}.
\end{align}

\subsubsection{Multiplet representation}
The fermions can be represented in the $\mathcal{N}=1$ multiplet representation spanned by the states $\ket{0}, \overline{\psi}\ket{0}$, in terms of the lowering and raising  lightcone combinations
\begin{align}
    \psi=\tfrac{1}{2}(\psi_1+i\psi_2), \qquad \overline{\psi}=\tfrac{1}{2}(\psi_1-i\psi_2),
\end{align}
satisfying the Dirac algebra $\{\psi,\overline{\psi}\}=1 $ and the vacuum defined to satisfy $\psi\ket{0}=0$. 
In terms of the matrix multiplet representation with  $\ket{0}\equiv\begin{pmatrix}
    1&0
\end{pmatrix}^t,\; \overline{\psi}\ket{0}\equiv \begin{pmatrix}
    0&1
\end{pmatrix}^t$, the fermions are represented using the Dirac algebra: 
\begin{align}
    \psi=\begin{pmatrix}
        0&1\\0&0
    \end{pmatrix},\qquad \overline{\psi}=\begin{pmatrix}
        0&0\\1&0
    \end{pmatrix},
\end{align} 
such that $\psi_1 = \sigma_x,\psi_2 = \sigma_y$ are realized as Pauli matrices. 
Decomposing the columns of the Whittaker functions then leads to sets of two linearly independent wavefunctions in the fermionic multiplet, where for the $\mathbf{R}$ states, we have \begin{align}
    \ket{F^-}=\chi_-\overline{\psi}\ket{0}, \qquad \ket{F^+}=\chi_+\ket{0}.
\end{align}
For the $\mathbf{NS}$ states, the relative signs are interchanged
\begin{align}
    \ket{L^-}=\chi_+\overline{\psi}\ket{0}, \qquad \ket{L^+}=\chi_-\ket{0}.
\end{align} 
The two wavefunctions are then given by 
\begin{align}
\label{eq:n1wavefunction}
&\chi_\pm^\theta(n)=q^{n/2}\int_{-\pi}^\pi ds e^{ins} \left(1-q^2\right)^{-\frac{4is}{\log q^2}}\times \\
&\times\left[\Gamma^{\text{NS}}_{q^2} \left( \frac{i\theta -is}{\log q^2} \right)\Gamma^{\text{R}}_{q^2} \left( -\frac{i\theta + is}{\log q^2}\right)  \pm \Gamma^{\text{R}}_{q^2} \left( \frac{i\theta -is}{\log q^2} \right)\Gamma^{\text{NS}}_{q^2} \left( -\frac{i\theta + is}{\log q^2} \right)\right] \nonumber.
\end{align}
The structure of these equations is to be compared to an analogous computation done for the modular double $U_q(\mathfrak{osp}(1|2,\mathbb{R})) \otimes U_{\tilde{q}}(\mathfrak{osp}(1|2,\mathbb{R}))$ in the context of $\mathcal{N}=1$ Liouville supergravity in \cite{Fan:2021bwt} (equation (3.46)), where the chief difference is replacing $q$-Gamma functions by double sine $S_b$ functions. 

These functions satisfy the recursion relations:  
 \begin{align}
  \boxed{ q^{1/2}\chi^\theta_\mp(n-1)-\left(q^{-1/2}\pm q^{n+1/2}\right)\;\chi_{\mp}^\theta(n+1)=-2i\sin\theta \;\chi_\pm^\theta(n),}
   \label{eq:recursionberkooz}
\end{align}
which can be identified with the transfer matrix for $\mathcal{N}=1$ DSSYK as studied in the chord formalism in \cite{Berkooz:2020xne}. 
A generic solution to the recursion relation \eqref{eq:recursionberkooz} is found by noting that the solution for $\chi_\pm^\theta(n)$ decomposes into two independent sectors classified by different choices of $n$. Since there is no relation relating $\chi^\theta_\pm(n)$ to $\chi^\theta_\pm(n + 1)$,  the two sets 
\begin{equation}
\{\chi^\theta_+(2m), \chi_-^\theta(2m + 1) \vert m \in  \mathbb{N}\}, \qquad  \{\chi^\theta_+(2m + 1), \chi_-^\theta(2m ) \vert m \in  \mathbb{N}\},
\end{equation}
are independent, and one view the system as defined on a graded lattice, illustrated by
\begin{align}
\raisebox{-0.42\height}{\includegraphics[height=3.3cm]{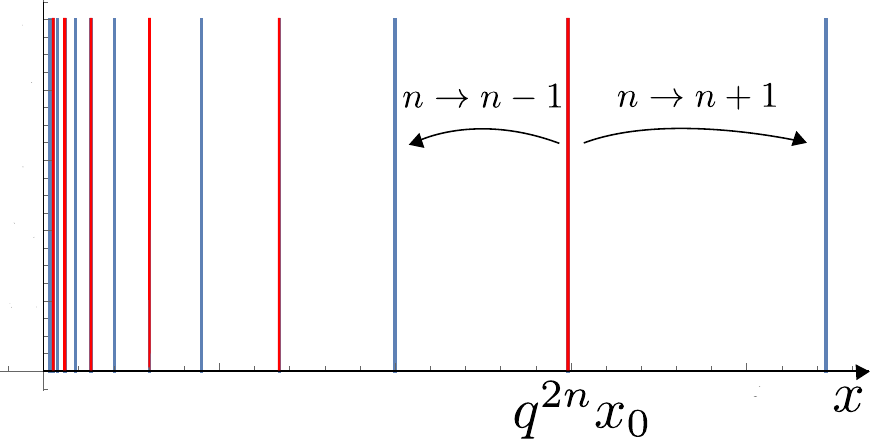}}
\end{align}
where the same eigenfunctions on the red and blue lines do not talk to each other.
Defining 
\begin{align}
      \xi^\theta_+(n) &= i(-1)^{\lfloor n/2\rfloor} q^{-n/2}((-1)^{n+1}q;-q)_{n}\chi^\theta_+(n), \\
    \xi_-^\theta(n) &= (-1)^{\lfloor n/2\rfloor}q^{-n/2} ((-1)^n q;-q)_{n} \chi_-^\theta(n),
\end{align}
and shifting $\theta \to \theta + \pi/2$ maps the recursion relations  
\eqref{eq:recursionberkooz} to
\begin{align}
    (1 - (-q)^n)\xi_\mp^\theta(n -1) + \xi_\mp^\theta(n+1) &= 2\cos\theta \xi_\pm(n)
    \label{eq:qhermiterel},\\
       (1 + (-q)^n)\xi_\mp^\theta(n -1) + \xi_\mp^\theta(n+1) &= 2\cos\theta \xi_\pm(n),
\end{align}
for the two sets respectively. 
The first of these can be identified as the defining recursion relation of the $q$-Hermite polynomials with $-q$ such that $\xi_+^\theta(2m) = H_{2m}(\cos\theta\vert-q)$ and $\xi_-^\theta(2m + 1) = H_{2m + 1}(\cos\theta\vert -q)$
leading to the wavefunctions
\begin{align}
    \chi_+^\theta(2m) &= -\frac{i(-q)^m}{(-q;-q)_{2m}} H_{2m}(\cos(\theta - \pi/2)\vert -q),\\
    \chi_-^\theta(2m+ 1) &= -\frac{(-q)^{m}q^{1/2}}{(-q;-q)_{2m + 1}} H_{2m +1}(\cos(\theta - \pi/2)\vert -q).
\end{align}
Note that the solution of this recursion is only unique upto a global scaling by a general function in $\theta$. I.e. for every $\rho(\theta)$ we find that $\tilde\chi_\pm^\theta(n) = \rho(\theta)\chi_{\pm}^\theta(n)$ solves the same recursion relation. This solution (upto a conventional rescaling) matches exactly with the wavefunctions derived from the chord formalism of the model \cite{Berkooz:2020xne}.  

The relation on the second set on the other hand differs from the defining relation of the $q$-Hermite polynomials by a relative sign in the coefficient of $\xi_\mp^\theta(n-1)$. This sign has the effect that the solutions can be extended to negative values of $n$ since this coefficient, $(1 + (-q))^n$, does not vanish at $n = 0$. In comparison evaluating \eqref{eq:qhermiterel} at $n = 0$ shows that the solutions for this equation for positive and negative $n$ completely decouple. Since the chord number $n$ is still positive by definition, a physical solution should follow this property. This can of course be implemented by hand, by choosing the initial value of the recursion relation to be trivial, i.e. $\rho(\theta) = 0$, for the unphysical solution. From the group theory perspective however, this arbitrary choice does not need to be made and is rather automatically implemented. Indeed, replacing $\Gamma_{q^2}^{\text{NS}}(.),\Gamma_{q^2}^{\text{R}}(.)$ in \eqref{eq:n1wavefunction} by their definitions \eqref{eq:NSgamma},\eqref{eq:Rgamma} and rewriting the resulting q-gamma functions in terms of q-Pochhammer symbols using \eqref{eq:qpochammergamma} the wavefunctions take the form 
\begin{align}
    &\chi^{\theta}_\pm(n) = q^{n/2} (q^2;q^2)_\infty^4(1-q^2)^{3 - \frac{2 \pi i }{\log q^2}}\int_{-\pi}^{\pi} ds e^{ins} \times \\
    &\times \bigg\{\left[(-qe^{i(\theta -s)};q^2)_\infty(e^{i(\theta -s)};q^2)_\infty(-e^{-i(\theta + s)};q^2)_\infty(qe^{-i(\theta +s)};q^2)_\infty\right]^{-1} \pm \nonumber \\
    &\hspace{0.39cm}\pm \left[(-e^{i(\theta -s)};q^2)_\infty(qe^{i(\theta -s)};q^2)_\infty(-qe^{-i(\theta + s)};q^2)_\infty(e^{-i(\theta +s)};q^2)_\infty\right]^{-1} \bigg\}. \nonumber
\end{align}
In this form it is easily checked that under the shift $s \mapsto  s - \pi$ the integrand transforms as  $f(e^{is},\theta,n) \mapsto f(e^{i(s - \pi)},\theta,n) = \pm (-1)^n f(e^{is},\theta,n)$. Using this property we find, dropping global prefactors,
\begin{align}
    \chi^\theta_\pm(n) &\sim   q^{n/2} \left[\int_{0}^{\pi} ds f(e^{is},\theta ,n)   +  \int_{-\pi}^{0} ds f(e^{ins},\theta ,n) \right] = \\
    &= q^{n/2} \left[\int_{0}^{\pi} ds f(e^{is},\theta ,n)   + f(e^{i(s-\pi)},\theta ,n) \right] = q^{n/2}\int_0^\pi \left[1 \pm (-1)^n \right]f(e^{is},\theta,n).
\end{align}
This implies $\chi_+^\theta(2m +1) = \chi^\theta_-(2m) = 0$ showing that vanishing of the unphysical solution automatically follows from the group theoretic definition.
\subsection{Edge states in $\mathcal{N}=1$ DSSYK}
Similarly to the bosonic case, the edge states are built from a complete set of hyperbolic eigenstates on the discretized superline $L^2(\mathbb{R}_{q^2}^{1|1})$. To construct a basis for this space, the edge labels now consist of the continuous labels $s$ and a (complex) Grassmann number $\gamma$ satisfying $\overline{\gamma}=-\gamma$. With respect to the inner product \eqref{eq:supermeasure}, the eigenstates diagonalize the commuting set of generators $K$ and $ \partial_\vartheta$ with eigenvalues $e^{is}$ and $\gamma$ respectively: \begin{align}
    \braket{x,\vartheta\mid s,\gamma}=\frac{x^{\frac{is}{\log q^2}}e^{-\gamma \vartheta}}{\sqrt{2\pi(1-q^2)}}=\frac{x^{\frac{is}{\log q^2}}}{\sqrt{2\pi(1-q^2)}}(1-\gamma\vartheta).
\end{align} 
One can check that in the first periodic zone $s\in[-\pi,\pi[$, the basis states are orthonormal with respect to the inner product \eqref{eq:supermeasure}\begin{align}
    \braket{s',\gamma'\mid s,\gamma}=\delta(s-s')\delta(\gamma'-\gamma),
\end{align}where the delta function of a Grassmann number acting on functions with definite parity is defined by the argument of that function\footnote{In the sense that $f(\vartheta)=\int d\vartheta' \;(\vartheta'-\vartheta)f(\vartheta')$, for a function $f$ of a Grassmann number $\vartheta$.} $\delta(\gamma'-\gamma)\equiv \gamma'-\gamma$. The hyperbolic matrix element can then be constructed as 
\begin{align}
    \gamma,\raisebox{-0.48\height}{\includegraphics[height=2.5cm]{Figures/hyperbolic.pdf}},\raisebox{-0.05cm}{$\gamma'$}\hspace{0.2cm}= \braket{s_1,\gamma\vert K^{-n}\vert s_2,\gamma'}&=e^{-is_2n}\;\delta(s_1-s_2)\delta(\gamma-\gamma').\label{eq:hyperbolicCartan}
\end{align}
They are also complete, leading to the Fourier decomposition of a generic $f(x,\vartheta)\in L^2(\mathbb{R}^{1|1}_{q^2})$: 
\begin{align}
    f(x,\vartheta)&=\frac{1}{\sqrt{2\pi(1-q^2)}}\int_{-\pi}^\pi ds \int d\gamma \,x^{\frac{is}{\log q^2}}e^{-\gamma\vartheta}F(s,\gamma), \\ F(s,\gamma)&=\frac{1}{\sqrt{2\pi(1-q^2)}}\int_{0}^\infty \frac{d_{q^2}x}{x}\int d\vartheta \,x^{-\frac{is}{\log q^2}}e^{\gamma \vartheta}f(x,\vartheta).
\end{align}
The one-sided wavefunction can then be constructed from the expressions of the Whittaker vectors as $\braket{s,\gamma\mid \phi_j^+}$ and $\braket{ \phi_j^-\mid s,\gamma}$ respectively: 
\begin{equation}
s,\gamma\,\raisebox{-0.46\height}{\includegraphics[height=2.8cm]{Figures/rightedge.pdf}}\hspace{0cm}=(1-q^2)^{\frac{is-i\theta}{\log q}+\frac{1}{2}} \left[c_1\gamma\Gamma^{\text{R}}_{q^2}\left(\frac{is-i\theta}{\log q^2}+\frac{1}{4}\right)-ic_2\psi_2 \Gamma^{\text{NS}}_{q^2}\left(\frac{is-i\theta}{\log q^2}+\frac{1}{4}\right)\right],\vspace{-0.3cm}
\end{equation}
and \vspace{-0.3cm}
\begin{equation}
     \raisebox{-0.46\height}{\includegraphics[height=2.8cm]{Figures/leftedge.pdf}}\hspace{0cm}\;s,\gamma=-(1-q^2)^{\frac{is+i\theta}{\log q}+\frac{1}{2}}\left[c_1\gamma \Gamma^{\text{R}}_{q^2}\left(\frac{is+i\theta}{\log q^2}+\frac{1}{4}\right)+c_2\psi_1 \Gamma^{\text{NS}}_{q^2}\left(\frac{is+i\theta}{\log q^2}+\frac{1}{4}\right)\right],\vspace{-0.2cm}
\end{equation}
where $c_1$ and $c_2$ were defined in \eqref{eq:defab}. These are the supersymmetric analogues of \eqref{eq:rightbosonicedge} and \eqref{eq:leftbosonicedge} respectively.

\section{Concluding remarks}
\label{s:concl}
In this work we explored the quantum group structure underlying DSSYK, leveraging this structure to construct an edge state factorization of the two-sided Hilbert space, as summarized in the Introduction. We end this work with some more speculative remarks. \\

\textbf{Von Neumann entanglement entropy} \\
With this edge state structure, which parallels the JT gravity and 3d gravity scenarios, we would immediately write down the usual continuum version of the entanglement entropy as a sum of two (three) terms:
\begin{equation}
\label{eq:JT_vN}
S= -\int_{0}^{\pi}  d \theta\,   P(\theta) \log P(\theta) \, + \, \int_{0}^{\pi}  d \theta\, P(\theta) \log \dim \theta \, + \, \log \frac{2\pi V_\text{G}}{V_T},
\end{equation}
mirroring analogous results as for compact groups \cite{Donnelly:2014gva,Lin:2018xkj}.
The first term is a classical Shannon entropy term on the Boltzmann probability distribution $P(\theta)$. The second term is the actual contribution to the entanglement entropy from the edge states, counting the number of microstates for each macrostate (specified by $\theta$). The final term is an additive state-independent contribution that is always present when dealing with continuous representations of non-compact group (or quantum groups) \cite{Mertens:2022ujr}, and that is signaling us that the model we are working with is not fully microscopic (or the algebra is still type II$_{1}$ and not type I). \\ 

\textbf{An even more refined algebraic structure than the semigroup SL$^{+}(2,\mathbb{R})$?} \\
 Since length positivity plays such a central role in DSSYK it is interesting to notice that mathematically this can be implemented, even in the bulk of JT gravity by further reducing the algebraic structure from the semigroup SL$^+(2,\mathbb{R})$ down to a further subset by looking at matrices:
\begin{equation}
\left(\begin{array}{cc}a & b \\ c & d \end{array}\right), \qquad ad-bc=1, \qquad a,b,c,d \geq 0, \qquad a\geq 1.
\end{equation}
This subset now has $\phi \geq 0$ or length positivity implemented, in addition to $\beta,\gamma \geq 0$. Intriguingly, this structure is once again a semigroup. In fact, we can readily catalog all natural semigroups (constructed by more stringent inequalities) that are subsets of SL$^+(2,\mathbb{R})$ by the following reasoning:\footnote{This classification can be extended to more generic subsemigroups of SL$(2,\mathbb{R})$, by repeating the discussion starting with the semigroup $a,d \geq 0,\quad b,c \leq 0 $. The results are qualitatively identical.}
\begin{itemize}
\item The identity element $\mathbb{1}$ should be included, so the highest possible lower bound on $a,d$ is 1 and for $b,c$ is 0.
\item We readily see that the choosing the lower bound on $a,d$ between $]0,1[$ is not preserved under composition.
\item This leaves us with four options in total: 
\begin{enumerate}
\item $a,d \geq 0$. This is just SL$^+(2,\mathbb{R})$.
\item $a\geq 1$, $d\geq 0$. This is the positive length semigroup of interest.
\item $a \geq 0$, $d\geq 1$. This is isomorphic to the previous case.
\item $a,d \geq 1$.
\end{enumerate}
\end{itemize}

This shows that length positivity is also a natural feature to implement in the group theoretical framework.
\\

\textbf{Towards fully microscopic factorization}\footnote{We thank J. Papalini for a discussion on this.} \\
We started this work with the ambition of finding out whether the DSSYK gravitational Hilbert space allows a bulk factorization that is more telling of an underlying microscopic system. This motivation stemmed from the fact that the energy spectrum of DSSYK is bounded from above as well.

Within the SSS matrix model of JT gravity \cite{Saad:2019lba}, it was recently argued in \cite{Boruch:2024kvv} that the bulk Hilbert space would fully factorize (without edge states) upon including wormhole and non-perturbative effects. This is still not fully microscopic since the SSS matrix model is still a double-scaled matrix model.
Recent results in DSSYK have also lead to a more detailed understanding of a candidate \emph{finite cut} matrix model dual for which DSSYK itself is the leading $N\to+\infty$ result. 
Combining these observations, it would be very interesting to explore how this calculation would work for the matrix model description that completes DSSYK, since one now has the potential of finding a truly microscopic factorization that in the $N \to +\infty$ limit descends to the continuous description we presented in this work. We leave this for future work.
\\

\textbf{Principal series character computation}\\
In section \ref{s:trumbran} we presented a direct interpretation of the single trumpet and brane in terms of characters of the quantum group. Our computation of these characters themselves was straightforward and exploited the relation between the different representations \eqref{eq:relrep}. However, it would have been useful to present a direct first principles derivation of the principal series character. In the classical case, a quick derivation uses directly the M\"obius transformed realization of the group representation matrix (see e.g. \cite{Vilenkin}). This motivated us around equation \eqref{eq:corep} to derive the analogous realization in the $q$-deformed setting. However, we do not know how to proceed with the character calculation in the current case of non-commutative geometry. It would be interesting to figure out a good direct mathematical line of attack on this problem. Further there exist multiple in-equivalent definitions of characters of quantum groups. It would be interesting to investigate how these definitions differ in our setting and what role they play in the gravitational setting.\\

\textbf{Bulk dual of DSSYK: the quantum disk?}\\
In applications of JT gravity, EOW brane wavefunctions are defined as worldline particles with mass $\mu$ are functionally equivalent to Wilson loops evaluated in the spin-$j$ highest weight representations \cite{Iliesiu:2019xuh, Belaey:2023jtr}: 
\begin{align}
\label{eq:pathintegralwilsonline}
    \text{Tr}_j\left(\mathcal{P}\exp-\oint_{\mathcal{C}}\mathbf{A}\right)  \simeq \oint_{\text{paths }\sim \, \mathcal{C}}\mathcal{D}x\;e^{-\mu \int ds\;\sqrt{g_{\alpha\beta}\dot{x}^\alpha\dot{x}^\beta}},
\end{align}
where the mass $\mu$ of the particle and spin $j$ of the representation are related through the quadratic Casimir eigenvalue $\mu^2 = j(j+1)$ via the standard holographic dictionary. The wavefunction of a closed EOW brane circling the defect can then indeed be identified as a highest weight character. This goes by the name of the Borel-Weil-Bott theorem, which rewrites a Wilson loop in terms of a coadjoint orbit path integral describing a particle moving on the PSL$(2,\mathbb{R})$ group manifold. Implementing the gravitational constraint identifies a coset of this group manifold with the hyperbolic disk geometry, in terms of the right coset under the Cartan generator. 

Since the highest weight character of the SL$_q(2,\mathbb{R})$ quantum group also describes the appropriate brane wavefunction in DSSYK, it is tempting to propose a similar identification as \eqref{eq:pathintegralwilsonline} for the bulk quantum geometry. The particles would then move on something akin to a coset of SL$_q(2,\mathbb{R})$, for example the discretized geometry known as the quantum disk \cite{Berkooz:2022mfk,Almheiri:2024ayc}. It would be interesting to see if this identification can be made concrete.

\section*{Acknowledgments}
We thank S. E. Aguilar-Gutierrez, A. Blommaert, A. Levine, J. Papalini, L. Tappeiner
and Q. Wu for discussions. AB acknowledges the UGent Special Research Fund (BOF) for
financial support. TT is supported by a Research Foundation — Flanders (FWO) doctoral
fellowship (grant 11A2925N). TM and TT acknowledge financial support from the European
Research Council (grant BHHQG-101040024). Funded by the European Union. Views and
opinions expressed are however those of the author(s) only and do not necessarily reflect
those of the European Union or the European Research Council. Neither the European Union
nor the granting authority can be held responsible for them.
\appendix
\section{ Corepresentation theory of SL$_q(2)$}
\label{sec:AppendixA}

In this appendix we will show that the space of functions defined in (\ref{eq:corep}) in terms of the generalized M\"obius transformation does indeed furnish a corepresentation of SL$_q(2)$\footnote{More precisely our definition of principal series corepresentation requires the formal inclusion of certain inverses in SL$_q(2)\otimes L^2(\mathbb{R}_q^2)$. This distinction will not affect our discussion however, as all notions can be extended in to these fractions in a natural manner. } and further derive that this corepresentation induces the correct representation on the dual Hopf algebra.

It turns out that these calculations take a simpler form in terms of the larger quantum group GL$_q(2)$ \cite{Ip:2013}. This quantum group can naturally be defined as the algebra generated by the non-commutative variables $z_{ij},\,\, i,j = 1,2$, subject to the modified relations
\begin{gather}
\left[ z_{11},  z_{12} \right] = \left[ z_{21},  z_{22} \right] = 0, \qquad  
\left[ z_{12},  z_{21} \right]  = \left[ z_{11},  z_{22} \right], \\
z_{11}z_{21} = q^2 z_{21} z_{11},  \qquad z_{12}z_{22} = q^2 z_{22}z_{12}, \qquad 
z_{12}z_{21} = q^2 z_{21} z_{12},
\end{gather}
where we can again view $z_{ij}$ as elements of a $2 \times 2$ matrix and the coproduct is again realized as a matrix product. We further note that the determinant of this matrix $N \equiv z_{11}z_{22} - z_{12} z_{21}$ satisfies 
\begin{equation}
    Nz_{ii} = z_{ii}N, \qquad N  z_{12} = q^{-2} z_{12} N, \qquad N z_{21} = q^{2} z_{21}  N.
    \label{eq:commN}
\end{equation}
The restriction to $\text{SL}_q(2)$ is then implemented via the identification
\begin{equation}
    \begin{pmatrix}
        a & b \\ c & d
    \end{pmatrix} \equiv  N^{-1/2}\begin{pmatrix}
z_{11} & q^{-1/2}z_{12} \\q^{1/2} z_{21} & z_{22}        
    \end{pmatrix}.
\end{equation} where it is easily checked that this produces the correct relations. Thus we can view $\text{SL}_q(2)$ as the subalgebra of $\text{GL}_q(2)$ generated by elements of the form $z_{11}^lz_{12}^mz_{21}^nz_{22}^o N^{-\left( l + m +n +o\right)/2}$. 
The Hopf duality between this quantum group and the quantum deformed universal enveloping algebra $U_q(\mathfrak{gl}(2))$ is then specified in terms of the bilinear $\braket{\cdot,\cdot}: U_q(\mathfrak{gl}(2)) \times \text{GL}_q(2) \to \mathbb{C}$ acting on the generators:
\begin{align}
    \braket{K,z_{11}} = q, &\quad \braket{K,z_{22}} = q^{-1},\\
    \braket{K_0,z_{11}} = q^{-1/2}, &\quad \braket{K_0,z_{22}} = q^{-1/2},\\
    \braket{E^+,z_{12}} = 1, &\quad \braket{E^-,z_{21}} = 1,  
\end{align}
where all other components vanish and the bilinear is extended to the algebra by requiring 
\begin{align}
    \braket{X,g_1 g_2} \equiv \braket{\Delta(X),g_1 \otimes g_2}, \qquad \braket{XY,g} \equiv \braket{X \otimes Y, \Delta(g)} ,
\end{align} 
for all $X,Y \in U_q(\mathfrak{gl}(2))$, and $g,g_1,g_2 \in \text{GL}_q(2) $. Here $U_q(\mathfrak{gl}(2))$ is defined to be the algebra generated by $K,K_0, E^+,E^-$ subject to the same relations as $U_q(\mathfrak{sl}(2))$ and $K_0$ is a grouplike element defined to strongly commute.
The coproduct is also slightly modified to accommodate the extra generator:\footnote{One can check that this choice of modified coproduct uniquely determines the action of $K_0$ on $z_{ij}$ by consistency with the algebra of $z_{ij}$. E.g. $\braket{E^-,[z_{11} + z_{21},z_{12} + z_{22} ] } = 0 \,\,\Rightarrow \,\, \braket{K_0,z_{22}} = \braket{K,z_{22}}^{1/2}$.}
\begin{align}
    \Delta(K) = K \otimes K, &\quad \Delta(E^+) = K_0 \otimes E^+ + E^+ \otimes K_0^{-1}K^{-1}, \\
    \Delta(K_0)= K_0 \otimes K_0,& \quad \Delta(E^-) =  K_0^{-1}K \otimes E^- + E^- \otimes K_0.
\end{align}
In essence $K_0$ can be understood to encode an abelian sector corresponding to the determinant. And indeed upon reducing to SL$_q(2)$ we find that the Hopf duality takes the form
Upon reducing to $U_q(\frak{sl}(2))$ we find that $K_0$ is dual to the identity element i.e. $\braket{K_0,g} = \mathbb{1}$ as expected. 
 
We now employ this notation to check that (\ref{eq:corep}) defines a corepresentation: 
Formally a (right) corepresentation of a coalgebra $\mathcal{A}$ is defined to be a vector space $V$ together with a linear map $\phi:V \to V\otimes \mathcal{A}$ such that\footnote{The classical analogue of this property is the familiar factorization property $\phi(g_1\cdot g_2)=\phi(g_1)\phi(g_2).$} 
\begin{align}
    (\text{id} \otimes \Delta) \circ \phi =(\phi \otimes \text{id}) \circ \phi, \qquad
    (\text{id} \otimes \epsilon) \circ \phi = \text{id}.
\end{align}
In our case the map $\phi$ is defined through its action
\begin{equation}
    f(x) \mapsto \phi(f(x)) \equiv  N^{k}\left( x z_{11} + z_{21} \right)^j \left( x z_{12} + z_{22} \right)^j x^{-j} f\left( \frac{xz_{11} + z_{21}}{ xz_{12} + z_{22}}\right) N^{l},
\label{eq:corepgl2}
\end{equation}
where $xz_{ij} + z_{kl}$ is defined as $x \otimes z_{ij} + 1 \otimes z_{kl}$, $N^{-j/2}$ is formally defined as $ 1 \otimes N^{-j/2}$  and, since $f(x)$ is defined to be a sum of monomials, its domain can be extended to $\bigoplus_n V \otimes \mathcal{A}^n$ by using the product of $\mathcal{A}$.  
Further $[xz_{11} + z_{21}, xz_{12} + z_{22}] = 0$ as $[z_{11}, z_{22}] = [z_{12}, z_{21}]$ such that the expression is independent of ordering.
 We will show shortly that this furnishes a corepresentation of GL$_q(2)$ for any $j,k,l \in \mathbb{C}$. Restricting $l + k = -j$ ensures that this descends to a corepresentation of SL$_q(2)$, while the choice $l = k = -j/2 $ is conventional. 
Noting that $(\text{id} \otimes \epsilon)(\phi(f(x))) = f(x)$ holds trivially since $\epsilon(z_{ij}) = \delta_{ij}$ we only need to check the second condition as linearity holds by definition. 
Now 
\begin{align}
    (\phi \otimes \text{id})(\phi(f(x))) &= (\phi \otimes \text{id})\left( N^{k}\left( x z_{11} + z_{21} \right)^j \left( x z_{12} + z_{22} \right)^j x^{-j} f\left( \frac{xz_{11} + z_{21}}{ xz_{12} + z_{22}}\right) N^{l} \right)\nonumber \\
    & = (N^k \otimes 1)(x (z_{11} \otimes 1) + (z_{21} \otimes 1))^j (x (z_{12} \otimes 1) + (z_{22} \otimes 1))^j x^{-j} \nonumber\\
&\qquad \times (1 \otimes N^k)\left(\frac{x (z_{11}  \otimes 1) + (z_{21} \otimes 1)}{x ( z_{12}\otimes 1 ) + (z_{22}  \otimes 1)}(1 \otimes z_{11}) + (1 \otimes z_{21}) \right)^j \nonumber \\
 &\qquad \times \left(\frac{x (z_{11}  \otimes 1) + (z_{21} \otimes 1)}{x ( z_{12}\otimes 1 ) + (z_{22}  \otimes 1)}(1 \otimes z_{12}) + (1 \otimes z_{22}) \right)^j \nonumber \\
 &\qquad\times \left(\frac{x (z_{11}  \otimes 1) + (z_{21} \otimes 1)}{x ( z_{12}\otimes 1 ) + (z_{22}  \otimes 1)}\right)^{-j}f(\tilde{x})(1\otimes N^l) (N^l \otimes 1),
\end{align}
where $\tilde{x}$ denotes the twice, successively M\"obius-transformed $x$ and we will show below that this does indeed satisfy the correct transformation property. Firstly notice that the term $(1 \otimes N^k)$ can be pulled to the front of the expression as all terms that we need to commute this with are of the form $(X \otimes 1)$.  Indeed further introducing a common denominator in the sums and combining tensor products we find
\begin{align}
    (\phi \otimes \text{id})&(\phi(f(x))) = \left((\phi \otimes \text{id}) N^{k}\left( x z_{11} + z_{21} \right)^j \left( x z_{12} + z_{22} \right)^j x^{-j} f\left( \frac{xz_{11} + z_{21}}{ xz_{12} + z_{22}}\right) N^{l}\right) \nonumber \\
    & = (N^k \otimes N^k)(x (z_{11} \otimes 1) + (z_{21} \otimes 1))^j (x (z_{12} \otimes 1) + (z_{22} \otimes 1))^j x^{-j} \nonumber \\
&\qquad\times \left(\frac{x (z_{11}  \otimes z_{11}) + (z_{21} \otimes z_{11})  + x(z_{12} \otimes z_{21}) + (z_{22}\otimes z_{21})}{x ( z_{12}\otimes 1 ) + (z_{22}  \otimes 1)} \right)^j \nonumber \\
 &\qquad\times \left(\frac{x (z_{11}  \otimes z_{12}) + (z_{21} \otimes z_{12}) + x(z_{12} \otimes z_{22}) + (z_{22} \otimes z_{22}) }{x ( z_{12}\otimes 1 ) + (z_{22}  \otimes 1)}\right)^j \nonumber \\
 &\qquad \times\left(\frac{x (z_{11}  \otimes 1) + (z_{21} \otimes 1)}{x ( z_{12}\otimes 1 ) + (z_{22}  \otimes 1)}\right)^{-j}f(\tilde{x})(N^l\otimes N^l) \nonumber \\
 &= (N^k \otimes N^k) ( x( z_{11} \otimes z_{11} + z_{12} \otimes z_{21}) + (z_{21} \otimes z_{11} + z_{22} \otimes z_{21}))^j \nonumber \\
 &\qquad\times ( x( z_{11} \otimes z_{12} + z_{12} \otimes z_{22}) + (z_{21} \otimes z_{12} + z_{22} \otimes z_{22}))^j x^{-j}f(\tilde{x}) (N^l \otimes N^l) \nonumber \\
 &= \Delta(N)^k (x\Delta(z_{11}) + \Delta(z_{21}))^j(x\Delta(z_{12}) + \Delta(z_{22}))^jx^{-j} f(\tilde{x}) \Delta(N)^l.
\end{align}
Here the second to last equality uses that $[xz_{11} + z_{21}, xz_{12} + z_{22}] = 0$ and we identify coproducts in the last line. Now 
\begin{align}
    \tilde{x} &= \left(\frac{x (z_{11}  \otimes 1) + (z_{21} \otimes 1)}{x ( z_{12}\otimes 1 ) + (z_{22}  \otimes 1)}(1 \otimes z_{11}) + (1 \otimes z_{21}) \right) \nonumber \\
    &\qquad\times \left(\frac{x (z_{11}  \otimes 1) + (z_{21} \otimes 1)}{x ( z_{12}\otimes 1 ) + (z_{22}  \otimes 1)}(1 \otimes z_{12}) + (1 \otimes z_{22}) \right)^{-1} \nonumber \\
    &=  \left(\frac{x (z_{11}  \otimes z_{11}) + (z_{21} \otimes z_{11})  + x(z_{12} \otimes z_{21}) + (z_{22}\otimes z_{21})}{x ( z_{12}\otimes 1 ) + (z_{22}  \otimes 1)} \right) \nonumber \\
 &\qquad \times \left(\frac{x (z_{11}  \otimes z_{12}) + (z_{21} \otimes z_{12}) + x(z_{12} \otimes z_{22}) + (z_{22} \otimes z_{22}) }{x ( z_{12}\otimes 1 ) + (z_{22}  \otimes 1)}\right)^{-1} \nonumber \\
 &=( x( z_{11} \otimes z_{11} + z_{12} \otimes z_{21}) + (z_{21} \otimes z_{11} + z_{22} \otimes z_{21}))\nonumber\\
 &\qquad \times ( x( z_{11} \otimes z_{12} + z_{12} \otimes z_{22}) + (z_{21} \otimes z_{12} + z_{22} \otimes z_{22}))^{-1} \nonumber \\
 &= \left( \frac{x \Delta(z_{11}) + \Delta(z_{21})}{x \Delta(z_{12}) + \Delta(z_{22})} \right),
\end{align}
such that indeed 
\begin{align}
    (\phi \otimes \text{id})(\phi(x)) = (\text{id} \otimes \Delta)(\phi(x)).
\end{align}
As such the proposed transformations do indeed furnish corepresentations of GL$_q(2,\mathbb{R})$. 
The reason that setting $l + k = -j$  directly produces a corepresentation of SL$_q(2)$ can be seen via a simple power counting argument. The projection map to SL$_q(2, \mathbb{R})$ pairs each $z_{ij}$ with a factor of $N^{-1/2}$. Now requiring $\phi(f(x))$ to be in $V \otimes \text{SL}_q(2,\mathbb{R})$, where $f(x) \in V$, can equivalently expressed as the condition that $\phi(f(x))$ should be writable solely in terms of $a,b,c,d$ and constants. As $\phi(f(x))$ is effectively $\sim z_{ij}^{2j}$ this requires a factor of $N^{-j}$ requiring $l +k = -j$. Indeed after dualizing the generators we will find that this condition is equivalent to imposing $K_0 = 1$ on the functional space which is equivalent to restricting $U_q(gl(2))$ to $U_q(sl(2))$.  

\subsection{Duality of principal series representations}
\label{sec:AppendixB}
Having shown that (\ref{eq:corep}) indeed produces a corepresentation we can now derive that the induced representation on $U_q(\frak{sl}_2)$ is indeed the principal series representations of the deformed universal enveloping algebra, defined through (\ref{eq:principalseriesgenerators2}). We will show that the induced form of the generators $E^+,E^-,K$ is exactly that of the principle series representation. 
The defining properties of the Hopf duality and the corepresentation then ensure that this consistently extends to the full algebra.  

To explicitly calculate the generators it is useful to first derive a few identities following from the Hopf pairing and coproduct of both algebras:

Let $\alpha \in \mathbb{C}$ and in a slight abuse of notation let us write 
$(\text{id} \otimes \braket{X,\cdot})(x\otimes z_{ij} + 1 \otimes z_{kl})^\alpha$ as $\braket{X, (xz_{ij} + z_{kl})^\alpha}$.

Firstly let $X$ be grouplike e.g $X \in \{K,K_0\}$ then  $\Delta(X) = X \otimes X$ and as such
\begin{align}
    \braket{X, (xz_{ij} + z_{kl})^\alpha} &= \braket{ X\otimes X, (xz_{ij} + z_{kl})^{\alpha-1} \otimes (xz_{ij} + z_{kl})} \nonumber \\&= \braket{X, (xz_{ij} + z_{kl})^{\alpha-1}} \braket{X, (xz_{ij} + z_{kl})} = \braket{X, (xz_{ij} + z_{kl})}^\alpha ,
\end{align}
where the first equality follows from the definition of the Hopf dual $\braket{X, a_1 a_2 } = \braket{\Delta(X), a_1 \otimes a_2}$, the second equality follows from the extension of the pairing to tensor products and the last equality follows from induction and we have analytically continued to $\alpha \in \mathbb{C}$.

It trivially follows that 
\begin{align}
    \braket{K, (xz_{11} + z_{21})^\alpha} = x^\alpha q^{\alpha}, \qquad& \braket{K, (xz_{12} + z_{22})^\alpha} = q^{-\alpha},\\
    \braket{K_0, (xz_{11} + z_{21})^\alpha} = x^\alpha q^{-\alpha/2}, \qquad& \braket{K_0, (xz_{12} + z_{22})^\alpha} = q^{-\alpha/2 },
\end{align}
and using the coproducts of $z_{ij}$ we find 
\begin{align}
    \braket{K_0^{-1} K^{-1}, (xz_{11} + z_{21})^\alpha} &=\braket{K_0^{-1}\otimes K^{-1} , \Delta(xz_{11} + z_{21})}^\alpha  \nonumber \\
    &=\braket{K_0^{-1}\otimes K^{-1} , x (z_{11} \otimes z_{11} + z_{12} \otimes z_{21} ) + (z_{21} \otimes z_{11} + z_{22} \otimes z_{21})}^\alpha \nonumber \\
    &= x^\alpha q^{-\alpha/2} ,
\end{align}
and similarly
\begin{align}
    \braket{K_0^{-1} K^{-1}, (xz_{11} + z_{21})^\alpha} = x^\alpha q^{- \alpha /2},&\qquad \braket{K_0 K, (xz_{11} + z_{21})^\alpha} = x^{\alpha}q^{ \alpha /2},\\
    \braket{K_0^{-1} K^{-1}, (xz_{12} + z_{22})^\alpha} = q^{3\alpha/2}, &\qquad
    \braket{K_0 K, (xz_{12} + z_{22})^\alpha} = q^{-3 \alpha/2},\\
    \braket{K_0^{-1} K, (xz_{11} + z_{21})^\alpha} = x^\alpha q^{3\alpha/2},  &\qquad \braket{K_0 K^{-1}, (xz_{11} + z_{21})^\alpha} = x^{\alpha}q^{- 3\alpha/2}, \\
    \braket{K_0^{-1} K, (xz_{12} + z_{22})^\alpha} = q^{-\alpha /2}, &\qquad
    \braket{K_0 K^{-1}, (xz_{12} + z_{22})^\alpha} = q^{ \alpha /2}.
\end{align}
Further, as $N$ is grouplike one finds that for any product of grouplike operators $X_1 ... X_n$:
\begin{equation}
    \braket{X_1 ... X_n ,N^\gamma} = \prod_{j = 1}^n \braket{X_j,N}^\gamma.
\end{equation}
Further $\braket{X,g} = 0 \implies \braket{X,g^\alpha} = 0$ if every summand in $\Delta(X)$ is either of the form $X \otimes Y$ or $ Y \otimes X$ for arbitrary $Y$ which again follows from induction and analytic continuation.
Explicitly this implies e.g.
\begin{align}
    \braket{E^+,N^\alpha} = 0, &\qquad \braket{E^-,N^\alpha} = 0,\\
    \braket{E^-,(x z_{12} + z_{22})^\alpha} = 0, &\qquad  \braket{E^+,(x z_{11} + z_{21})^\alpha} = 0.
\end{align}
Finally let us calculate $\braket{E^+,(x z_{12} + z_{22})^\alpha}$ and  $ \braket{E^-,(x z_{11} + z_{21})^\alpha} $. We will do so by induction.  Using the coproduct of $E^-$ one finds
\begin{align}
    \braket{E^-,(x z_{11} + z_{21})^\alpha} &= \braket{E^- \otimes K_0   + K_0^{-1} K \otimes E^- ,(x z_{11} + z_{21})\otimes (x z_{11} + z_{21})^{\alpha -1}} \nonumber \\
    &=  x^{\alpha -1} q^{-(\alpha -1)/2} \braket{E^-,(x z_{11} + z_{21})} + xq^{3/2}\braket{E^-,(x z_{11} + z_{21})^{\alpha-1}}  \nonumber \\
    &=  x^{\alpha -1} q^{-(\alpha -1)/2} + xq^{3/2}\braket{E^-,(x z_{11} + z_{21})^{\alpha-1}},
\end{align}
denoting $\braket{E^-,(x z_{11} + z_{21})^\alpha} \equiv f^\alpha $ to shorten notation we find the recursion relation and initial condition 
\begin{align}
    f^\alpha &= x^{\alpha -1 } q^{-(\alpha -1)/2} + xq^{3/2} f^{\alpha -1},\\
    f^1 &= \braket{E^-,(x z_{11} + z_{21})} = 1.
\end{align}
for natural numbers $\alpha$. It is easily checked that setting 
\begin{align}
    f^\alpha = x^{\alpha -1}q^{-(\alpha-1)/2} \left[ \sum_{j = 0}^{\alpha-1} q^{2j}\right] =  x^{\alpha -1}q^{-(\alpha -1)/2} \frac{1 - q^{2\alpha}}{1 - q^{2}},
\end{align}
solves the above recursion relation for all $\alpha$.
Similarly 
\begin{align}
   e^\alpha \equiv \braket{E^+,(x z_{12} + z_{22})^\alpha} &= \braket{K_0 \otimes E^++  E^+\otimes K_0^{-1} K^{-1},(x z_{12} + z_{22})^{\alpha-1} \otimes (x z_{12} + z_{22})} \nonumber \\
    &= e^{\alpha -1} \braket{K_0^{-1}K^{-1},(xz_{12} + z_{22})} + e^1 \braket{K_0,(xz_{12} + z_{22})}^{\alpha -1} \nonumber \\
    &= e^{\alpha -1} q^{3/2} +  x q^{-(\alpha -1)/2},
\end{align}
resulting in the solution 
\begin{align}
    e^\alpha =  x q^{-(\alpha -1)/2}\frac{1 - q^{2 \alpha}}{1 - q^{2}}.
\end{align}
It follows from the above considerations that 
\begin{align}
    \braket{K, N^{\gamma}\left( x z_{11} + z_{21} \right)^\alpha \left( x z_{12} + z_{22} \right)^\beta N^\delta } &=  \braket{K,\left( x z_{11} + z_{21} \right)}^\alpha \braket{K, \left( x z_{12} + z_{22} \right)}^\beta \braket{K, N}^{\gamma + \delta} \nonumber \\
    & = x^\alpha q^{\alpha -\beta} ,
\end{align}
and
\begin{align}
    \braket{K_0, N^\gamma\left( x z_{11} + z_{21} \right)^\alpha \left( x z_{12} + z_{22} \right)^\beta N^\delta } &= \braket{K_0,\left( x z_{11} + z_{21} \right)}^\alpha \braket{K_0, \left( x z_{12} + z_{22} \right)}^\beta \braket{K_0, N}^{\gamma +\delta} \nonumber \\
    &= x^\alpha q^{-(\alpha + \beta)/2} q^{-(\gamma + \delta)},
\end{align}
where we have used the grouplike coproduct of $K,K_0$. Further using the coproduct and $\braket{E^-,N} =\braket{E^-, (xz_{12} + z_{22})} = 0 $ one finds 
\begin{align}
    \braket{E^-, N^\gamma \left( x z_{11} + z_{21} \right)^\alpha \left( x z_{12} + z_{22} \right)^\beta N^\delta } 
    &= \braket{E^-,N^\gamma (x z_{11} + z_{21})^\alpha}\braket{K_0 ,(x z_{12} + z_{22})^{\beta} N^\delta}  \nonumber\\
    &\hspace{-2.5cm}= q^{\gamma - \delta -\beta/2} f^{\alpha} = q^{-1/2}x^{\alpha -1} q^{\gamma - \delta } \frac{q^{-(\alpha + \beta)/2} - q^{3\alpha/2 - \beta/2}}{q^{-1} - q}.
    \end{align}
    In an identical manner we find:
    \begin{align}
    \braket{E^+, N^\gamma \left( x z_{11} + z_{21} \right)^\alpha \left( x z_{12} + z_{22} \right)^\beta N^\delta } &=  x^\alpha q^{\delta - \gamma - \alpha/2} e^\beta = q^{-1/2} x^{\alpha + 1} q^{\delta - \gamma} \frac{q^{-(\alpha + \beta)/2} - q^{3\beta/2 -\alpha/2}}{q^{-1} - q}. \end{align}
    The generators can now be determined on the space of functions spanned by $x^s$ as 
    \begin{align}
        x^s = f(x) &\mapsto  \tilde{f}(x) = N^{-j/2}\left( x z_{11} + z_{21} \right)^j \left( x z_{12} + z_{22} \right)^j x^{-j} f\left( \frac{xz_{11} + z_{21}}{ xz_{12} + z_{22}}\right)N^{-j/2},
        \end{align}
        implying
        \begin{align}
        X(f(x)) &\equiv \braket{X,\tilde{f}(x)}  = \left\langle X,N^{-j/2} \left( x z_{11} + z_{21} \right)^j \left( x z_{12} + z_{22} \right)^j x^{-j} \left( \frac{xz_{11} + z_{21}}{ xz_{12} + z_{22}}\right)^s N^{-j/2}\right\rangle  \nonumber \\
        &=x^{-j}\braket{X,N^\gamma(xz_{11} +z_{21})^\alpha (xz_{12} + z_{22})^\beta N^\delta}\vert_{\alpha = j +s, \beta = j -s , \gamma = -j/2, \delta = -j/2}.
    \end{align}
    Explicitly for $X = K,K_0, E^+, E^-$ this leads to
    \begin{align}
        K(x^s)  &= x^{s}q^{2s} = R_{q^{2}} f(x), \\
        K_0(x^s) &= x^{s} = f(x),\\
        E^+(x^s) &= - q^{1/2} x^{s+1} \frac{q^{-j} - q^{j-2s}}{q - 
        q^{-1}} = -x q^{-1/2} \frac{q^{-j} - q^{j}R_{q^{-2}}}{q - q^{-1}} f(x),\\
        E^-(x^s) &= x^{s-1} q^{-1/2}\frac{q^{j + 2s} - q^{-j}}{q - q^{-1}} = \frac{q^{-1/2}}{x}\frac{q^jR_{q^2} - q^{- j}}{q - q^{-1}} f(x),
    \end{align}
    which produces the explicit form of the principal series generators of the deformed universal enveloping algebra.

 \subsection{Unitarity and star structures on the dual Hopf algebra}
 \label{app:A2}
In the main text we mostly focused on describing the quantum group structure from the perspective of the deformation of the universal enveloping algebra, finding the twisted star to be a convenient choice for imposing positivity on the quantum group in a consistent manner. Combined with the insights of how relevant representations are realized on the dual group it is of course natural to ask what imprint the twisting of the star structure leaves on the dual quantum group. It is known that introducing twisted stars leads to a range of subtleties especially related to unitarity of corepresentations see e.g. \cite{Coquereaux:1999va}. Most results of the current work, with the exception of \ref{sec:Hopfduality}, are formulated directly in terms of properties of the deformed universal enveloping algebra. Here unitarity of representations was explicitly checked and as such the results are expected to be independent of the subtleties arising on the dual that we discuss here.

Following \cite{Coquereaux:1999va}, a twisted Hopf star naturally induces a twisting of the star on the ``group manifold'' as well. This star is induced from the pairing by requiring 
\begin{equation}
    \langle f^*,g\rangle = \overline{\langle f,g^*\rangle },
\end{equation}
which differs from the usual consistency condition of Hopf stars
\begin{equation}\label{eq:usualstar}
    \langle f^*,g\rangle = \overline{\langle f,S(g)^*\rangle } \qquad    \langle f,g^*\rangle = \overline{\langle S(f)^*,g \rangle},
\end{equation}
by the absence of the antipode in the definition.
Using this definition the twisted star on the dual algebra is found to act consistently via the involution
\begin{equation}
    a^* = d, \qquad d^* =a, \qquad b^* = -b,\qquad c^* = -c.
    \label{eq:twistedstarSL}
\end{equation}
It is easily checked that this action indeed furnishes a twisted Hopf star on the dual consistent with the defining relations. For example, the twisting is consistent with the coproduct 
\begin{align*}
    \Delta(a)^* &=  (a\otimes a)^* +(b  \otimes c)^* = d\otimes d+ c\otimes b = \Delta(d) =\Delta(a^*),  \\ 
    \Delta(d)^* &=  (d\otimes d)^* +(c  \otimes b)^* = a\otimes a+ b\otimes c = \Delta(a) =\Delta(d^*), \\
    \Delta(b)^* &=  (a\otimes b)^* +(b  \otimes d)^* = - b\otimes d - a\otimes b = -\Delta(b) =\Delta(b^*),  \\ 
    \Delta(c)^* &=  (c\otimes a)^* +(d \otimes c)^* = -d\otimes c - c\otimes a = -\Delta(c) = \Delta(c^*).  
\end{align*}
The restriction to unit $q$-determinant is also preserved
\begin{equation}
   (\det \{\}_q)^* = (ad - qbc)^* = d^*a^* -qb^*c^* =  ad - qcb = ad - qbc = 1 = 1^*.
\end{equation}
The other algebraic relations \eqref{eq:coordinatealgebra} can be checked in a similar manner. It is also interesting to note that the relations \eqref{eq:twistedstarSL} are equivalent to sending $g$ to its classical ($q = 1$) matrix inverse $g^{-1}$.  This should not be surprising since the action of the twisted star on the generators is, up to a $q$ dependent scaling that ensures $* \circ * = \text{id}$, equivalent to that of the antipode $S$. It is exactly this $q$ dependent scaling that deforms the classical inverse to the inverse in the quantum group where $S^2 \neq \text{id}$.  Further, in classical Lie groups, this property is exactly the defining property of unitarity on representations i.e. $T(g)^\dagger = T(g^{-1})$ for a representation $T$. 

We emphasize that we did not need this star in any of the main calculations in this work, since we imposed positivity directly on the deformed universal enveloping algebra. Requiring consistency with the Hopf duality then imposes the correct notion of positivity on the dual Hopf algebra. 

\section{Representations of $U_q(\mathfrak{sl}_2)$}
\label{sec:AppendixD}
We have previously realized representations of $U_q(\mathfrak{sl}(2,\mathbb{R}))$ in terms of operators on certain functional spaces as in \eqref{eq:principalseriesgenerators2}. However, for certain applications a different realization in terms of ladder operators is beneficial. This is also the language in which these representations appeared recently in the DSSYK literature, see e.g. \cite{Berkooz:2018jqr,Jafferis:2022wez}. We will briefly review this construction, originally described in \cite{Burban:1992ys,groenevelt} and relate it to our construction.\footnote{A similar construction for the discrete series representation was conducted in \cite{stovicek2000}.}
Let $E^+,E^+,K,K^{-1}$ be the generators of $U_q(\mathfrak{sl}_2)$ as before, and $V$ be the complex vector space spanned by the states $\{\ket{m}\}_{m \in \mathbb{Z}}$. Then the action 
\begin{align}
\label{eq:modulessu111}
    K \ket{m} &= q^{2m} \ket{m}, \\
    E^+ \ket{m} &= -\frac{q^{m+1/2}}{q - q^{-1}} \sqrt{ \left(1- q^{-2(m + 1 + j )}\right)\left(1 - q^{2 (j  - m)}\right)} \ket{m+ 1}\label{eq:modulesu1E},\\
    E^- \ket{m} &= \frac{q^{-m + 1/2}}{q - q^{-1}}\sqrt{\left( 1 - q^{2 (m - 1 - j)} \right) \left( 1 -  q^{2(m  + j)}\right) }\ket{m- 1},
    \label{eq:modulesu1F}
\end{align}
defines a representation $T_j$ for every complex $j$ as is easily checked.
Thus $E^\pm$ are realized as raising and lowering operators respectively while $K$ acts diagonally. Note that the representation $T_j$ is infinite-dimensional while its irreducibility depends on the choice of $j$. 
Importantly it turns out that all irreducible representations of $U_q(\frak{sl}_2)$ are realized as $T_j$ for some $j \in \mathbb{C}$ or a quotient of such a module.\footnote{To be precise the characterization of all irreducible representations requires an extra parameter $\epsilon$ and replacing $n \to n + \epsilon$ in all expressions. This parameter also shows up in the classical case and is related to whether the representations exponentiate to representations of SL$_q(2,\mathbb{R})$ or its universal cover. }   

This realization \eqref{eq:modulessu111} - \eqref{eq:modulesu1F} can be related back to our previous realization
\begin{align}
\label{eq:principalseriesgeneratorsa}
    \hat{K}&=R_{q^2},\qquad \hat{E}^-= q^{-1/2}\frac{q^jR_{q^2}-q^{-j}}{x(q-q^{-1})},\qquad \hat{E}^+=-xq^{-1/2} \frac{q^{-j}-q^jR_{q^{-2}}}{q-q^{-1}},
\end{align}
by introducing a set of states $\ket{x}$ for which  
\begin{equation}
\braket{x\vert m} = x^{m}f(m),
\end{equation}
with the function $f(m)$ to be fixed via algebraic relations as follows. 
Using the two separate definition of the representation i.e. the action \eqref{eq:modulesu1E}, as well as the explicit form of the generators as operators on the discretized line (\ref{eq:principalseriesgeneratorsa}), we find
\begin{align}
    \braket{x\vert E^+ \vert m} &= -\frac{q^{m+1/2}}{q - q^{-1}} \sqrt{ \left(1- q^{-2(m + 1 + j )}\right)\left(1 - q^{2 (j  - m)})\right)} f(m +1) x^{m+ 1},\\
   \hat{E}^+\braket{x\vert m } &= -q^{-1/2} x \frac{q^{-j} - q^jR_{q^{-2}}}{q - q^{-1}}f(m)x^{m} = -q^{-1/2}\frac{q^{-j} - q^{j -2m}}{q-q^{-1}}f(m)x^{m+1}.
\end{align}
Equating these two and explicitly introducing the quantum numbers $[n]_{q^2} =  \frac{1 - q^{2n}}{1- q^2}$ produces a recursion relation for $f(m)$ of the form
\begin{equation}
    f(m + 1) = \sqrt{q^{-2(m + j  + 1)}\frac{[j-m]_{q^2}}{[-(j+ m + 1)]_{q^2}}} f(m),
\end{equation}
with solution
\begin{equation}
    f(m) = f(0) q^{-m(j+1) - \frac{m(m+1)}{2}} \sqrt{\frac{\Gamma_{q^2}[-(j+m)]}{\Gamma_{q^2}[j-m +1]}}.
\end{equation}
With this, one easily shows that 
\begin{align}
   \hat{E}^-\braket{x\vert m } &= \frac{q^{-1/2}}{ x}\frac{q^jR_{q^2} - q^{- j}}{q - q^{-1}}f(m)x^{m} = q^{-1/2}\frac{q^{j+2m} - q^{-j }}{q-q^{-1}}f(m)x^{m-1}, \\
   \hat{K}\braket{x\vert m } &= f(m)R_{q^2}x^m = q^{2m}f(m)x^m,
\end{align}
consistent with the same $f(m)$.

Note that the inner product on this space will not be what is chosen in \cite{Burban:1992ys} since we are interested in unitarity with respect to the twisted star defined in \eqref{eq:ourstar} instead of $U_q(\frak{su}(1,1))$. The correct inner product can be derived from the inner product on the discretized line specified in \eqref{eq:innerproduct} and the relation between $g(x)$ and $\vert n \rangle$ defined above, however this inner product will not play any role in this work. 

Explicitly one finds that the (unitary) class of principal series representations are realized as $T_j$ if $j = -\frac{1}{2} + ik \in -\frac{1}{2} + i \mathbb{R}$ while the positive/negative discrete series representations and finite-dimensional representations are defined as quotients of the above module at integer values of $j$.
At these values the space $V_j^{\text{FD}} \equiv\{\ket{m} : -j \leq m \leq  j  \}$ forms an invariant subspace since $E \ket{j} = \left( q^{-2j + 1}[ 1 + 2j ]_{q^2} [0]_{q^2}  \right)^{1/2} \ket{m+ 1} = 0$ as $[0]_q = 0$ and similarly $E^-\ket{-j} = 0$. The representation on this space is denoted $T_j^{\text{FD}}$ and is the $q$-deformed analogue of the well-known $2j+1$ dimensional spin $j$ representation of $\text{SU}(2)$.
The quotient $V_j/V_j^\text{FD}$ further decomposes into two invariant subspaces $V_j^+ = \{\ket{m} : l < m  \}$ and $V_j^- = \{\ket{m} : -l > m  \} $. The representations acting on these spaces are denoted by $T_j^\pm$ and are called positive/negative discrete series representations. They are respectively lowest/highest weight representations of the quantum group. 
Since this construction is readily generalized to negative $j$ \cite{Burban:1992ys} we find that in summary 
\begin{equation}
\boxed{
    T_j \simeq T^+_j \oplus T^\text{FD}_j \oplus T^-_j.}
\end{equation}
and as a consequence that principal series representations can be viewed as a analytic continuation (in $j$) of the direct sum of a positive, negative discrete, and finite dimensional representation. 

\section{Hamiltonian reduction towards $q$-Liouville quantum mechanics }
\label{sec:Hamiltonianreduction}
We have implemented the Hopf duality between the coordinate algebra SL$_q(2,\mathbb{R})$ and algebra of generators $U_q(\mathfrak{sl}(2,\mathbb{R}))$ via the Gauss-Euler exponentiation \eqref{eq:generalGE}. Evaluating the generators in the principal series representation \eqref{eq:alegbragenerators2}, we can determine the action of the parabolic exponentials on the Whittaker vectors $\ket{\phi_j^\pm}$. Since the right adjoint action on the left Whittaker vector is exactly diagonalized through \eqref{eq:adjointleftaction}, we can immediately exponentiate
\begin{align}
    \bra{\phi_j^-}e_{q^{-2}}(\beta E^-)=e_{q^{-2}}\left(\frac{q^{1/2-j}}{1-q^2}\beta\right)\bra{\phi_j^-}.
\end{align} 
On the other hand, since the right Whittaker vector is diagonalized by the right parabolic generator up to the action of $K$ as in \eqref{eq:rightaction}, we instead have to use successively the algebra relation $KE^+=q^2E^+K$, leading to:
\begin{align}
    (E^+)^n\ket{\phi_j^+}=\left(-\frac{q^{5/2+j}}{1-q^2}\right)^nq^{n(n-1)}K^{-n}\ket{\phi_j^-}.
\end{align}
Subsequently using the non-commutativity of the coordinates $e^{2\phi H}\gamma=K\gamma e^{2\phi H}$, it follows that the $e_{q^2}(.)$ exponential becomes a $E_{q^2}(.)$ exponential after diagonalization by the right Whittaker vector 
\begin{align}
    e^{2\phi H}e_{q^2}(\gamma E^+)\ket{\phi_j^+}&=\sum_{n=0}^\infty\frac{\gamma^n}{[n]_{q^2}!}e^{2\phi H}K^n(E^+)^n \ket{\phi_j^+}\\&=\sum_{n=0}^\infty \frac{\gamma^n}{[n]_{q^2}!}\left(-\frac{q^{5/2+j}}{1-q^2}\right)^n q^{n(n-1)}e^{2\phi H} \ket{\phi_j^+}\\&=E_{q^2}\left(-\frac{q^{5/2+j}}{1-q^2}\gamma\right)e^{2\phi H}\ket{\phi_j^+}.
\end{align} 
Consequently, the matrix element reduces to the Whittaker function $\psi_j(\phi)\equiv\braket{\phi_j^-\vert e^{2\phi H}\vert \phi_j^+}$, multiplied on the left by the parabolic exponential prefactors 
\begin{align}
\label{eq:GEfactorization}
    \braket{\phi_j^-\vert e_{q^{-2}}(\beta E^-)e^{2\phi H}e_{q^2}(\gamma E^+)\vert \phi_j^+}=e_{q^{-2}}\left(\frac{q^{1/2-j}}{1-q^2}\beta\right)E_{q^2}\left(-\frac{q^{5/2+j}}{1-q^2}\gamma\right)\;\psi_j(\phi).
\end{align}
Using this, it is now straightforward to show that the regular representation Casimir $L_{\mathcal{C}}$ (i.e. the Laplacian on the quantum group manifold) when acting on the representation matrix element reduces to its action on the Whittaker function as \cite{Mertens:2022aou,Blommaert:2023opb}:
\begin{align}
    &(q-q^{-1})^2\;L_{\mathcal{C}}\braket{\phi_j^-\vert g\vert \phi_j^+}\\&=e_{q^{-2}}\left(\frac{q^{1/2-j}}{1-q^2}\beta\right)E_{q^2}\left(-\frac{q^{5/2+j}}{1-q^2}\gamma\right)\left(-qe^{-2\phi}T^\phi_{-\log q}+q^{-1}T^\phi_{-\log q}+qT^\phi_{\log q}\right)\;\braket{\phi_j^-\vert e^{2\phi H}\vert \phi_j^+}.\nonumber
\end{align}
Using the principal series Casimir eigenvalue $\frac{2\cos\theta}{(q-q^{-1})^2}$ and removing the parabolic eigenvalues from the left, the Whittaker function is consistently solved by the gravitational $q$-Liouville Schrödinger equation: 
\begin{align}
\label{eq:bosonicrecursion}
    \left(qT^\phi_{\log q}+(q^{-1}-qe^{-2\phi})T^\phi_{-\log q}\right)\braket{\phi_j^-\vert e^{2\phi H}\vert \phi_j^+}=2\cos\theta\; \braket{\phi_j^-\vert e^{2\phi H}\vert \phi_j^+}.
\end{align} 
Relating the hyperbolic coordinate with the chord number through \eqref{eq:dictionary}, the Liouville Hamiltonian of the two-sided gravitational system becomes the transfer matrix of the auxiliary quantum mechanical system of DSSYK \cite{Lin:2022rbf} 
\begin{align}
    \left(q T^n_{-1}+\left(q^{-1}-q^{2n+1}\right)T^n_1\right)\braket{\phi_j^-\vert K^{-n}\vert \phi_j^+}=2\cos\theta \braket{\phi_j^-\vert K^{-n}\vert \phi_j^+},
\end{align} 
whose solutions are the Whittaker functions \eqref{eq:finalchordwhitt}, using the defining recurrence relation of the $q$-Hermite polynomials: 
\begin{align}
    (1-q^{2n})H_{n-1}(\cos\theta\vert q^2)+H_{n+1}(\cos\theta\vert q^2)=2\cos\theta H_n(\cos\theta\vert q^2).
\end{align}
Therefore, implementing the gravitational boundary conditions by choosing mixed parabolic matrix elements in the construction of the principal series matrix element is indeed consistent with solving the Schrödinger equation of DSSYK.

\newpage 
\bibliographystyle{ourbst}
\bibliography{DSSYK.bib}

\providecommand{\href}[2]{#2}\begingroup\raggedright\begin{thebibliography}{10}

\bibitem{Buividovich:2008gq}
P.~V. Buividovich and M.~I. Polikarpov, ``{Entanglement entropy in gauge theories and the holographic principle for electric strings},'' \href{http://dx.doi.org/10.1016/j.physletb.2008.10.032}{{\em Phys. Lett. B} {\bfseries 670} (2008) 141--145}, \href{http://arxiv.org/abs/0806.3376}{{\ttfamily arXiv:0806.3376 [hep-th]}}.

\bibitem{Casini:2013rba}
H.~Casini, M.~Huerta, and J.~A. Rosabal, ``{Remarks on entanglement entropy for gauge fields},'' \href{http://dx.doi.org/10.1103/PhysRevD.89.085012}{{\em Phys. Rev. D} {\bfseries 89} no.~8, (2014) 085012}, \href{http://arxiv.org/abs/1312.1183}{{\ttfamily arXiv:1312.1183 [hep-th]}}.

\bibitem{Donnelly:2014gva}
W.~Donnelly, ``{Entanglement entropy and nonabelian gauge symmetry},'' \href{http://dx.doi.org/10.1088/0264-9381/31/21/214003}{{\em Class. Quant. Grav.} {\bfseries 31} no.~21, (2014) 214003}, \href{http://arxiv.org/abs/1406.7304}{{\ttfamily arXiv:1406.7304 [hep-th]}}.

\bibitem{Lin:2018bud}
J.~Lin and D.~Radi\v{c}evi\'c, ``{Comments on defining entanglement entropy},'' \href{http://dx.doi.org/10.1016/j.nuclphysb.2020.115118}{{\em Nucl. Phys. B} {\bfseries 958} (2020) 115118}, \href{http://arxiv.org/abs/1808.05939}{{\ttfamily arXiv:1808.05939 [hep-th]}}.

\bibitem{Ghosh:2015iwa}
S.~Ghosh, R.~M. Soni, and S.~P. Trivedi, ``{On The Entanglement Entropy For Gauge Theories},'' \href{http://dx.doi.org/10.1007/JHEP09(2015)069}{{\em JHEP} {\bfseries 09} (2015) 069}, \href{http://arxiv.org/abs/1501.02593}{{\ttfamily arXiv:1501.02593 [hep-th]}}.

\bibitem{Harlow:2015lma}
D.~Harlow, ``{Wormholes, Emergent Gauge Fields, and the Weak Gravity Conjecture},'' \href{http://dx.doi.org/10.1007/JHEP01(2016)122}{{\em JHEP} {\bfseries 01} (2016) 122}, \href{http://arxiv.org/abs/1510.07911}{{\ttfamily arXiv:1510.07911 [hep-th]}}.

\bibitem{McGough:2013gka}
L.~McGough and H.~Verlinde, ``{Bekenstein-Hawking Entropy as Topological Entanglement Entropy},'' \href{http://dx.doi.org/10.1007/JHEP11(2013)208}{{\em JHEP} {\bfseries 11} (2013) 208}, \href{http://arxiv.org/abs/1308.2342}{{\ttfamily arXiv:1308.2342 [hep-th]}}.

\bibitem{Lin:2018xkj}
J.~Lin, ``{Entanglement entropy in Jackiw-Teitelboim Gravity},'' \href{http://arxiv.org/abs/1807.06575}{{\ttfamily arXiv:1807.06575 [hep-th]}}.

\bibitem{Blommaert:2018iqz}
A.~Blommaert, T.~G. Mertens, and H.~Verschelde, ``{Fine Structure of Jackiw-Teitelboim Quantum Gravity},'' \href{http://dx.doi.org/10.1007/JHEP09(2019)066}{{\em JHEP} {\bfseries 09} (2019) 066}, \href{http://arxiv.org/abs/1812.00918}{{\ttfamily arXiv:1812.00918 [hep-th]}}.

\bibitem{Mertens:2022ujr}
T.~G. Mertens, J.~Sim\'on, and G.~Wong, ``{A proposal for 3d quantum gravity and its bulk factorization},'' \href{http://dx.doi.org/10.1007/JHEP06(2023)134}{{\em JHEP} {\bfseries 06} (2023) 134}, \href{http://arxiv.org/abs/2210.14196}{{\ttfamily arXiv:2210.14196 [hep-th]}}.

\bibitem{Wong:2022eiu}
G.~Wong, ``{A note on the bulk interpretation of the quantum extremal surface formula},'' \href{http://dx.doi.org/10.1007/JHEP04(2024)024}{{\em JHEP} {\bfseries 04} (2024) 024}, \href{http://arxiv.org/abs/2212.03193}{{\ttfamily arXiv:2212.03193 [hep-th]}}.

\bibitem{Berkooz:2018qkz}
M.~Berkooz, P.~Narayan, and J.~Simon, ``{Chord diagrams, exact correlators in spin glasses and black hole bulk reconstruction},'' \href{http://dx.doi.org/10.1007/JHEP08(2018)192}{{\em JHEP} {\bfseries 08} (2018) 192}, \href{http://arxiv.org/abs/1806.04380}{{\ttfamily arXiv:1806.04380 [hep-th]}}.

\bibitem{Berkooz:2018jqr}
M.~Berkooz, M.~Isachenkov, V.~Narovlansky, and G.~Torrents, ``{Towards a full solution of the large N double-scaled SYK model},'' \href{http://dx.doi.org/10.1007/JHEP03(2019)079}{{\em JHEP} {\bfseries 03} (2019) 079}, \href{http://arxiv.org/abs/1811.02584}{{\ttfamily arXiv:1811.02584 [hep-th]}}.

\bibitem{Lin:2022rbf}
H.~W. Lin, ``{The bulk Hilbert space of double scaled SYK},'' \href{http://dx.doi.org/10.1007/JHEP11(2022)060}{{\em JHEP} {\bfseries 11} (2022) 060}, \href{http://arxiv.org/abs/2208.07032}{{\ttfamily arXiv:2208.07032 [hep-th]}}.

\bibitem{Jafferis:2022wez}
D.~L. Jafferis, D.~K. Kolchmeyer, B.~Mukhametzhanov, and J.~Sonner, ``{JT gravity with matter, generalized ETH, and Random Matrices},'' \href{http://arxiv.org/abs/2209.02131}{{\ttfamily arXiv:2209.02131 [hep-th]}}.

\bibitem{Susskind:2022bia}
L.~Susskind, ``{De Sitter Space, Double-Scaled SYK, and the Separation of Scales in the Semiclassical Limit},'' \href{http://arxiv.org/abs/2209.09999}{{\ttfamily arXiv:2209.09999 [hep-th]}}.

\bibitem{Bhattacharjee:2022ave}
B.~Bhattacharjee, P.~Nandy, and T.~Pathak, ``{Krylov complexity in large q and double-scaled SYK model},'' \href{http://dx.doi.org/10.1007/JHEP08(2023)099}{{\em JHEP} {\bfseries 08} (2023) 099}, \href{http://arxiv.org/abs/2210.02474}{{\ttfamily arXiv:2210.02474 [hep-th]}}.

\bibitem{Blommaert:2023opb}
A.~Blommaert, T.~G. Mertens, and S.~Yao, ``{Dynamical actions and q-representation theory for double-scaled SYK},'' \href{http://arxiv.org/abs/2306.00941}{{\ttfamily arXiv:2306.00941 [hep-th]}}.

\bibitem{Blommaert:2023wad}
A.~Blommaert, T.~G. Mertens, and S.~Yao, ``{The q-Schwarzian and Liouville gravity},'' \href{http://dx.doi.org/10.1007/JHEP11(2024)054}{{\em JHEP} {\bfseries 11} (2024) 054}, \href{http://arxiv.org/abs/2312.00871}{{\ttfamily arXiv:2312.00871 [hep-th]}}.

\bibitem{Susskind:2023hnj}
L.~Susskind, ``{De Sitter Space has no Chords. Almost Everything is Confined.},'' \href{http://dx.doi.org/10.22128/jhap.2023.661.1043}{{\em JHAP} {\bfseries 3} no.~1, (2023) 1--30}, \href{http://arxiv.org/abs/2303.00792}{{\ttfamily arXiv:2303.00792 [hep-th]}}.

\bibitem{Mukhametzhanov:2023tcg}
B.~Mukhametzhanov, ``{Large p SYK from chord diagrams},'' \href{http://dx.doi.org/10.1007/JHEP09(2023)154}{{\em JHEP} {\bfseries 09} (2023) 154}, \href{http://arxiv.org/abs/2303.03474}{{\ttfamily arXiv:2303.03474 [hep-th]}}.

\bibitem{Berkooz:2023cqc}
M.~Berkooz, Y.~Jia, and N.~Silberstein, ``{Parisi\textquoteright{}s Hypercube, Fock-Space Frustration, and Near-AdS2/Near-CFT1 Holography},'' \href{http://dx.doi.org/10.1103/PhysRevLett.132.081601}{{\em Phys. Rev. Lett.} {\bfseries 132} no.~8, (2024) 081601}, \href{http://arxiv.org/abs/2303.18182}{{\ttfamily arXiv:2303.18182 [hep-th]}}.

\bibitem{Okuyama:2023bch}
K.~Okuyama and K.~Suzuki, ``{Correlators of double scaled SYK at one-loop},'' \href{http://dx.doi.org/10.1007/JHEP05(2023)117}{{\em JHEP} {\bfseries 05} (2023) 117}, \href{http://arxiv.org/abs/2303.07552}{{\ttfamily arXiv:2303.07552 [hep-th]}}.

\bibitem{Lin:2022nss}
H.~Lin and L.~Susskind, ``{Infinite Temperature's Not So Hot},'' \href{http://arxiv.org/abs/2206.01083}{{\ttfamily arXiv:2206.01083 [hep-th]}}.

\bibitem{Berkooz:2022mfk}
M.~Berkooz, M.~Isachenkov, M.~Isachenkov, P.~Narayan, and V.~Narovlansky, ``{Quantum groups, non-commutative AdS$_{2}$, and chords in the double-scaled SYK model},'' \href{http://dx.doi.org/10.1007/JHEP08(2023)076}{{\em JHEP} {\bfseries 08} (2023) 076}, \href{http://arxiv.org/abs/2212.13668}{{\ttfamily arXiv:2212.13668 [hep-th]}}.

\bibitem{Goel:2023svz}
A.~Goel, V.~Narovlansky, and H.~Verlinde, ``{Semiclassical geometry in double-scaled SYK},'' \href{http://arxiv.org/abs/2301.05732}{{\ttfamily arXiv:2301.05732 [hep-th]}}.

\bibitem{Narovlansky:2023lfz}
V.~Narovlansky and H.~Verlinde, ``{Double-scaled SYK and de Sitter Holography},'' \href{http://arxiv.org/abs/2310.16994}{{\ttfamily arXiv:2310.16994 [hep-th]}}.

\bibitem{Verlinde:2024zrh}
H.~Verlinde and M.~Zhang, ``{SYK Correlators from 2D Liouville-de Sitter Gravity},'' \href{http://arxiv.org/abs/2402.02584}{{\ttfamily arXiv:2402.02584 [hep-th]}}.

\bibitem{Berkooz:2024evs}
M.~Berkooz, N.~Brukner, Y.~Jia, and O.~Mamroud, ``{From Chaos to Integrability in Double Scaled Sachdev-Ye-Kitaev Model via a Chord Path Integral},'' \href{http://dx.doi.org/10.1103/PhysRevLett.133.221602}{{\em Phys. Rev. Lett.} {\bfseries 133} no.~22, (2024) 221602}, \href{http://arxiv.org/abs/2403.01950}{{\ttfamily arXiv:2403.01950 [hep-th]}}.

\bibitem{Lin:2023trc}
H.~W. Lin and D.~Stanford, ``{A symmetry algebra in double-scaled SYK},'' \href{http://dx.doi.org/10.21468/SciPostPhys.15.6.234}{{\em SciPost Phys.} {\bfseries 15} no.~6, (2023) 234}, \href{http://arxiv.org/abs/2307.15725}{{\ttfamily arXiv:2307.15725 [hep-th]}}.

\bibitem{Verlinde:2024znh}
H.~Verlinde, ``{Double-scaled SYK, Chords and de Sitter Gravity},'' \href{http://arxiv.org/abs/2402.00635}{{\ttfamily arXiv:2402.00635 [hep-th]}}.

\bibitem{Almheiri:2024ayc}
A.~Almheiri and F.~K. Popov, ``{Holography on the Quantum Disk},'' \href{http://arxiv.org/abs/2401.05575}{{\ttfamily arXiv:2401.05575 [hep-th]}}.

\bibitem{Almheiri:2024xtw}
A.~Almheiri, A.~Goel, and X.-Y. Hu, ``{Quantum gravity of the Heisenberg algebra},'' \href{http://arxiv.org/abs/2403.18333}{{\ttfamily arXiv:2403.18333 [hep-th]}}.

\bibitem{Bossi:2024ffa}
L.~Bossi, L.~Griguolo, J.~Papalini, L.~Russo, and D.~Seminara, ``{Sine-dilaton gravity vs double-scaled SYK: exploring one-loop quantum corrections},'' \href{http://arxiv.org/abs/2411.15957}{{\ttfamily arXiv:2411.15957 [hep-th]}}.

\bibitem{Xu:2024hoc}
J.~Xu, ``{Von Neumann Algebras in Double-Scaled SYK},'' \href{http://arxiv.org/abs/2403.09021}{{\ttfamily arXiv:2403.09021 [hep-th]}}.

\bibitem{Xu:2024gfm}
J.~Xu, ``{On Chord Dynamics and Complexity Growth in Double-Scaled SYK},'' \href{http://arxiv.org/abs/2411.04251}{{\ttfamily arXiv:2411.04251 [hep-th]}}.

\bibitem{Heller:2024ldz}
M.~P. Heller, J.~Papalini, and T.~Schuhmann, ``{Krylov spread complexity as holographic complexity beyond JT gravity},'' \href{http://arxiv.org/abs/2412.17785}{{\ttfamily arXiv:2412.17785 [hep-th]}}.

\bibitem{Tietto:2025oxn}
D.~Tietto and H.~Verlinde, ``{A microscopic model of de Sitter spacetime with an observer},'' \href{http://arxiv.org/abs/2502.03869}{{\ttfamily arXiv:2502.03869 [hep-th]}}.

\bibitem{Berkooz:2024lgq}
M.~Berkooz and O.~Mamroud, ``{A cordial introduction to double scaled SYK},'' \href{http://dx.doi.org/10.1088/1361-6633/ada889}{{\em Rept. Prog. Phys.} {\bfseries 88} no.~3, (2025) 036001}, \href{http://arxiv.org/abs/2407.09396}{{\ttfamily arXiv:2407.09396 [hep-th]}}.

\bibitem{Collier:2023cyw}
S.~Collier, L.~Eberhardt, B.~M\"uhlmann, and V.~A. Rodriguez, ``{The Virasoro minimal string},'' \href{http://dx.doi.org/10.21468/SciPostPhys.16.2.057}{{\em SciPost Phys.} {\bfseries 16} no.~2, (2024) 057}, \href{http://arxiv.org/abs/2309.10846}{{\ttfamily arXiv:2309.10846 [hep-th]}}.

\bibitem{Collier:2024kmo}
S.~Collier, L.~Eberhardt, B.~M\"uhlmann, and V.~A. Rodriguez, ``{The complex Liouville string},'' \href{http://arxiv.org/abs/2409.17246}{{\ttfamily arXiv:2409.17246 [hep-th]}}.

\bibitem{Blommaert:2024ydx}
A.~Blommaert, T.~G. Mertens, and J.~Papalini, ``{The dilaton gravity hologram of double-scaled SYK},'' \href{http://dx.doi.org/10.1007/JHEP06(2025)050}{{\em JHEP} {\bfseries 06} (2025) 050}, \href{http://arxiv.org/abs/2404.03535}{{\ttfamily arXiv:2404.03535 [hep-th]}}.

\bibitem{Blommaert:2024whf}
A.~Blommaert, A.~Levine, T.~G. Mertens, J.~Papalini, and K.~Parmentier, ``{An entropic puzzle in periodic dilaton gravity and DSSYK},'' \href{http://arxiv.org/abs/2411.16922}{{\ttfamily arXiv:2411.16922 [hep-th]}}.

\bibitem{Blommaert:2025avl}
A.~Blommaert, A.~Levine, T.~G. Mertens, J.~Papalini, and K.~Parmentier, ``{Wormholes, branes and finite matrices in sine dilaton gravity},'' \href{http://arxiv.org/abs/2501.17091}{{\ttfamily arXiv:2501.17091 [hep-th]}}.

\bibitem{Schaller:1994es}
P.~Schaller and T.~Strobl, ``{Poisson structure induced (topological) field theories},'' \href{http://dx.doi.org/10.1142/S0217732394002951}{{\em Mod. Phys. Lett. A} {\bfseries 9} (1994) 3129--3136}, \href{http://arxiv.org/abs/hep-th/9405110}{{\ttfamily arXiv:hep-th/9405110}}.

\bibitem{Ikeda:1993aj}
N.~Ikeda and K.~I. Izawa, ``{General form of dilaton gravity and nonlinear gauge theory},'' \href{http://dx.doi.org/10.1143/PTP.90.237}{{\em Prog. Theor. Phys.} {\bfseries 90} (1993) 237--246}, \href{http://arxiv.org/abs/hep-th/9304012}{{\ttfamily arXiv:hep-th/9304012}}.

\bibitem{Ikeda:1993fh}
N.~Ikeda, ``{Two-dimensional gravity and nonlinear gauge theory},'' \href{http://dx.doi.org/10.1006/aphy.1994.1104}{{\em Annals Phys.} {\bfseries 235} (1994) 435--464}, \href{http://arxiv.org/abs/hep-th/9312059}{{\ttfamily arXiv:hep-th/9312059}}.

\bibitem{Cattaneo:2001bp}
A.~S. Cattaneo and G.~Felder, ``{Poisson sigma models and deformation quantization},'' \href{http://dx.doi.org/10.1142/S0217732301003255}{{\em Mod. Phys. Lett. A} {\bfseries 16} (2001) 179--190}, \href{http://arxiv.org/abs/hep-th/0102208}{{\ttfamily arXiv:hep-th/0102208}}.

\bibitem{Grumiller:2003ad}
D.~Grumiller and W.~Kummer, ``{The Classical solutions of the dimensionally reduced gravitational Chern-Simons theory},'' \href{http://dx.doi.org/10.1016/S0003-4916(03)00138-6}{{\em Annals Phys.} {\bfseries 308} (2003) 211--221}, \href{http://arxiv.org/abs/hep-th/0306036}{{\ttfamily arXiv:hep-th/0306036}}.

\bibitem{Mertens:2020hbs}
T.~G. Mertens and G.~J. Turiaci, ``{Liouville quantum gravity -- holography, JT and matrices},'' \href{http://dx.doi.org/10.1007/JHEP01(2021)073}{{\em JHEP} {\bfseries 01} (2021) 073}, \href{http://arxiv.org/abs/2006.07072}{{\ttfamily arXiv:2006.07072 [hep-th]}}.

\bibitem{Mertens:2020pfe}
T.~G. Mertens, ``{Degenerate operators in JT and Liouville (super)gravity},'' \href{http://dx.doi.org/10.1007/JHEP04(2021)245}{{\em JHEP} {\bfseries 04} (2021) 245}, \href{http://arxiv.org/abs/2007.00998}{{\ttfamily arXiv:2007.00998 [hep-th]}}.

\bibitem{Fan:2021bwt}
Y.~Fan and T.~G. Mertens, ``{From quantum groups to Liouville and dilaton quantum gravity},'' \href{http://dx.doi.org/10.1007/JHEP05(2022)092}{{\em JHEP} {\bfseries 05} (2022) 092}, \href{http://arxiv.org/abs/2109.07770}{{\ttfamily arXiv:2109.07770 [hep-th]}}.

\bibitem{Kyono:2017pxs}
H.~Kyono, S.~Okumura, and K.~Yoshida, ``{Comments on 2D dilaton gravity system with a hyperbolic dilaton potential},'' \href{http://dx.doi.org/10.1016/j.nuclphysb.2017.07.013}{{\em Nucl. Phys. B} {\bfseries 923} (2017) 126--143}, \href{http://arxiv.org/abs/1704.07410}{{\ttfamily arXiv:1704.07410 [hep-th]}}.

\bibitem{Mertens:2025ydx}
T.~G. Mertens and Q.-F. Wu, ``{Minimal Factorization of Chern-Simons Theory -- Gravitational Anyonic Edge Modes},'' \href{http://arxiv.org/abs/2505.00501}{{\ttfamily arXiv:2505.00501 [hep-th]}}.

\bibitem{Blommaert:2018oro}
A.~Blommaert, T.~G. Mertens, and H.~Verschelde, ``{The Schwarzian Theory - A Wilson Line Perspective},'' \href{http://dx.doi.org/10.1007/JHEP12(2018)022}{{\em JHEP} {\bfseries 12} (2018) 022}, \href{http://arxiv.org/abs/1806.07765}{{\ttfamily arXiv:1806.07765 [hep-th]}}.

\bibitem{Iliesiu:2019xuh}
L.~V. Iliesiu, S.~S. Pufu, H.~Verlinde, and Y.~Wang, ``{An exact quantization of Jackiw-Teitelboim gravity},'' \href{http://dx.doi.org/10.1007/JHEP11(2019)091}{{\em JHEP} {\bfseries 11} (2019) 091}, \href{http://arxiv.org/abs/1905.02726}{{\ttfamily arXiv:1905.02726 [hep-th]}}.

\bibitem{Faddeev:1999fe}
L.~D. Faddeev, ``{Modular double of quantum group},'' in {\em {Conference Moshe Flato}}, pp.~149--156.
\newblock 2000.
\newblock \href{http://arxiv.org/abs/math/9912078}{{\ttfamily arXiv:math/9912078}}.

\bibitem{de2003integral}
A.~De~Sole and V.~Kac, ``On integral representations of q-gamma and q-beta functions,'' \href{http://arxiv.org/abs/math/0302032}{{\ttfamily arXiv:math/0302032 [math.QA]}}.

\bibitem{KC02}
V.~Kac and P.~Cheung, {\em Quantum calculus}.
\newblock Universitext. Springer-Verlag, New York, 2002.

\bibitem{Klimyk:1997eb}
A.~Klimyk and K.~Schmudgen, {\em {Quantum groups and their representations}}.
\newblock 1997.

\bibitem{Ponsot:1999uf}
B.~Ponsot and J.~Teschner, ``{Liouville bootstrap via harmonic analysis on a noncompact quantum group},'' \href{http://arxiv.org/abs/hep-th/9911110}{{\ttfamily arXiv:hep-th/9911110}}.

\bibitem{Ponsot:2000mt}
B.~Ponsot and J.~Teschner, ``{Clebsch-Gordan and Racah-Wigner coefficients for a continuous series of representations of U(q)(sl(2,R))},'' \href{http://dx.doi.org/10.1007/PL00005590}{{\em Commun. Math. Phys.} {\bfseries 224} (2001) 613--655}, \href{http://arxiv.org/abs/math/0007097}{{\ttfamily arXiv:math/0007097}}.

\bibitem{Kharchev:2001rs}
S.~Kharchev, D.~Lebedev, and M.~Semenov-Tian-Shansky, ``{Unitary representations of $U_q (\mathfrak{sl}(2, \mathbb{R}))$, the modular double, and the multiparticle q deformed Toda chains},'' \href{http://dx.doi.org/10.1007/s002200100592}{{\em Commun. Math. Phys.} {\bfseries 225} (2002) 573--609}, \href{http://arxiv.org/abs/hep-th/0102180}{{\ttfamily arXiv:hep-th/0102180}}.

\bibitem{Burban:1992ys}
I.~M. Burban and A.~U. Klimyk, ``{On representations of the quantum algebra $U_q(su(1,1))$},'' {\em J. Phys. A} {\bfseries 26} (1993) 2139--2152.

\bibitem{groenevelt}
W.~Groenevelt, ``Wilson function transforms related to racah coefficients,'' \href{http://arxiv.org/abs/math/0501511}{{\ttfamily arXiv:math/0501511 [math.CA]}}.

\bibitem{Fan:2021wsb}
Y.~Fan and T.~G. Mertens, ``{Supergroup structure of Jackiw-Teitelboim supergravity},'' \href{http://dx.doi.org/10.1007/JHEP08(2022)002}{{\em JHEP} {\bfseries 08} (2022) 002}, \href{http://arxiv.org/abs/2106.09353}{{\ttfamily arXiv:2106.09353 [hep-th]}}.

\bibitem{Belaey:2024dde}
A.~Belaey, F.~Mariani, and T.~G. Mertens, ``{Gravitational wavefunctions in JT supergravity},'' \href{http://dx.doi.org/10.1007/JHEP10(2024)037}{{\em JHEP} {\bfseries 10} (2024) 037}, \href{http://arxiv.org/abs/2405.09289}{{\ttfamily arXiv:2405.09289 [hep-th]}}.

\bibitem{Vilenkin}
N.~Y. Vilenkin and A.~U. Klimyk, ``{Representation of Lie groups and Special Functions: Volume 1},'' {\em Kluwer Academic Publishers} (1991) .

\bibitem{Bytsko:2002br}
A.~G. Bytsko and J.~Teschner, ``{R operator, coproduct and Haar measure for the modular double of U(q)(sl(2,R))},'' \href{http://dx.doi.org/10.1007/s00220-003-0894-5}{{\em Commun. Math. Phys.} {\bfseries 240} (2003) 171--196}, \href{http://arxiv.org/abs/math/0208191}{{\ttfamily arXiv:math/0208191}}.

\bibitem{Bytsko:2006ut}
A.~G. Bytsko and J.~Teschner, ``{Quantization of models with non-compact quantum group symmetry: Modular XXZ magnet and lattice sinh-Gordon model},'' \href{http://dx.doi.org/10.1088/0305-4470/39/41/S11}{{\em J. Phys. A} {\bfseries 39} (2006) 12927--12981}, \href{http://arxiv.org/abs/hep-th/0602093}{{\ttfamily arXiv:hep-th/0602093}}.

\bibitem{Coquereaux:1999va}
R.~Coquereaux, A.~O. Garcia, and R.~Trinchero, ``{Hopf stars, twisted Hopf stars and scalar products on quantum spaces},'' \href{http://dx.doi.org/10.1016/S0393-0440(00)00013-9}{{\em J. Geom. Phys.} {\bfseries 36} (2000) 22--59}, \href{http://arxiv.org/abs/math-ph/9904037}{{\ttfamily arXiv:math-ph/9904037}}.

\bibitem{Buffenoir:1999}
E.~Buffenoir and P.~Roche, ``Harmonic analysis on the quantum lorentz group,'' \href{https://doi.org/10.1007/s002200050736}{{\em Communications in Mathematical Physics} {\bfseries 207} no.~3, (Nov, 1999) 499--555}.

\bibitem{Majid:1992bz}
S.~Majid, ``{The Quantum double as quantum mechanics},'' \href{http://arxiv.org/abs/hep-th/9210044}{{\ttfamily arXiv:hep-th/9210044}}.

\bibitem{Ip:2013}
I.~C.-H. Ip, ``{Representation of the quantum plane, its quantum double and harmonic analysis on $GL_q^+(2,R)$},'' \href{http://dx.doi.org/10.1007/s00029-012-0112-4}{{\em Selecta Mathematica New Series} {\bfseries Vol 19 (4)} (2013) 987--1082}, \href{http://arxiv.org/abs/1108.5365}{{\ttfamily arXiv:1108.5365 [math.QA]}}.

\bibitem{Fronsdal:1991gf}
C.~Fronsdal and A.~Galindo, ``{The Dual of a quantum group},'' \href{http://dx.doi.org/10.1007/BF00739590}{{\em Lett. Math. Phys.} {\bfseries 27} (1993) 59--72}.

\bibitem{Belaey:2023jtr}
A.~Belaey, F.~Mariani, and T.~G. Mertens, ``{Branes in JT (super)gravity from group theory},'' \href{http://dx.doi.org/10.1007/JHEP02(2024)058}{{\em JHEP} {\bfseries 02} (2024) 058}, \href{http://arxiv.org/abs/2310.04245}{{\ttfamily arXiv:2310.04245 [hep-th]}}.

\bibitem{Okuyama:2023byh}
K.~Okuyama, ``{End of the world brane in double scaled SYK},'' \href{http://dx.doi.org/10.1007/JHEP08(2023)053}{{\em JHEP} {\bfseries 08} (2023) 053}, \href{http://arxiv.org/abs/2305.12674}{{\ttfamily arXiv:2305.12674 [hep-th]}}.

\bibitem{Mertens:2019tcm}
T.~G. Mertens and G.~J. Turiaci, ``{Defects in Jackiw-Teitelboim Quantum Gravity},'' \href{http://dx.doi.org/10.1007/JHEP08(2019)127}{{\em JHEP} {\bfseries 08} (2019) 127}, \href{http://arxiv.org/abs/1904.05228}{{\ttfamily arXiv:1904.05228 [hep-th]}}.

\bibitem{Alvarez-Gaume:1988izd}
L.~Alvarez-Gaume, C.~Gomez, and G.~Sierra, ``{Quantum Group Interpretation of Some Conformal Field Theories},'' \href{http://dx.doi.org/10.1016/0370-2693(89)90027-0}{{\em Phys. Lett. B} {\bfseries 220} (1989) 142--152}.

\bibitem{Alvarez-Gaume:1989blj}
L.~Alvarez-Gaume, C.~Gomez, and G.~Sierra, ``{Duality and Quantum Groups},'' \href{http://dx.doi.org/10.1016/0550-3213(90)90116-U}{{\em Nucl. Phys. B} {\bfseries 330} (1990) 347--398}.

\bibitem{Mudrov2007}
A.~Mudrov, ``Quantum conjugacy classes of simple matrix groups,'' \href{https://doi.org/10.1007/s00220-007-0222-6}{{\em Communications in Mathematical Physics} {\bfseries 272} no.~3, (Jun, 2007) 635--660}.

\bibitem{Isaev:1991pg}
A.~P. Isaev and Z.~Popowicz, ``{q trace for the quantum groups and q deformed Yang-Mills theory},'' \href{http://dx.doi.org/10.1016/0370-2693(92)91140-5}{{\em Phys. Lett. B} {\bfseries 281} (1992) 271--278}.

\bibitem{Blommaert:2021etf}
A.~Blommaert and M.~Usatyuk, ``{Microstructure in matrix elements},'' \href{http://dx.doi.org/10.1007/JHEP09(2022)070}{{\em JHEP} {\bfseries 09} (2022) 070}, \href{http://arxiv.org/abs/2108.02210}{{\ttfamily arXiv:2108.02210 [hep-th]}}.

\bibitem{Mertens:2022aou}
T.~G. Mertens, ``{Quantum exponentials for the modular double and applications in gravity models},'' \href{http://dx.doi.org/10.1007/JHEP09(2023)106}{{\em JHEP} {\bfseries 09} (2023) 106}, \href{http://arxiv.org/abs/2212.07696}{{\ttfamily arXiv:2212.07696 [hep-th]}}.

\bibitem{Donnelly:2018ppr}
W.~Donnelly and G.~Wong, ``{Entanglement branes, modular flow, and extended topological quantum field theory},'' \href{http://dx.doi.org/10.1007/JHEP10(2019)016}{{\em JHEP} {\bfseries 10} (2019) 016}, \href{http://arxiv.org/abs/1811.10785}{{\ttfamily arXiv:1811.10785 [hep-th]}}.

\bibitem{Hung:2019bnq}
L.~Y. Hung and G.~Wong, ``{Entanglement branes and factorization in conformal field theory},'' \href{http://dx.doi.org/10.1103/PhysRevD.104.026012}{{\em Phys. Rev. D} {\bfseries 104} no.~2, (2021) 026012}, \href{http://arxiv.org/abs/1912.11201}{{\ttfamily arXiv:1912.11201 [hep-th]}}.

\bibitem{Donnelly:2016jet}
W.~Donnelly and G.~Wong, ``{Entanglement branes in a two-dimensional string theory},'' \href{http://dx.doi.org/10.1007/JHEP09(2017)097}{{\em JHEP} {\bfseries 09} (2017) 097}, \href{http://arxiv.org/abs/1610.01719}{{\ttfamily arXiv:1610.01719 [hep-th]}}.

\bibitem{Donnelly:2020teo}
W.~Donnelly, Y.~Jiang, M.~Kim, and G.~Wong, ``{Entanglement entropy and edge modes in topological string theory. Part I. Generalized entropy for closed strings},'' \href{http://dx.doi.org/10.1007/JHEP10(2021)201}{{\em JHEP} {\bfseries 10} (2021) 201}, \href{http://arxiv.org/abs/2010.15737}{{\ttfamily arXiv:2010.15737 [hep-th]}}.

\bibitem{Kulish:1989sv}
P.~P. Kulish and N.~Y. Reshetikhin, ``{Universal R matrix of the quantum superalgebra osp(2 | 1)},'' \href{http://dx.doi.org/10.1007/BF00401868}{{\em Lett. Math. Phys.} {\bfseries 18} (1989) 143--149}.

\bibitem{Hadasz:2013bwa}
L.~Hadasz, M.~Pawelkiewicz, and V.~Schomerus, ``{Self-dual Continuous Series of Representations for $\mathcal{U}_q(sl(2))$ and $\mathcal{U}_q(osp(1|2))$},'' \href{http://dx.doi.org/10.1007/JHEP10(2014)091}{{\em JHEP} {\bfseries 10} (2014) 091}, \href{http://arxiv.org/abs/1305.4596}{{\ttfamily arXiv:1305.4596 [hep-th]}}.

\bibitem{Pawelkiewicz:2013wga}
M.~Pawelkiewicz, V.~Schomerus, and P.~Suchanek, ``{The universal Racah-Wigner symbol for U$_q$(osp(1|2))},'' \href{http://dx.doi.org/10.1007/JHEP04(2014)079}{{\em JHEP} {\bfseries 04} (2014) 079}, \href{http://arxiv.org/abs/1307.6866}{{\ttfamily arXiv:1307.6866 [hep-th]}}.

\bibitem{Berkooz:2020xne}
M.~Berkooz, N.~Brukner, V.~Narovlansky, and A.~Raz, ``{The double scaled limit of Super--Symmetric SYK models},'' \href{http://dx.doi.org/10.1007/JHEP12(2020)110}{{\em JHEP} {\bfseries 12} (2020) 110}, \href{http://arxiv.org/abs/2003.04405}{{\ttfamily arXiv:2003.04405 [hep-th]}}.

\bibitem{Saad:2019lba}
P.~Saad, S.~H. Shenker, and D.~Stanford, ``{JT gravity as a matrix integral},'' \href{http://arxiv.org/abs/1903.11115}{{\ttfamily arXiv:1903.11115 [hep-th]}}.

\bibitem{Boruch:2024kvv}
J.~Boruch, L.~V. Iliesiu, G.~Lin, and C.~Yan, ``{How the Hilbert space of two-sided black holes factorises},'' \href{http://arxiv.org/abs/2406.04396}{{\ttfamily arXiv:2406.04396 [hep-th]}}.

\bibitem{stovicek2000}
P.~Stovicek, ``Discrete series of representations for $u_q(\frak{sl}(2,\mathbb{R}))$,'' \href{http://arxiv.org/abs/math/0008111}{{\ttfamily arXiv:math/0008111 [math.QA]}}.

\end{thebibliography}\endgroup
\end{document}